%% file: paper_main.tex
\documentclass[acmtog]{acmart}
\usepackage{textcomp}

\usepackage[noend]{algorithmic}
\usepackage{algorithm}
\usepackage{subcaption}
\usepackage{multirow}

\usepackage[titletoc,title]{appendix}
\usepackage[T1]{fontenc}
\usepackage{enumitem}

\usepackage{soul}

\usepackage{tikz}
\usetikzlibrary{decorations.pathreplacing}
\usetikzlibrary{arrows}

\newcommand\widgetone[2]{
\draw[fill=red!40] (#1,#2) rectangle (#1+1.000000,#2+1.000000);
\draw[very thick, black] (#1+0.500000,#2+0.000000) -- (#1+0.500000,#2+1.000000);
\draw[fill=yellow!30] (#1+1.000000,#2+0.000000) rectangle (#1+2.000000,#2+1.000000);
\draw[very thick, black] (#1+1.000000,#2+0.500000) -- (#1+2.000000,#2+0.500000);
\draw[fill=yellow!30] (#1+0.000000,#2+1.000000) rectangle (#1+1.000000,#2+2.000000);
\draw[very thick, black] (#1+0.000000,#2+1.500000) -- (#1+1.000000,#2+1.500000);
\draw[fill=red!40] (#1+1.000000,#2+1.000000) rectangle (#1+2.000000,#2+2.000000);
\draw[very thick, black] (#1+1.500000,#2+1.000000) -- (#1+1.500000,#2+2.000000);
}

\newcommand\widgettwo[2]{
\draw[fill=yellow!30] (#1+0.000000,#2+0.000000) rectangle (#1+1.000000,#2+1.000000);
\draw[very thick, black] (#1+0.000000,#2+0.500000) -- (#1+1.000000,#2+0.500000);
\draw[fill=red!40] (#1+1.000000,#2+0.000000) rectangle (#1+2.000000,#2+1.000000);
\draw[very thick, black] (#1+1.500000,#2+0.000000) -- (#1+1.500000,#2+1.000000);

\draw[fill=red!40] (#1+0.000000,#2+1.000000) rectangle (#1+1.000000,#2+2.000000);
\draw[very thick, black] (#1+0.500000,#2+1.000000) -- (#1+0.500000,#2+2.000000);

\draw[fill=yellow!30] (#1+1.000000,#2+1.000000) rectangle (#1+2.000000,#2+2.000000);
\draw[very thick, black] (#1+1.000000,#2+1.500000) -- (#1+2.000000,#2+1.500000);
}

\setcopyright{rightsretained}

\begin{document}

\title{Procedural Wang Tile Algorithm for Stochastic Wall Patterns}

\author{Alexandre Derouet-Jourdan}
\affiliation{%
	\institution{OLM Digital, Inc.}
	\streetaddress{Setagaya Ku, Wakabayashi 1-18-10}
	\city{Tokyo} 
	\state{Japan} 
	\postcode{154-0023}
}
\email{alex@olm.cp.jp}

\author{Marc Salvati}
\affiliation{%
	\institution{OLM Digital, Inc.}
	\streetaddress{Setagaya Ku, Wakabayashi 1-18-10}
	\city{Tokyo} 
	\state{Japan} 
	\postcode{154-0023}
}
\email{salvati.marc@olm.cp.jp}
\author{Theo Jonchier}
\affiliation{%
	\institution{OLM Digital, Inc.}
	\streetaddress{Setagaya Ku, Wakabayashi 1-18-10}
	\city{Tokyo} 
	\state{Japan} 
	\postcode{154-0023}
}
\email{jonchier@olm.cp.jp}

\begin{abstract}
	The game and movie industries always face the challenge of reproducing materials.
	This problem is tackled by combining illumination models and various textures (painted or procedural patterns).
	Generating stochastic wall patterns is crucial in the creation of a wide range of backgrounds (castles, temples, ruins...).
	A specific Wang tile set was introduced previously to tackle this problem, in a non-procedural fashion. 
	Long lines may appear as visual artifacts.
	We use this tile set in a new procedural algorithm to generate stochastic wall patterns.
	For this purpose, we introduce specific hash functions implementing a constrained Wang tiling.
	This technique makes possible the generation of boundless textures while giving control over the maximum line length.
	The algorithm is simple and easy to implement, and the wall structure we get from the tiles allows to achieve visuals that reproduce all the small details of artist painted walls.
\end{abstract}


%
%

\ccsdesc[500]{Computing methodologies~Texturing}

\keywords{Procedural methods, Texture synthesis, Wang Tiling, Structured patterns}

\begin{teaserfigure}
	\centering
	\begin{subfigure}{0.40\linewidth}
		\centering
	\includegraphics[height=1.5in]{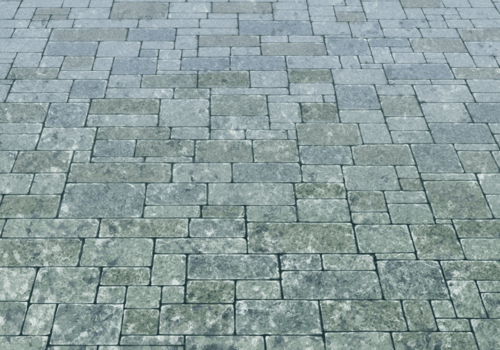}
	\caption{}
	\end{subfigure}
	\begin{subfigure}{0.40\linewidth}
		\centering
	\includegraphics[height=1.5in]{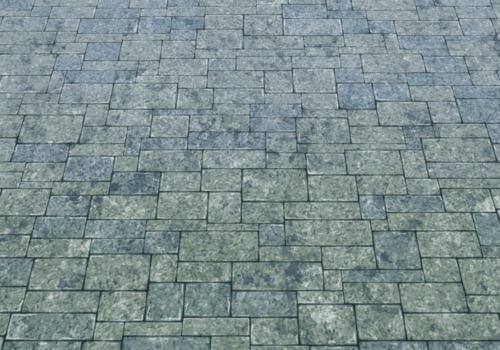}
	\caption{}
	\end{subfigure}
	\caption{Wall painted by an artist (a). Stochastic wall generated using our procedural algorithm (b). Our algorithm generates a line structure similar to the one created by the artist. This structure is used to generate brick colors and details around the edges. The global material appearance has been generated using texture bombing techniques. }
	\label{fig:wangtileResults} 
\end{teaserfigure}

\maketitle

\section{Introduction}
The final look in movies and games is the result of the combination of textures, 3D models and illumination models (BSDF).
It is usually more efficient to increase the level of details through a texture rather than directly into the 3D model.
With the continuous improvement in display technology (4k and 8k) and increase of CPU/GPU power, always higher resolution textures are required.
The cost of painting textures by hand is then increasing.
Generating textures at render time that preserve the organic feeling of hand painted ones is the challenge that all shading artists face everyday.
They typically combine noises and patterns (fractal, Perlin, Gabor, cellular, flakes, Voronoi...) with some well designed BSDF to re-create a material appearance.

\begin{figure}[htb]
	\centering
	\begin{subfigure}{0.45\linewidth}
	\centering	
	\includegraphics[width=\linewidth]{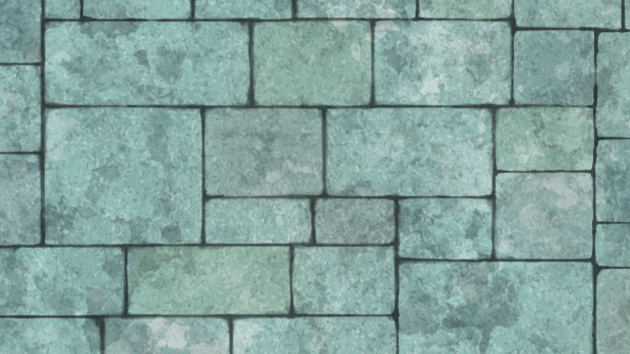}
	\caption{}
	\end{subfigure}
	\begin{subfigure}{0.45\linewidth}
	\centering	
	\includegraphics[width=\linewidth]{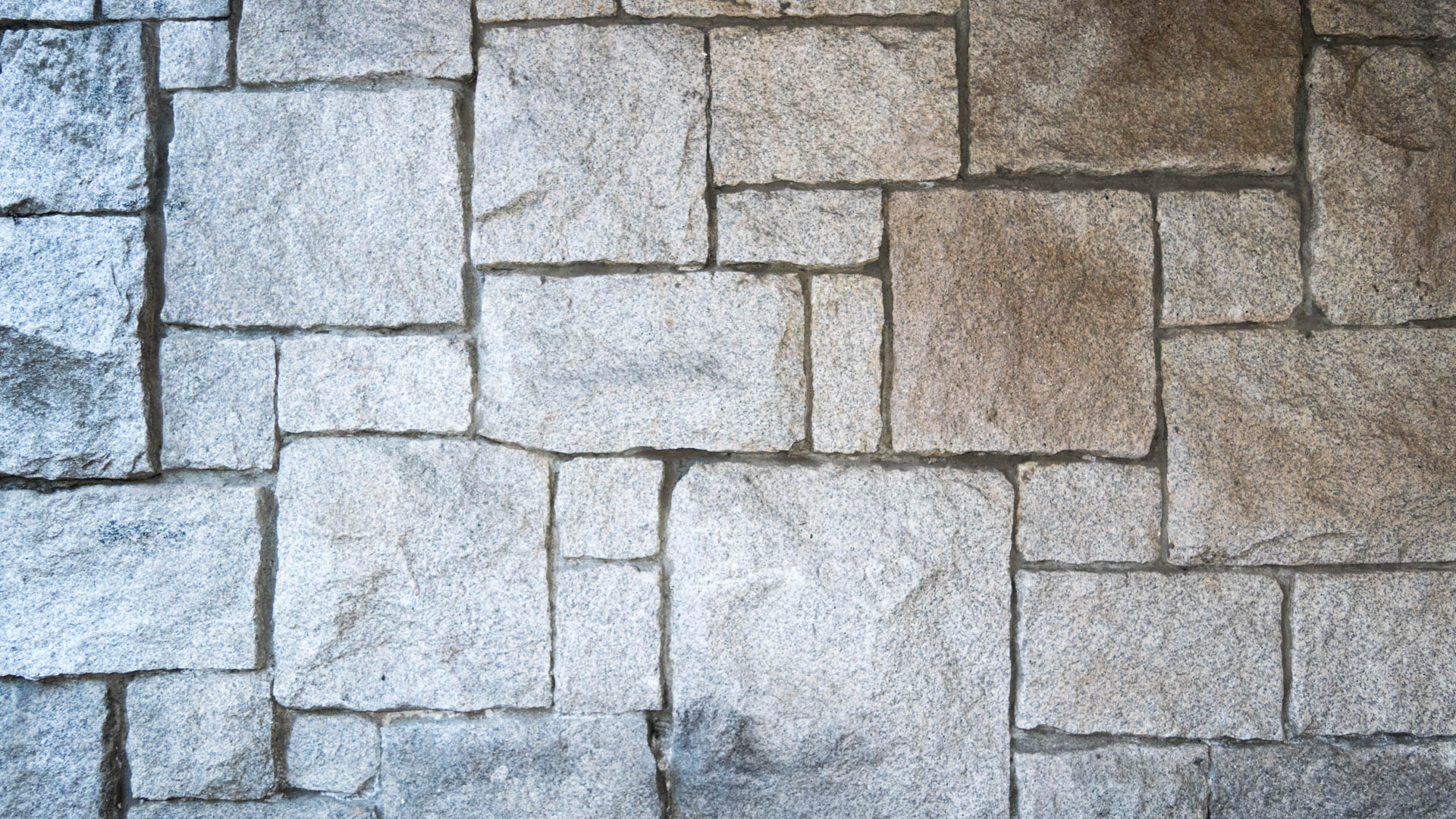}
	\caption{}
	\end{subfigure}
	\caption{\label{fig:WallArtist}  Examples of stochastic walls. (a) Painting. (b) Photography.}
\end{figure}

\begin{figure}[htb]
	\centering
\includegraphics[width=0.45\linewidth]{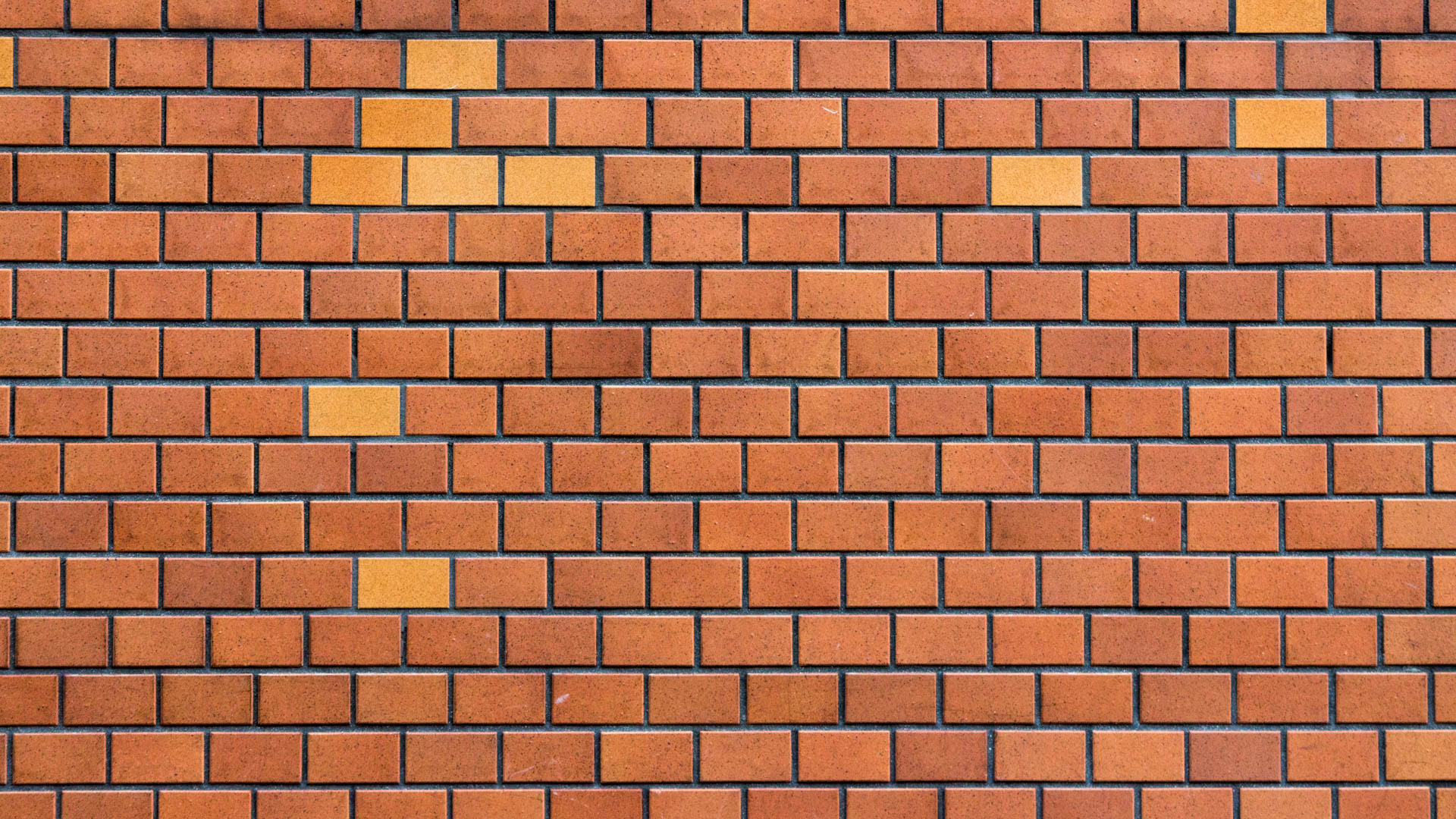}
\includegraphics[width=0.45\linewidth]{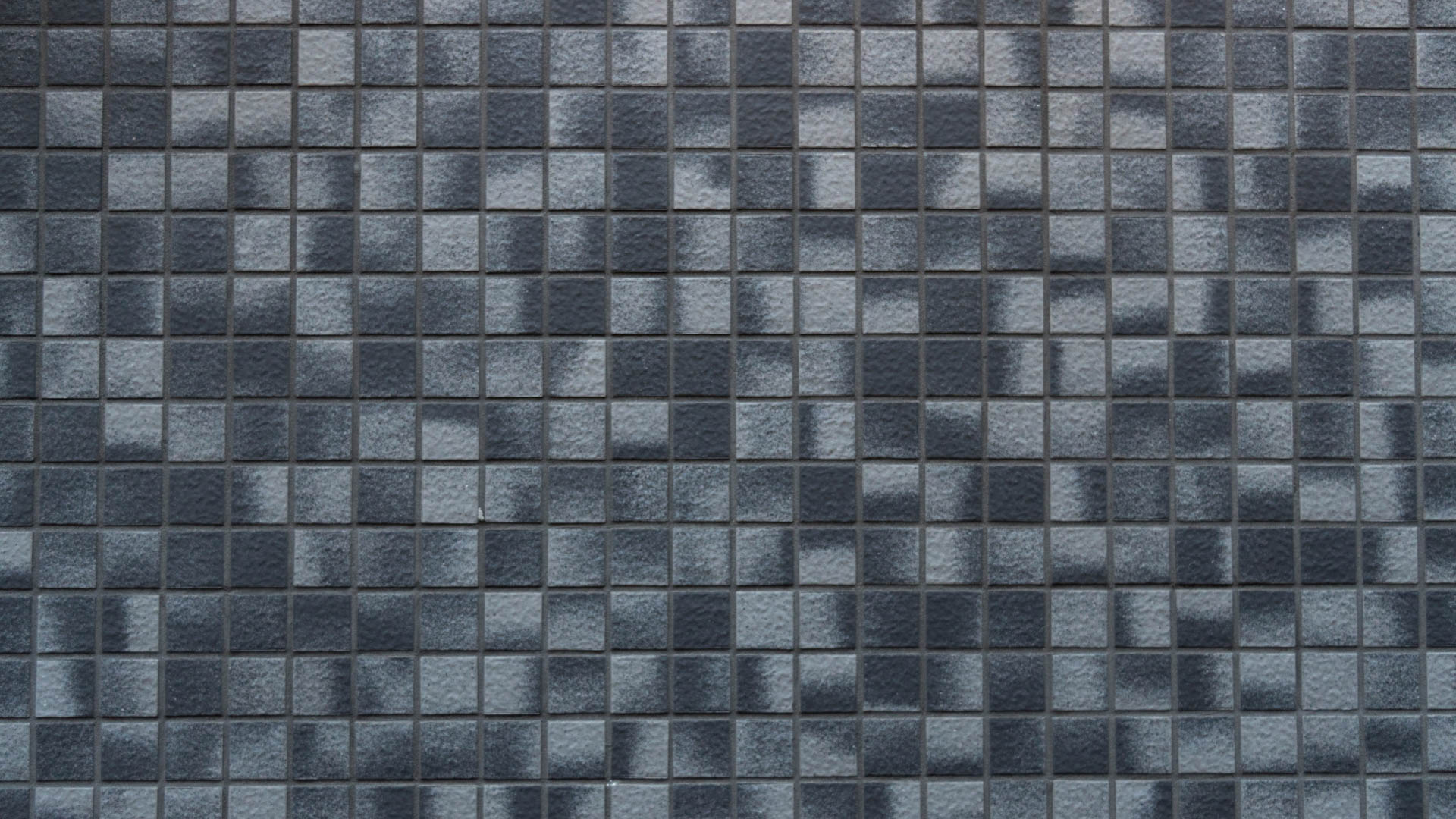}
\caption{\label{fig:WallRegularPhoto} Photographies of regular wall patterns. Left: alternate pattern. Right:square pattern.}
\end{figure}

Every movie and cartoon involves the creation of patterns to generate textures for backgrounds.
In this context, we face the problem of generating stochastic wall patterns for stone walls and paved grounds as shown in Figure~\ref{fig:WallArtist}.
These patterns appear in various constructions such as castles, temples, ruins...
They are different from a regular wall pattern with unique size of bricks (see Figure~\ref{fig:WallRegularPhoto}) in many ways.

\begin{itemize}
\item No cross: pattern where 4 bricks share a corner.
\item No long lines: brick with edges aligned in sequences.
\item Irregular space between bricks.
\end{itemize}

Also, it is important to be able to generate unbounded textures that give total freedom on the wall space size. 
This can be achieved by using fully procedural techniques (defined on a boundless domain), i.e aperiodic, parameterized, random-accessible and compact function as defined in \cite{Lagae2010}.
Also procedural method are GPU friendly.
On the other hand non-procedural approaches can generate unbounded textures but only by using periodic patterns and then producing visual artifacts.

Wang tiles have become a standard tool in texture synthesis.
They have the advantage of producing stochastic patterns, while providing access to the structure of the pattern.
This feature is crucial to provide control over the rendering and shading of the structure.
Also, they already have been used in procedural algorithms \cite{Lagae2005}.

In this paper we use the model of Wang tile from \cite{Derouet-Jourdan2015} that has been used to create walls with no cross patterns, albeit using non-procedural algorithm and generating long lines artifacts. 
We create a new and simple yet general procedural algorithm that generates stochastic wall patterns while avoiding cross patterns.
In addition, our algorithm provides full control over the maximum length of the lines.
We design a procedural solution to the dappling problem introduced in \cite{Kaji2016}, that produces better distribution of long lines.
The procedural nature of the algorithm makes its GPU implementation straightforward (we provide one as additional material to the present paper).

After an overview of the related work, we explain the Wang tile model we use and the limitations of current algorithms. 
We then introduce the procedural algorithm. 
Finally we present some result analysis and discussion about non-procedural flavor of the pattern generation. 
We compare our results with state of the art texture synthesis methods and another wall generation algorithm before concluding about the future work.

Although we illustrate our method with the reproduction of hand painted walls with cartoon style,
there is no restrictions to use our stochastic wall structure to render photo-realistic walls.

\section{Related work}
\label{section:relatedWork}
To create material appearance, artists use a combination of raster textures and procedural textures.
There is no limit but the skill of the artist to what he may paint in a raster texture.
Because of the need of very high resolution textures, multiple methods have been proposed to generate them.

Texture bombing \cite{Glanville2004} has been used successfully \cite{Salvati2014} to generate on the fly the equivalent of  100k textures while preserving the hand painted feeling.
However, such a method cannot reproduce highly structured patterns such as wall patterns.

Texture synthesis generates large size textures based on an exemplar of limited size \cite{Diamanti2015,Kwatra2003,Lefebvre2006,Yeh2013}.
Those methods aggregate patches of the exemplar locally, satisfying local continuity of the structure in the larger texture.
Global constraints can be added on top of that to reproduce more advanced structure of the material \cite{Ramanarayanan2007}.
Large textures can be generated on the fly \cite{Vanhoey2013}, without storing at any time the result in memory.
However, the control over the structure is limited.
It remains hard to control locally the features. 
For example it would be very hard to control the color of individual bricks or the space between bricks of a wall pattern produced by such a method.
The reason is that even if the visual structure is reproduced, the algorithm does not have the knowledge of the underlying structure of the pattern.

On the other hand, procedural textures generate the pattern from a small set of parameters.
Procedural textures are algorithms devised to generate the specific structure of patterns (cellular/Voronoi \cite{Worley1996}, regular bricks, star shapes, various noises \cite{Ebert2003}).
We did not find any algorithm in the existing literature to reproduce procedurally the stochastic wall patterns. 

Some dedicated techniques have been proposed \cite{Legakis2001,Miyata1990} to create wall patterns.
However the control over the size of the bricks and the occurrence of cross and long lines remain an issue.
Also the iterative nature of those methods prevents a procedural implementation.

\begin{figure}[htb]
\centering
	\input{tile}
	\caption{\label{fig:Wang_tile} Model of Wang tile. In a valid tiling, the color of shared edge is the same for the two tiles sharing the edge. The tile $(i,j)$ is constituted of the colors $H_{i,j+1}$, $V_{i,j}$, $H_{i,j}$ and $V_{i+1,j}$}
\end{figure}
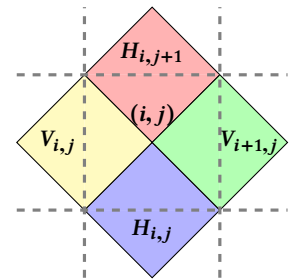

We have been inspired by a specific technique called Wang tiles \cite{Wang1961}. 
Wang tiles have been introduced in computer graphics to generate aperiodic stochastic textures \cite{Cohen2003,Stam1997}.
Wang tiles are square tiles with colors on the edges as shown in Figure~\ref{fig:Wang_tile}. 
Tiles are placed edges to edges in the tiling space.
A tiling is valid when every two tiles sharing an edge have the same color on this edge.
We identify the problem of tiling to a problem of edge coloring like in \cite{Lagae2005}.
In that configuration, for a tile in $(i,j)$, we denote $H_{i,j+1}$, $V_{i,j}$, $H_{i,j}$ and $V_{i+1,j}$ the top, left, bottom and right edges, as well as their colors.
Wang tiles have the characteristics of providing local continuity on the edges of the tiles.
Due to their local nature, it is possible to create procedural functions to build textures on the fly based on Wang tiles \cite{Schloemer2010,Wei2004}.
However all those methods rely on the use of complete Wang tile sets.

Wang tiles have already been used to generate non-procedural and repeatable wall brick patterns \cite{Derouet-Jourdan2015}.
To avoid visual artifacts like cross in the visual result, they use a reduced tile set.
However they do not account for long lines.
Dappling algorithms solve 2-colorization of grid with constraints on the number of consecutive same color grid cells.
Each color represents horizontal/vertical tiles of our model (as detailed in \ref{subsection:WangTileModel}).
By combining Wang tiles with dappling coloring of \cite{Kaji2016}, it is possible to design a non-procedural algorithm to generate repeatable wall patterns without long lines (as we show in Appendix \ref{appendix:nonProcedural}).

\section{Walls with Wang tiles}
\label{section:model}
In this section we describe how we use Wang tiles to generate the visual goal.
First we recall the Wang tile model as introduced in \cite{Derouet-Jourdan2015} and then we explain how to use the generated structure to render the walls.

\subsection{Wang tile model}
\label{subsection:WangTileModel}
The stochastic wall patterns we generate consist of a set of various size of rectangular bricks.
They don't contain cross patterns as these break the randomness and attract the eye.
That is the reason why when painting by hands, artists avoid such patterns (see Figure~\ref{fig:WallArtist}~(a)).

\begin{figure}[htb]
	\centering
	\input{tile_in_wall}
	\caption{\label{fig:WangTileModel} Modeling a wall pattern with Wang tiles. Tile colors are mapped to the bricks edges positions.}
\end{figure}
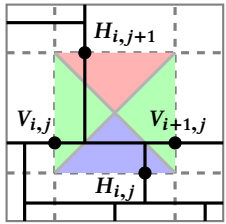

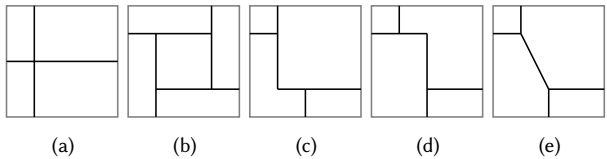
\begin{figure}[htb]
	\centering
	\begin{subfigure}{0.18\linewidth}
		\resizebox{\linewidth}{!}{\input{invalid_tile_cross}}
		\caption{}	
	\end{subfigure}
	\begin{subfigure}{0.18\linewidth}
		\resizebox{\linewidth}{!}{\input{invalid_tile_small_brick}}
		\caption{}	
	\end{subfigure}
	\begin{subfigure}{0.18\linewidth}
		\resizebox{\linewidth}{!}{\input{invalid_tile_bottom}}
		\caption{}	
	\end{subfigure}
	\begin{subfigure}{0.18\linewidth}
		\resizebox{\linewidth}{!}{\input{invalid_tile_top}}
		\caption{}	
	\end{subfigure}
	\begin{subfigure}{0.18\linewidth}
		\resizebox{\linewidth}{!}{\input{invalid_tile_diagonal}}
		\caption{}	
	\end{subfigure}
	\caption{\label{fig:excludedTiles} The tiles removed from the Wang tile set. (a) cross tiles ($H_{i,j+1} = H_{i,j}$ and $V_{i+1,j} = V_{i,j}$). (c) (d) (e) non rectangular tiles/(b) extra brick tiles ($H_{i,j+1}\neq H_{i,j}$ and $V_{i+1,j}\neq V_{i,j}$).}
\end{figure}

\begin{figure}[htb]
	\centering
	\begin{minipage}{\linewidth}
		\centering
		\resizebox{0.4\linewidth}{!}{\input{orientation_vertical}}
	\end{minipage}
	\begin{minipage}{\linewidth}
		\centering
		\resizebox{0.4\linewidth}{!}{\input{orientation+horizontal}}
	\end{minipage}
	\caption{\label{fig:tileWeUse} Correspondence between the brick pattern(left), the Wang tile (center) and the  orientation (right). Top line describes horizontal tiles, and bottom line vertical ones.}
\end{figure}
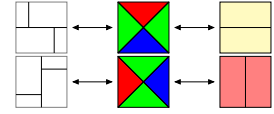

Expressing stochastic wall patterns with Wang tiles is straightforward.
Each tile models the corner junctions of four bricks.
This is done by mapping the four colors of each tile to the bricks edges placement as shown in Figure~\ref{fig:WangTileModel}.
However to avoid non rectangular bricks (see Figure~\ref{fig:excludedTiles} (c) (d) (e)) and avoid cross patterns (see Figure~\ref{fig:excludedTiles} (a)), the tile set is reduced.
To enforce rectangular bricks and avoid cross patterns, only "vertical" and "horizontal" tiles are considered. These constraints are shown in Figure~\ref{fig:tileWeUse} and explicited in Equations \eqref{constraintsV} and \eqref{constraintsH} . 
This avoids all the tiles introducing non rectangular bricks (such that $H_{i,j+1}\neq H_{i,j}$ and $V_{i+1,j}\neq V_{i,j}$).

\begin{align}
	\label{constraintsV} \text{Vertical constraint } & H_{i,j+1}=H_{i,j} \text{ and } V_{i+1,j}\neq V_{i,j}\\
	\label{constraintsH} \text{Horizontal constraint } & H_{i,j+1}\neq H_{i,j} \text{ and } V_{i+1,j}= V_{i,j}	
\end{align}

\begin{figure}[htb]
	\centering
	\resizebox{0.7\linewidth}{!}{\input{tile_set}}
	\caption{\label{fig:fullTileSetWeUse} Tile set example with 3 colors, 36 possible tiles.}
\end{figure}
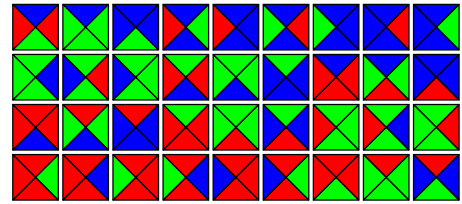

This also excludes tiles introducing an extra brick in the center (Figure~\ref{fig:excludedTiles} (b)). 
It would not be complicated to consider them, but they introduce tiny bricks that stand out.
For $n_c$ colors, we obtain a tile set of size $2*n_c^2*(n_c-1)$ (see Figure~\ref{fig:fullTileSetWeUse} for the tile set with 3 colors, 36 tiles)

\subsection{Rendering and shading}
\begin{figure}[htb]
 	\centering
 	\begin{subfigure}{0.99\linewidth} 
 			\centering		
 	\includegraphics[width=0.45\linewidth]{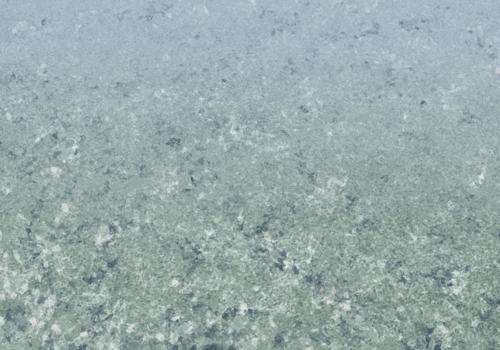}
  	\includegraphics[width=0.45\linewidth]{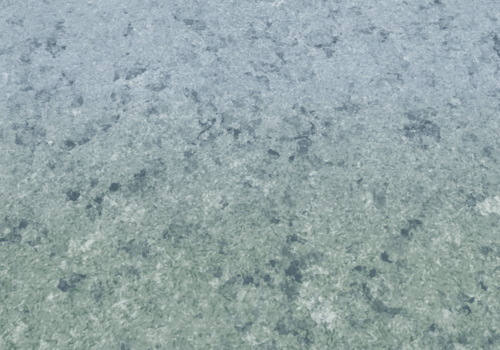}	
  	\caption{The global appearance: artist (left), texture bombing (right).}
  	\end{subfigure}
	\begin{subfigure}{0.99\linewidth} 
			\centering
  	\includegraphics[width=0.45\linewidth]{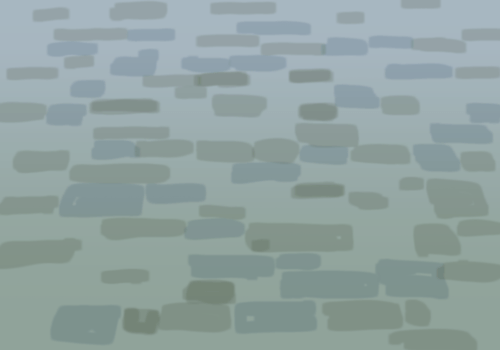}
   	\includegraphics[width=0.45\linewidth]{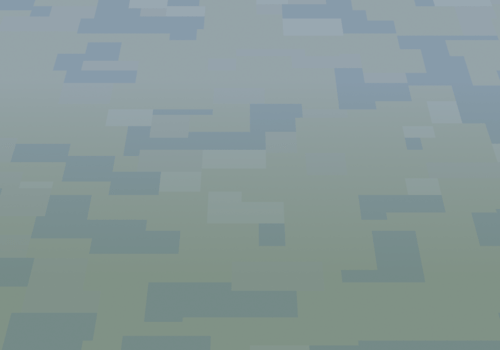}
  	\caption{The brick color painted: artist (left), from wall structure (right).}
	\end{subfigure}
	\begin{subfigure}{0.99\linewidth} 
			\centering
   	\includegraphics[width=0.45\linewidth]{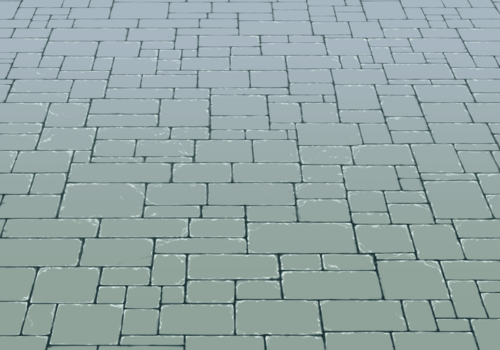}
   	\includegraphics[width=0.45\linewidth]{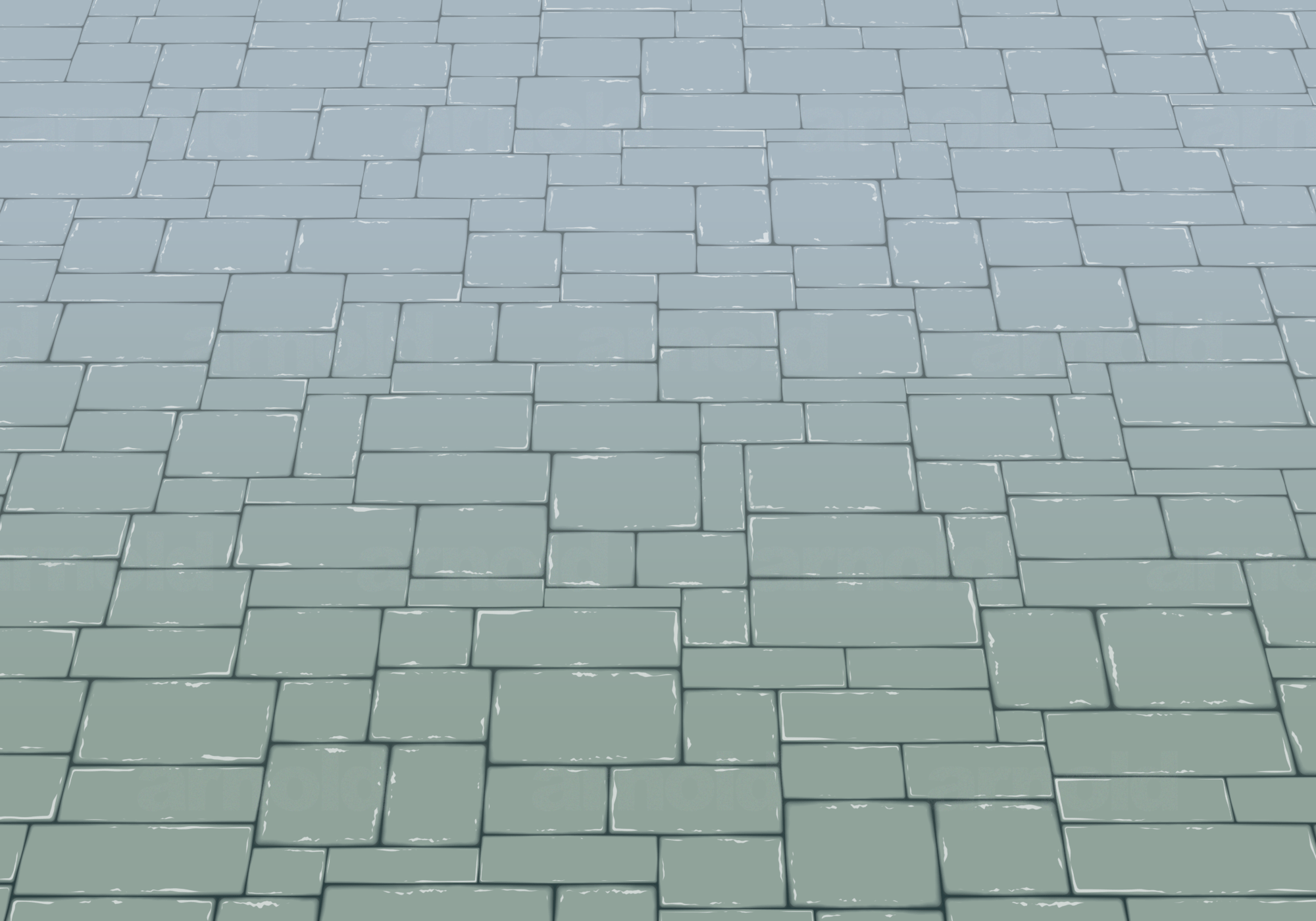}
  	\caption{The lines and edge highlights: artist (left), from brick parameterization (right).}
	\end{subfigure}
	\caption{\label{fig:cgAppearance}
		Comparison of the features of both artist painted wall (left) and the generated wall rendering (right).		
	}	
\end{figure}

In this section we focus on the visual appearance of the bricks, given a stochastic wall structure of rectangular bricks without cross patterns.
Observing the hand painted wall (see Figure~\ref{fig:WallArtist}), we notice the following features created by the artist:
\begin{itemize}
\item The global rock material.
\item Color variation for each brick.
\item Variable space between the bricks and corner roundness.
\item Highlights and scratches near the edges of the bricks.
\end{itemize}
In Figure~\ref{fig:cgAppearance}, we decompose the painting and our rendering in those features.
We then explain how we use the stochastic wall pattern to reproduce all those features.

\begin{figure}[htb]
	\centering
	\includegraphics[width=0.4455\linewidth]{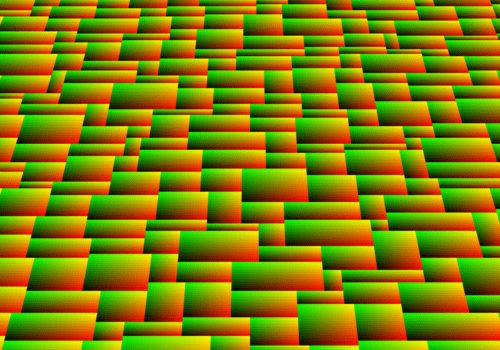}
	\includegraphics[width=0.4455\linewidth]{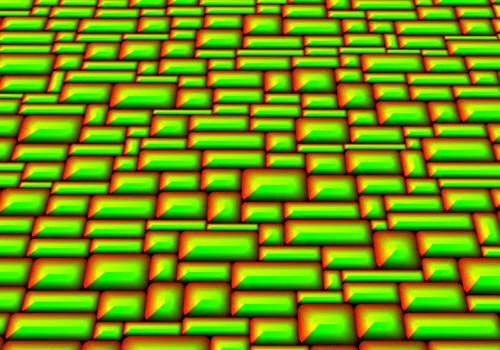}
	\caption{\label{fig:perBrickParameterization}UV parameterization of brick.  Cartesian (left),  polar parameterization with corner roundness (right).}
\end{figure}

Space between bricks, corner roundness and edge scratches are features localized near brick edges. 
Thanks to the brick structure we can generate a Cartesian parameterization as well as a polar parameterization that includes corner roundness (see Figure~\ref{fig:perBrickParameterization}).
From that structure, it's easy to generate normal maps to be used in a photo-realistic context.

Distance to brick edges can be combined with some noise function to create irregular space between bricks (see Figure~\ref{fig:cgAppearance} (c)). 
The same distance can also be used to localize and apply highlights and scratch textures (see Figure~\ref{fig:cgAppearance} (c)).
The color variation is simply enforced by generating a color based on the brick id or center position (see Figure~\ref{fig:cgAppearance} (b)). 
We can also use a texture map to determine this color.
As for the global material appearance it may be obtained by a combination of hand-painted and procedural textures. 
In our case we use texture bombing techniques as in the brush shader \cite{Salvati2014} to generate all those details while keeping the hand-painted touch (see Figure~\ref{fig:cgAppearance} (a)).

Without the wall/brick structure, such visual features would be really difficult to reproduce.

\section{Algorithm}
\label{section:procedural}

This section presents our procedural algorithm for wall pattern generation.
As seen in various procedural texture generation papers like~\cite{Worley1996}, we make a correspondence between a given request point $P(x,y)$ and an underlying "virtual" grid cell $(i,j)$ (for example integer part of scaled coordinates). 
In our algorithm, we build two functions $h'(i,j)$ and $v'(i,j)$ implementing procedural tiling while avoiding long lines:
\begin{equation*}
	H_{i,j} = h'(i,j)\quad V_{i,j} = v'(i,j).
\end{equation*}
In the following sections, we start by explaining the long line problem inherent to \cite{Derouet-Jourdan2015}.
We start by giving a general procedural solution Section \ref{section:solgeneral} that do not consider the long line problem. 
Then we explain how to build $h'$ and $v'$ on top of the general solution to avoid long lines procedurally in Section \ref{section:soldappling}.

\subsection{The long line problem}
\label{section:limitations}
\label{subsection:existingAlgorithms}
\begin{figure}[htb]
	\centering
	\resizebox{0.6\linewidth}{!}{\input{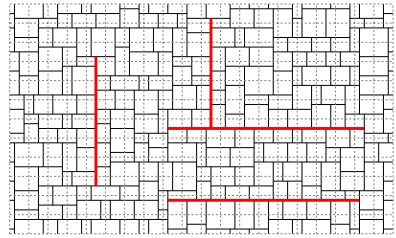}}
	\caption{\label{fig:LongLines} Long lines stand out and attract the eye in a wall pattern.}
\end{figure}

\begin{figure}[htb]
	\centering
	\resizebox{0.32\linewidth}{!}{\input{long_lines_border_3}}
	\resizebox{0.32\linewidth}{!}{\input{long_lines_border_5}}
	\resizebox{0.32\linewidth}{!}{\input{long_lines_border_10}}
	\caption{\label{fig:longLineArtifactSide} Notice the long lines on the side when the number of connections is increasing (from left to right: 3, 5, 10 connections).}
\end{figure}
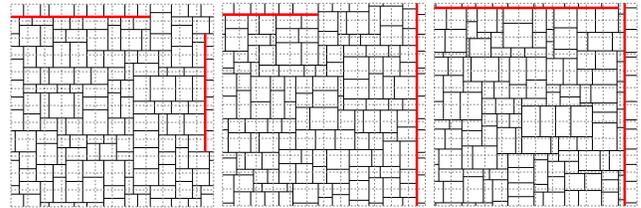

The Wang tile set we use has been introduced along with an algorithm to generate a repeatable wall pattern in~\cite{Derouet-Jourdan2015}.
However you can see long lines occurring in walls generated by this algorithm (see Figure~\ref{fig:LongLines} and \ref{fig:longLineArtifactSide}).

The reason of the occurrence of the long lines in Figure~\ref{fig:LongLines} is simple.
Tiles are separated in two categories: the vertical and horizontal one. 
So the probability to have an horizontal/vertical line of more than three tiles length is $1/2^3 =0.125$. 
The occurrence of long lines on the side in Figure~\ref{fig:longLineArtifactSide} is inherent to the sequential algorithm of \cite{Derouet-Jourdan2015}.
We can see that the length of the lines on the side is increasing with the number of connections. 
In a sequential algorithm, the last tile of a row is solved with 3 constraints.
With $n_c$ connections the probability that the last constraints match is roughly $1/n_c$.
This means that the probability the last tile of a row is horizontal is roughly $1/n_c$.
The more connections you have, the longer the vertical lines on the side are.
For the same reasons long horizontal lines occur on the top.
The stochastic variation of Wang tiling \cite{Kopf2006} does not suffer from this limitation.
However the stochastic nature of the algorithm makes difficult the control of the length of the straight lines shown in Figure~\ref{fig:LongLines}.
The dappling algorithm of \cite{Kaji2016} solves  the issue of long lines for non-repeatable walls. 
We adapted that method to generate repeatable walls by solving the Wang tile problem on top of a dappling with border constraint.
We use that algorithm as a general reference to compare to our procedural approach in Section \ref{section::results}.
Details of the algorithm can be found in Appendix \ref{appendix:nonProcedural}.

\subsection{General solution}
\label{section:solgeneral}
The general procedural tiling method is based on a property of our Wang tiles set, enunciated in \cite{Derouet-Jourdan2015}, that a $2\times 2$ square can always be tiled, for any boundary coloring.

The idea is to cut the grid into $2\times 2$ squares.
We need to determine the color of their outer edges.

For each cell $(i,j)$, we compute the bottom left cell index $(i', j')$, which identify a $2\times 2$ square, with
\begin{equation*}
i' = i - i \% 2\quad j' = j - j \% 2,
\end{equation*}
where $x \% y$ is the positive remainder in the Euclidean division of $x$ by $y$.

\begin{figure}[htb]
	\centering
	\resizebox{0.9\linewidth}{!}{\input{2x2_problem}}
	\caption{\label{fig:exampleSolution} Given a set of colors (a), an example solution (b).}
\end{figure}
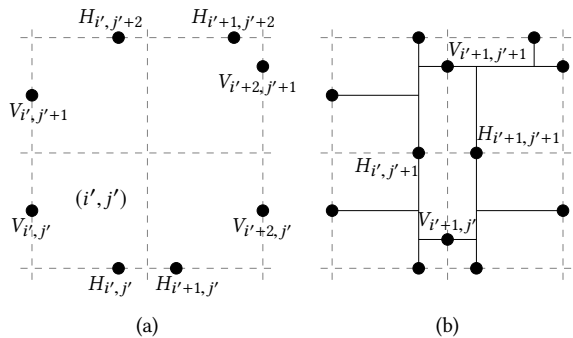

We use $h$ and $v$ to associate pseudo-random values to the outer edge color ($H_{i',j'+2}$, $H_{i', j'}$, $H_{i'+1,j'}$, $H_{i'+1, j'+2}$, $V_{i', j'+1}$, $V_{i', j'}$, $V_{i'+2, j'}$, $V_{i'+2, j'+1}$) as shown in Figure \ref{fig:exampleSolution} (a):
\begin{equation*}
	h(i,j)=\mathcal{H}(i,j)\% n_c\quad
	v(i,j)=\mathcal{V}(i,j)\% n_c,
\end{equation*}
 where $\mathcal{H}$ and $\mathcal{V}$ are hash function and associate "random" integer values to $(i,j)$.
In practice we use a FNV hash scheme~\cite{Fowler1991} to seed a xorshift random number generator from which we peek a value.

However to satisfy the constraints (Equations \eqref{constraintsV} and \eqref{constraintsH}) of our Wang tile model, we cannot use pseudo-random values for the inner edge.
We compute their colors using the result from \cite{Derouet-Jourdan2015}: we solve the interior of the $2\times 2$ square by considering the different cases on the border.
This gives us the colors $H_{i', j'+1}, H_{i'+1, j}$ and $V_{i', j'+1}, V_{i'+1, j'+1}$.
Denoting $h_2$ and $v_2$ the function combining the two coloring of the edges (both outer and inner edges of $2\times 2$), we have
\begin{equation*}
\begin{array}{l}
H_{i,j}=h_2(i,j) =
	\left\{ \begin{array}{ll}
		h(i,j) & \text{ if } j\%2 = 0\\
		\text{solved by $2\times 2$ solver} & \text{ otherwise,}
	\end{array} \right.\\
V_{i,j}=v_2(i,j) =
	\left\{ \begin{array}{ll}
		v(i,j) & \text{ if } i\%2 = 0\\
		\text{solved by $2\times 2$ solver} & \text{ otherwise.}
	\end{array} \right.
\end{array}
\end{equation*}

An example of $2\times 2$ solution is shown in Figure \ref{fig:exampleSolution} (b).
\cite{Derouet-Jourdan2015} proves that there is a solution, without explicitly giving it. 
Solution can be built by constraints based or backtracking algorithms.
In our implementation, we choose to consider all possible cases of color matching for opposite edges.
It leads to 16 possible cases that are listed in appendix \ref{appendix:2x2solutions}. 

\subsection{Restricting the line length with dappling}
\label{section:soldappling}
\begin{figure}[htb]
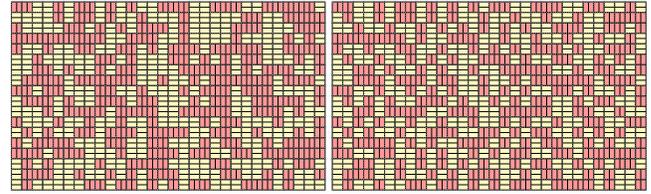

	\centering
	\resizebox{0.49\linewidth}{!}{\input{orientation_pre_dappling}}
	\resizebox{0.49\linewidth}{!}{\input{orientation_post_dappling}}
	\caption{\label{fig:noDapplingVsPerfect} Left: without dappling, we can notice the long continuous horizontal and vertical lines. Right: non procedural dappling algorithm from~\protect\cite{Kaji2016} (maximum line length of 2)}	
\end{figure}

In Figure~\ref{fig:noDapplingVsPerfect}, we can see how the dappling algorithm of \cite{Kaji2016} can produce random distributions while avoiding the long lines.
The vertical tiles are represented in red and the horizontal ones in yellow. 
Vertical successions of red tiles and horizontal successions of yellow tiles result in long lines.
By combining this dappling algorithm with the previous tiling algorithm of~\cite{Derouet-Jourdan2015}, it is possible to generate repeatable stochastic patterns with a given maximum length of lines (see Appendix~\ref{appendix:nonProcedural}). However this approach is non procedural.
We propose a procedural method to achieve similar results and restrict the longest line length to an arbitrary value $n$. 

The idea behind the dappling in the paper \cite{Kaji2016} is to traverse the configuration diagonally and correct the dappling when the number of consecutive horizontal or vertical tiles is too large.
It is not possible to use this method in a procedural fashion as it would necessitate the construction the whole configuration until the requested cell.
What we propose is to force the correction, that is, we "preemptively" correct the dappling automatically, whether the correction is necessary or not.
In the following, we explain the core idea behind the procedural dappling and then we explain how to build a tiling with a procedural dappling directly, without explicitely building the dappling for $n>2$ and $n=2$.

\begin{figure}[htb]
	\centering
	\resizebox{0.3\linewidth}{!}{\input{widgets}}
	\caption{\label{fig:checkerboardPatterns} 2 possible checkerboard patterns.}
\end{figure}
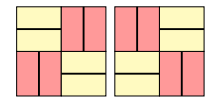

\paragraph{$\mathbf{n>2}$}
For a maximum number of $n$ consecutive tiles of a same orientation, we use $2\times 2$ checkerboards (see Figure \ref{fig:checkerboardPatterns}) on diagonals.

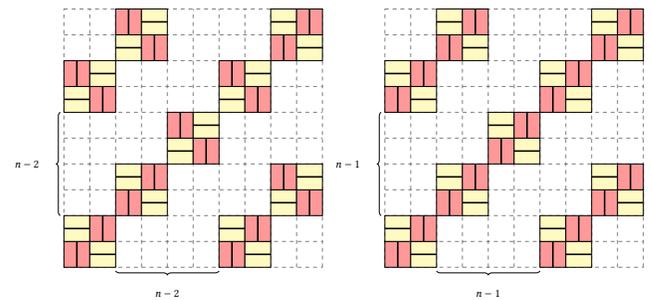
\begin{figure}[htb]
	\centering
	\begin{minipage}{0.49\linewidth}
		\resizebox{\linewidth}{!}{\input{dappling_even}}
	\end{minipage}
	\begin{minipage}{0.49\linewidth}
		\resizebox{\linewidth}{!}{\input{dappling_odd}}
	\end{minipage}
	\caption{\label{fig:diagonalWidgets}Dappling pattern for even $n$ (left), odd $n$ (right)}
\end{figure}

Each diagonal is separated from the other one by $n-2$ cells horizontally and vertically (see Figure \ref{fig:diagonalWidgets}).

The cells inbetween the $2\times 2$ checkerboards are not constrained and can be oriented in any way.
It is easy to see that in such a case, there are no alignment of more than $n$ consecutive tiles with the same orientation ($n-2$ for the cells inbetween and $+2$ for the checkerboards).
Because the checkerboards are of size $2\times 2$, this technique only works for even values of $n$.
It is possible to solve the same problem for odd values of $n$: we use only one type of checkerboard and  separate each diagonal from the other one by $n-1$ cells horizontally and vertically.

To include it in the general solution of Section \ref{section:procedural}, we need to adjust the hash functions $h_2$ and $v_2$ and replace them by dappling enabled ones $h'$ and $v'$.

$2\times 2$ checkerboards are aligned on the diagonal: $i'\%n = j'\%n$, with $i'=i-i\%2, j'=j-j\%2$.
Considering a checkerboard with its bottom left cell in $(i',j')$, there is exactly one horizontal tile between two horizontal edges opposed on the outer border, $H_{i',j'}$ and $H_{i', j'+2}$.
It means that we necessarily have $h'(i',j') \neq h'(i',j'+2)$.
The same reasoning applies for  $v'$.
So we need to create  $h'$ and $v'$ to enforce that condition.

\begin{equation*}
	\begin{array}{r}
	h'(i,j) = \left\{
		\begin{array}{ll}
			h_{d1}(i, j, h(i,j+2)) & \text{if } j\%2 = 0 \text{ and } i'\%n=j'\%n\\
			h(i,j) & \text{if } j\%2 = 0  \text{ and } i'\%n\neq j'\%n\\
			\text{solved by $2\times 2$ solver} &\text{if }j\%2 \neq 0
		\end{array}
		\right.\\	
	v'(i,j) = \left\{
		\begin{array}{ll}
			v_{d1}(i, j, v(i+2,j)) & \text{if } i\%2=0\text{ and } i'\%n=j'\%n\\
			v(i,j) & \text{if } i\%2=0\text{ and } i'\%n\neq j'\%n\\
			\text{solved by $2\times 2$ solver} &\text{if }i\%2\neq 0
		\end{array}
		\right. 
	\end{array}
\end{equation*}
This defines all the outer edges of the $2\times 2$ squares, and then with use the general solution to solve each $2\times 2$ locally. $h_{d1}$ and $v_{d1}$ compute random colors different from their input and are defined in Appendix \ref{appendix:2x2solutions}.

\paragraph{$\mathbf{n=2}$}
In the special case of $n=2$, we just need to generate a dappling with randomly chosen $2\times 2$ checkerboards (from Figure \ref{fig:checkerboardPatterns}).

The idea to create a hash function that generate such a dappling is to consider the grid by pack of four cells, enforcing in the middle the checkerboard condition (opposite outer edge color are different).
Similarly to the case n>2, we get the condition $h'(i',j') \neq h'(i',j'+2)$.
Since in this case $n-2 = 0$, there is another checkerboard with its bottom left corner in $(i', j'+2)$, meaning $h'(i',j'+2) \neq h'(i',j'+4)$.
Combining the two constraints we get $h'(i', j'+2) = h_{d2}(i', j'+2, h'(i', j'), h'(i', j'+4)$.
If we choose $h'(i', j')$ and $h'(i', j'+4)$ arbitrarily, we can compute $h'(i', j'+2)$.
The same reasoning apply to $v'$ and can be summed up as

\begin{equation*}
	\begin{array}{r}
	h'(i,j) = \left\{
		\begin{array}{ll}
			h_{d2}(i, j, h(i,j-2), h(i,j+2)) & \text{ if } j\%4 = 2\\
			h(i,j) & \text{ if } j \%4 = 0 \\
			\text{solved by $2\times 2$ solver} & \text{ otherwise.}
		\end{array}
		\right.\\
	v'(i,j) = \left\{
		\begin{array}{ll}
			v_{d2}(i, j, v(i-2,j), v(i+2,j)) & \text{ if } i\%4 = 2\\
			v(i,j) & \text{ if } i \%4 = 0 \\
			\text{solved by $2\times 2$ solver} & \text{ otherwise.}
		\end{array}
		\right.	
	\end{array}
\end{equation*}
$h_{d2}$ and $v_{d2}$ compute random colors different from their input and are defined in Appendix \ref{appendix:2x2solutions}.
\section{Results and Discussion}
\label{section::results}

\subsection{Visual results and comparison}
\begin{figure}[htb]
	\centering
	\begin{subfigure}{0.27\linewidth}
		\resizebox{0.99\linewidth}{!}{\input{result_n_1}}
		\caption*{$n=1$}
	\end{subfigure}
	\begin{subfigure}{0.27\linewidth}
		\resizebox{0.99\linewidth}{!}{\input{result_n_3}}
		\caption*{$n=3$}
	\end{subfigure}
	\begin{subfigure}{0.27\linewidth}
		\resizebox{0.99\linewidth}{!}{\input{result_n_5}}
		\caption*{$n=5$}
	\end{subfigure}
	\caption{\label{fig:variousLength} Output of our algorithm for  $n=1$, $n=3$, $n=5$}
\end{figure}
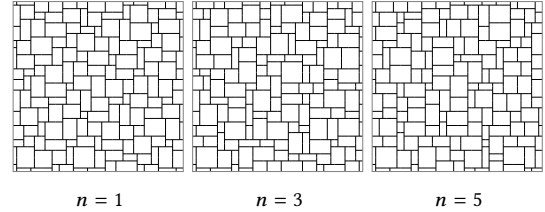

Thanks to our algorithm, we can control over the maximum length of the lines of the stochastic wall pattern. 
Output result of the pattern for maximum line length $n=1$, $n=3$, $n=5$ are shown in Figure \ref{fig:variousLength}.
We could also reproduce the artist painted wall patten as shown in Figure \ref{fig:wangtileResults}.

\begin{figure}[htb]
	\centering
	\begin{subfigure}{0.45\linewidth}
		\includegraphics[width=\linewidth]{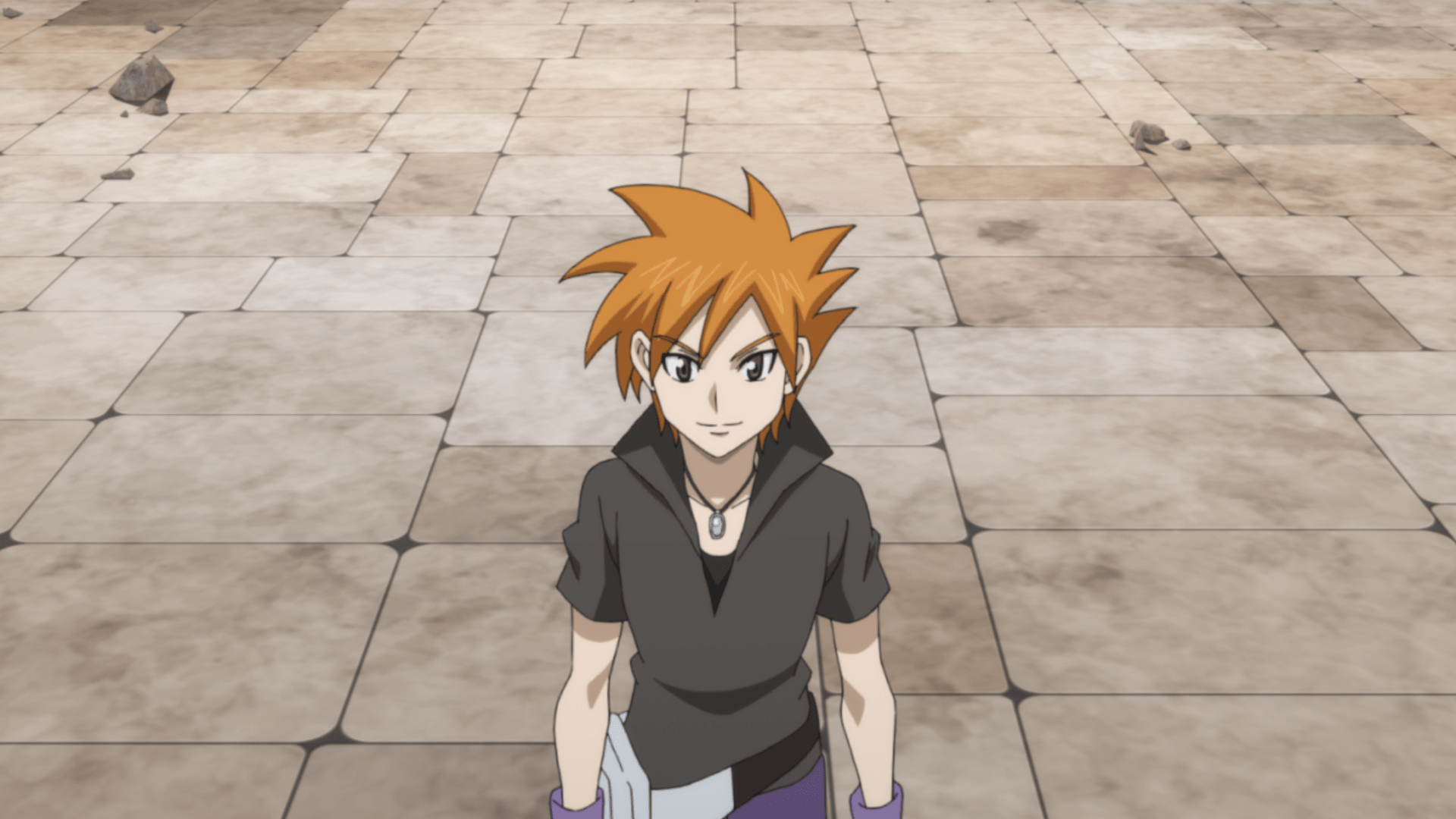}	
	\end{subfigure}
	\begin{subfigure}{0.45\linewidth}
		\includegraphics[width=\linewidth]{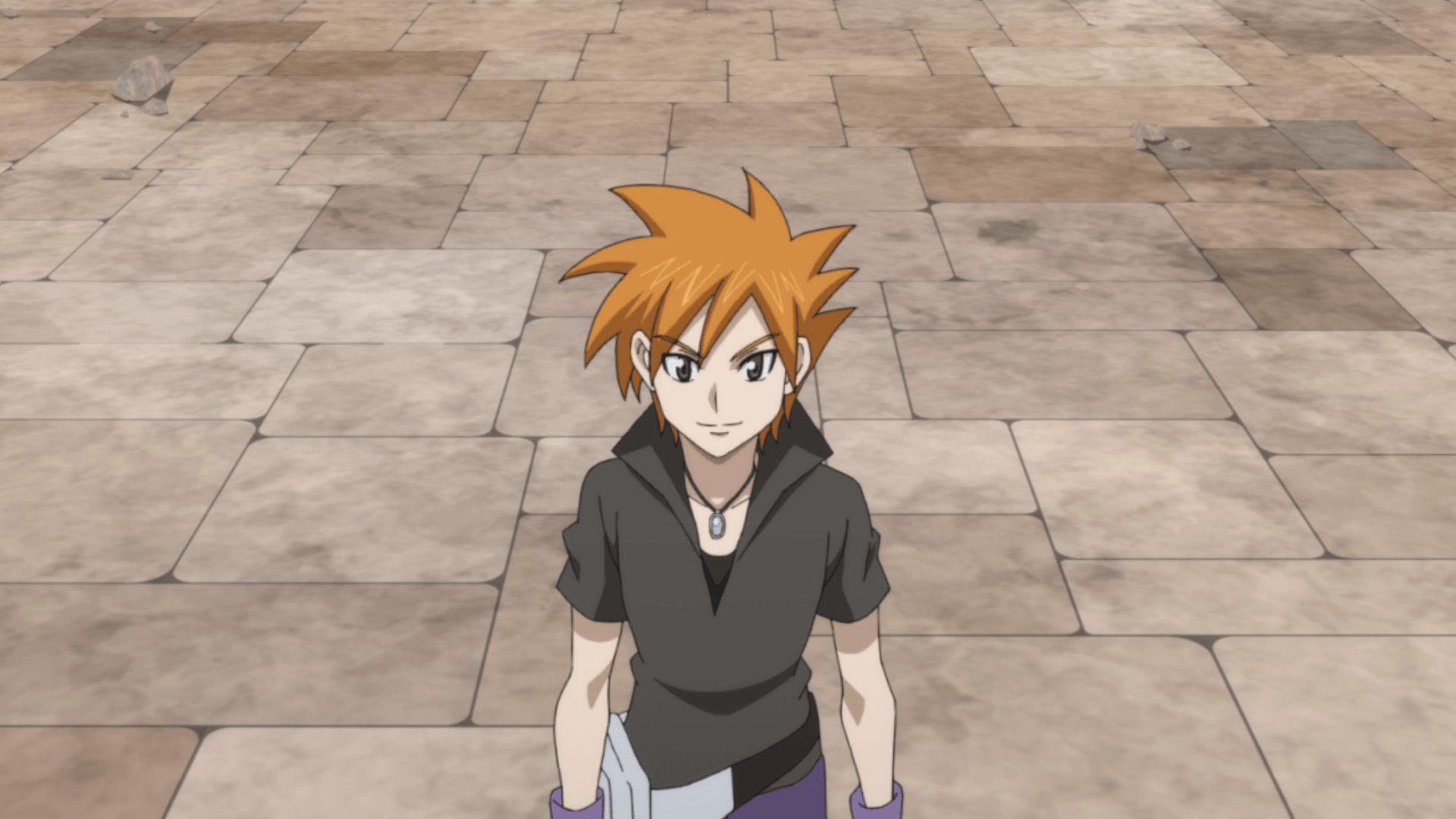}	
	\end{subfigure}
	\caption{\label{fig:PMPVresults} Production result using  \cite{Derouet-Jourdan2015} (left), using our algorithm ($n=2$) (right). Our method breaks the long lines artifacts behind the character. These lines attract the eye, breaking them restores the focus on the character.}
\end{figure}

Algorithm from \cite{Derouet-Jourdan2015} has been used in production and long lines were standing out (Figure \ref{fig:PMPVresults} (left)). 
But limiting the maximum length to $n=2$ with our new algorithm, the output looks much more natural (Figure \ref{fig:PMPVresults} (right)).

\begin{figure}[htb]
	\centering
	\begin{subfigure}{0.24\linewidth}
		\includegraphics[width=\linewidth]{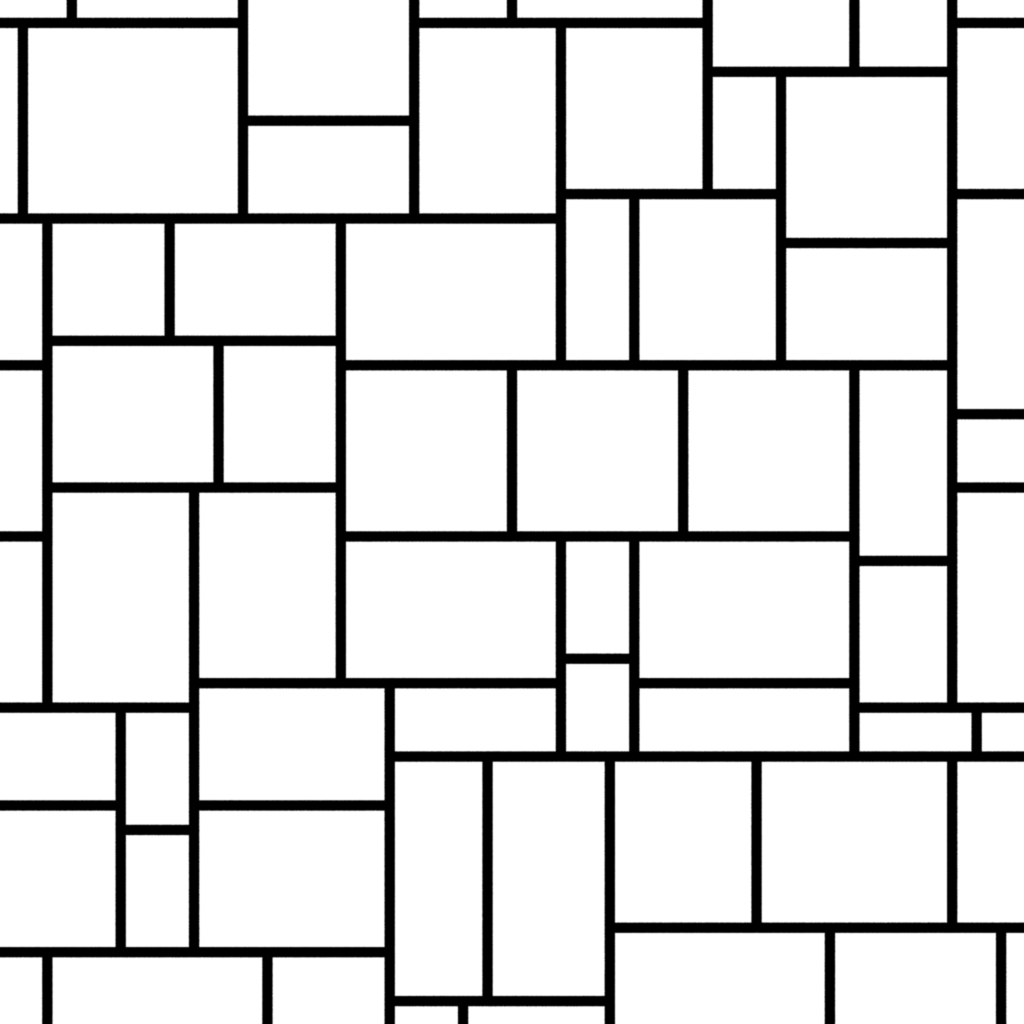}	
		\caption*{wall pattern}
	\end{subfigure}
	\begin{subfigure}{0.24\linewidth}
	\includegraphics[width=\linewidth]{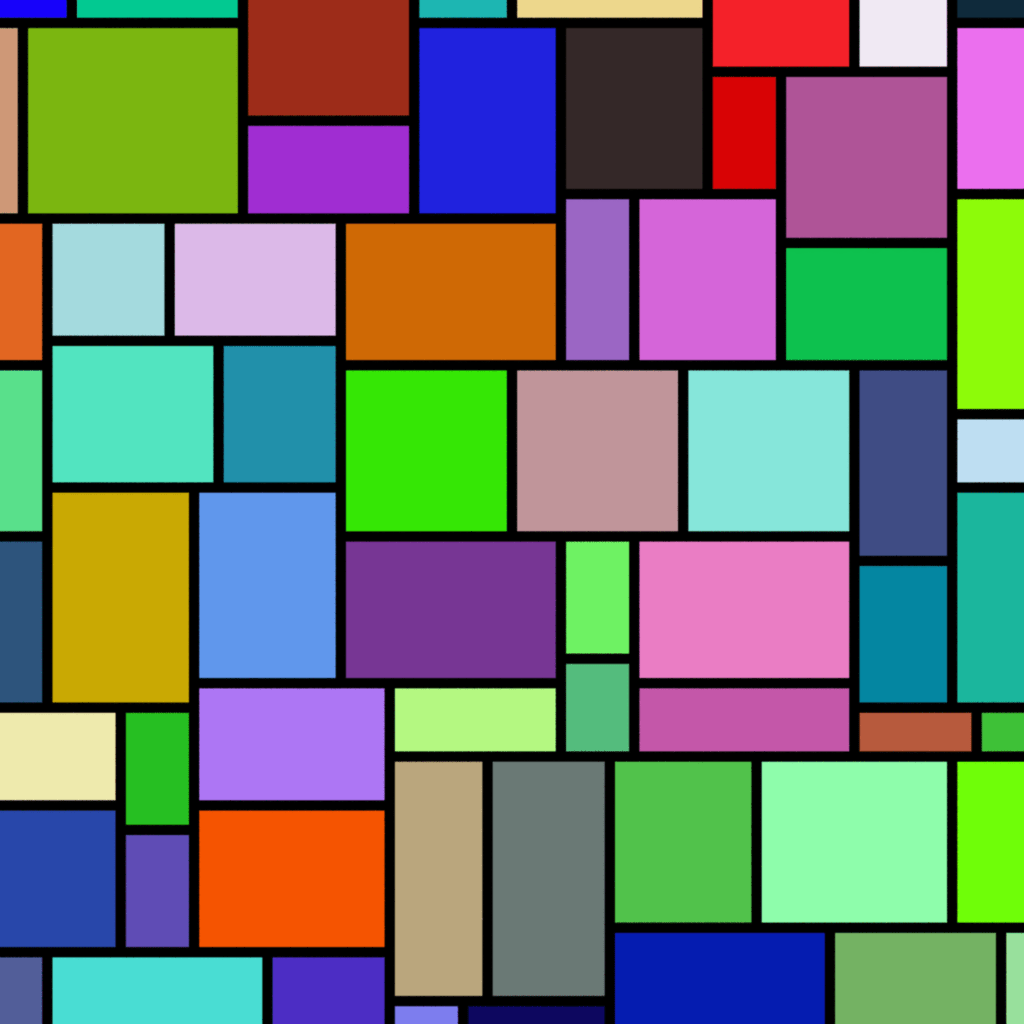}	
	\caption*{color}
\end{subfigure}
	\begin{subfigure}{0.24\linewidth}
	\includegraphics[width=\linewidth]{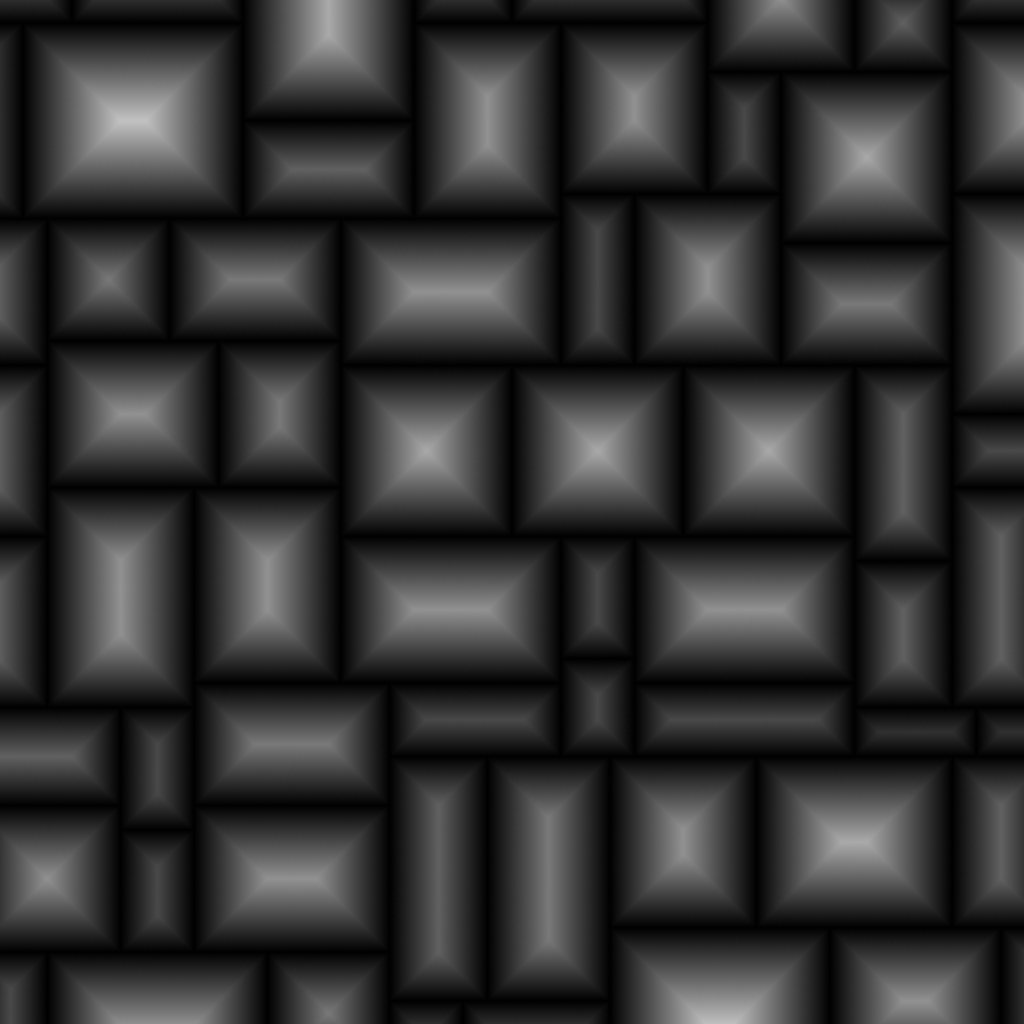}	
	\caption*{distance to edge}
\end{subfigure}
	\begin{subfigure}{0.24\linewidth}
	\includegraphics[width=\linewidth]{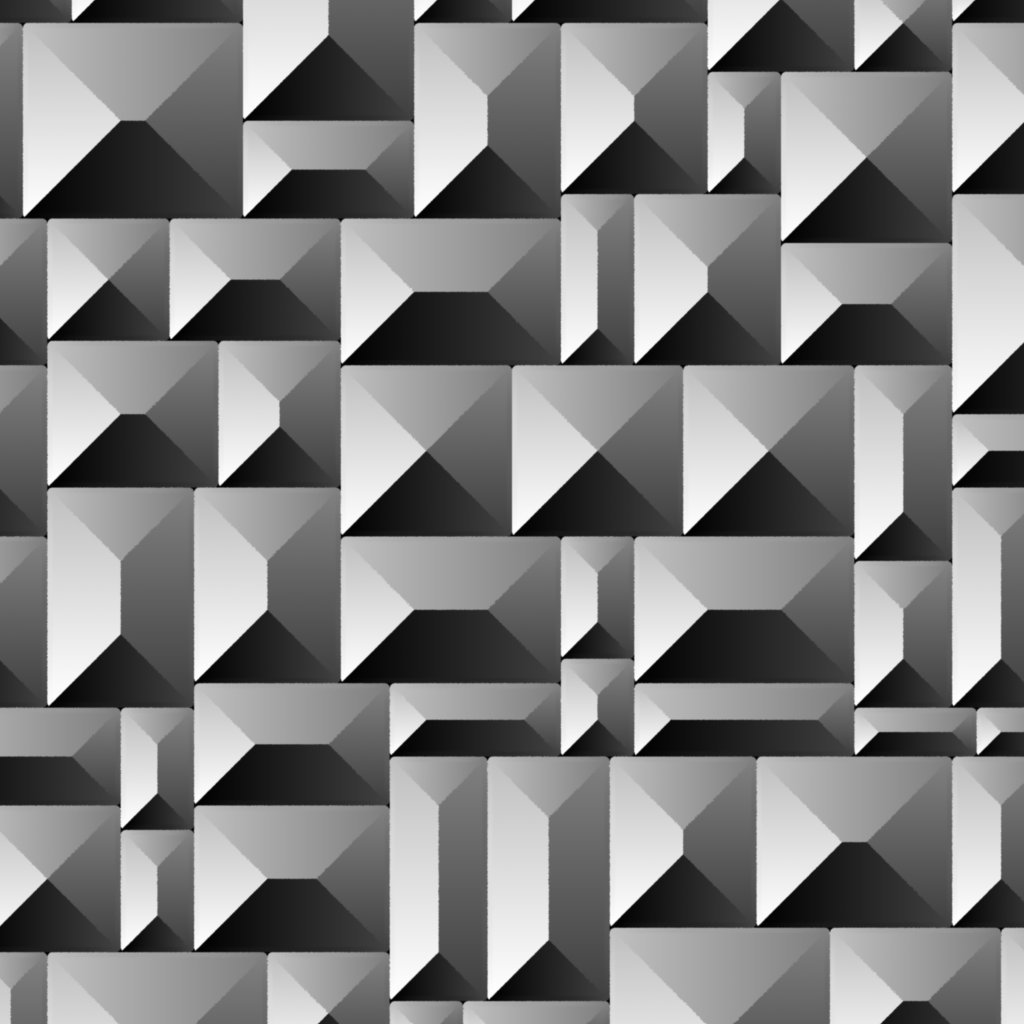}	
	\caption*{phase}
\end{subfigure}

\begin{subfigure}{0.24\linewidth}
\includegraphics[width=\linewidth]{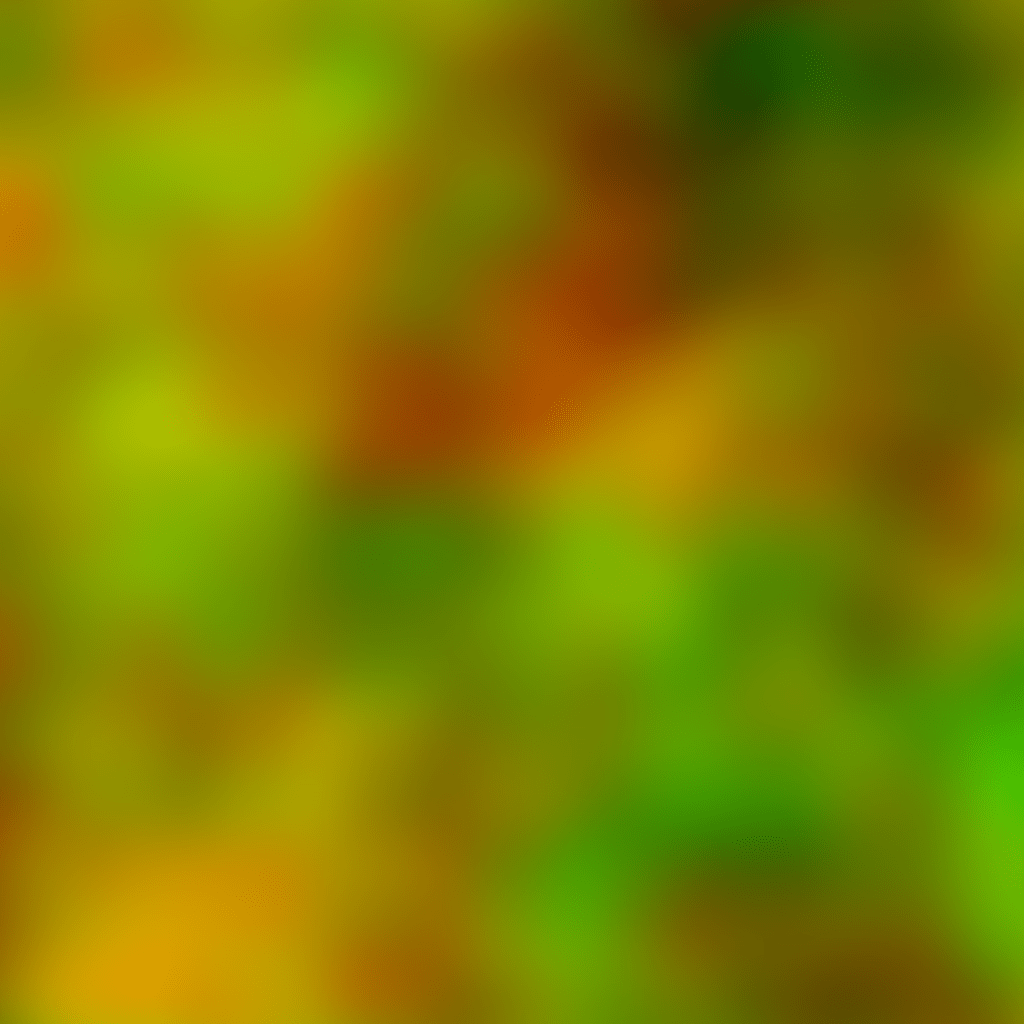}	
\caption*{input noise}
\end{subfigure}
\begin{subfigure}{0.24\linewidth}
	\includegraphics[width=\linewidth]{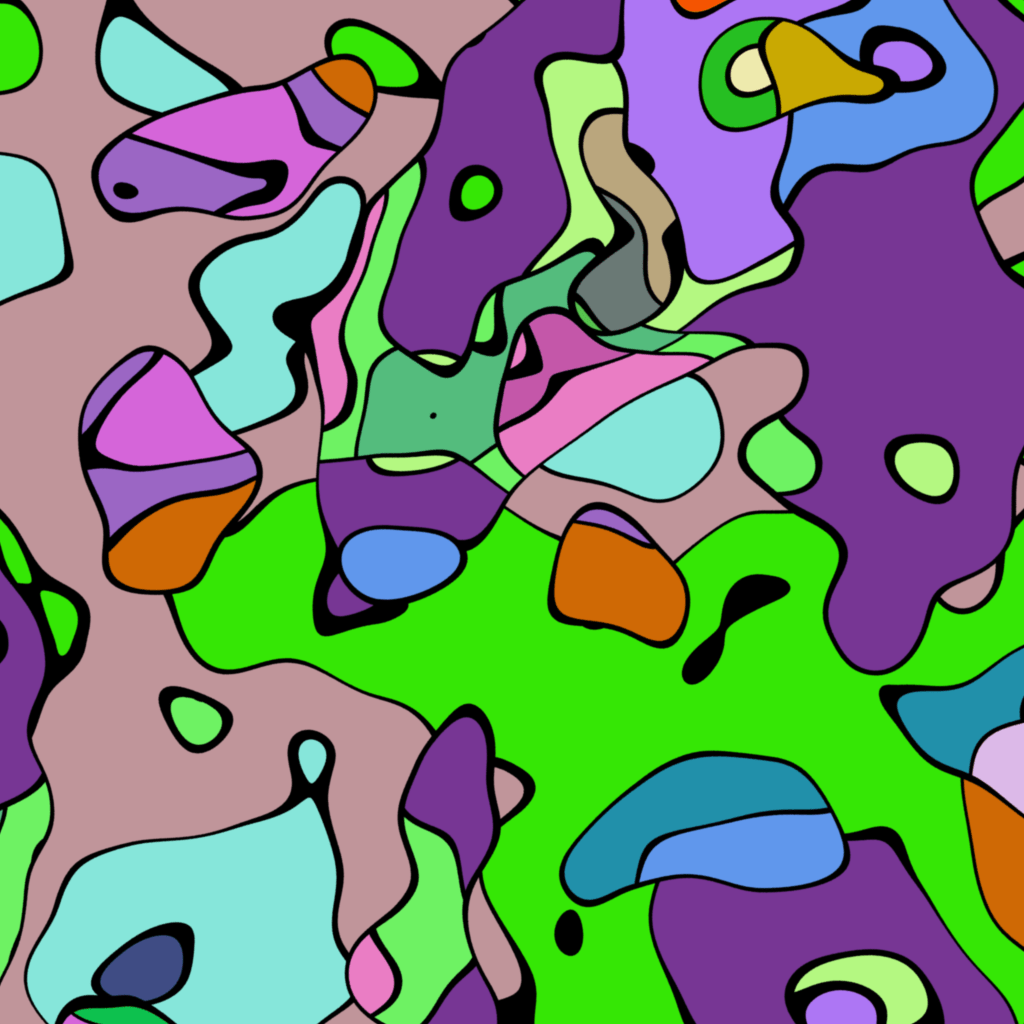}
	\caption*{color}
\end{subfigure}
\begin{subfigure}{0.24\linewidth}
	\includegraphics[width=\linewidth]{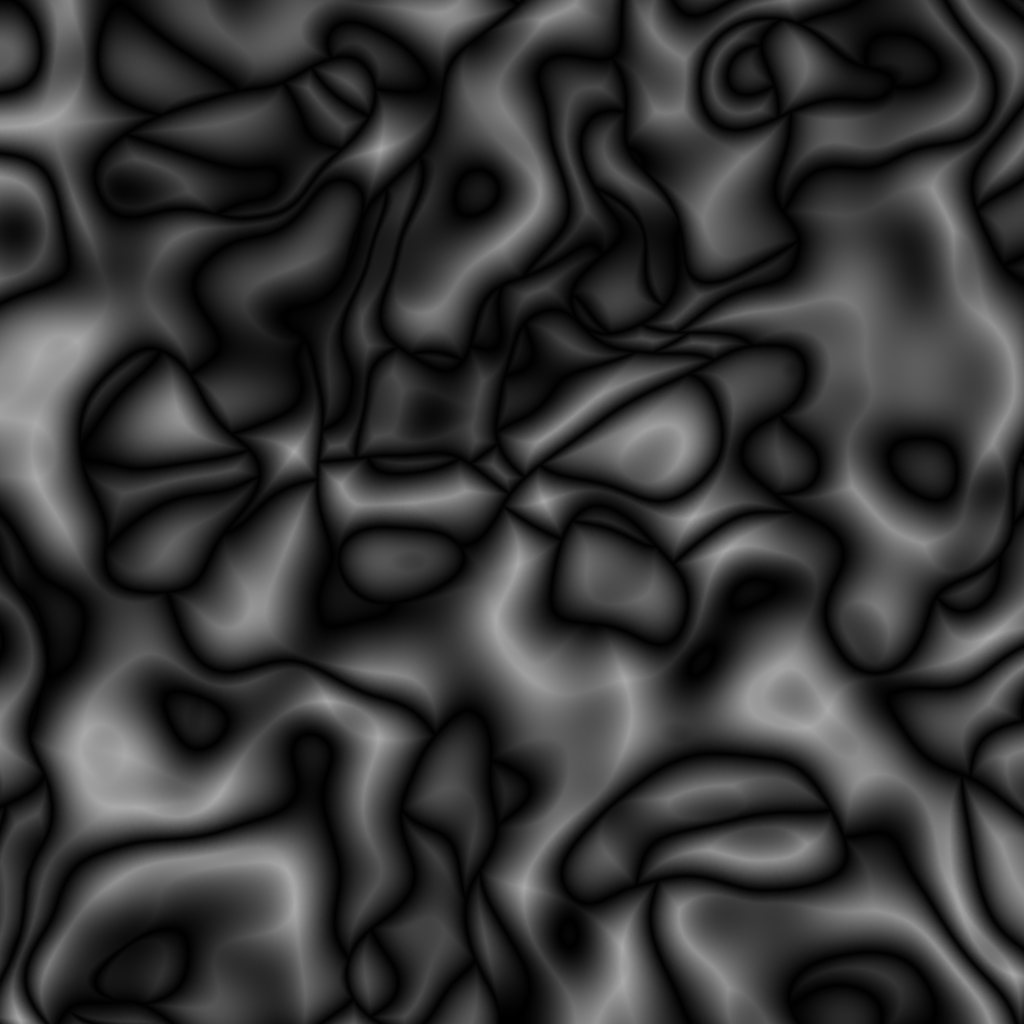}
	\caption*{distance to edge}
\end{subfigure}
\begin{subfigure}{0.24\linewidth}
	\includegraphics[width=\linewidth]{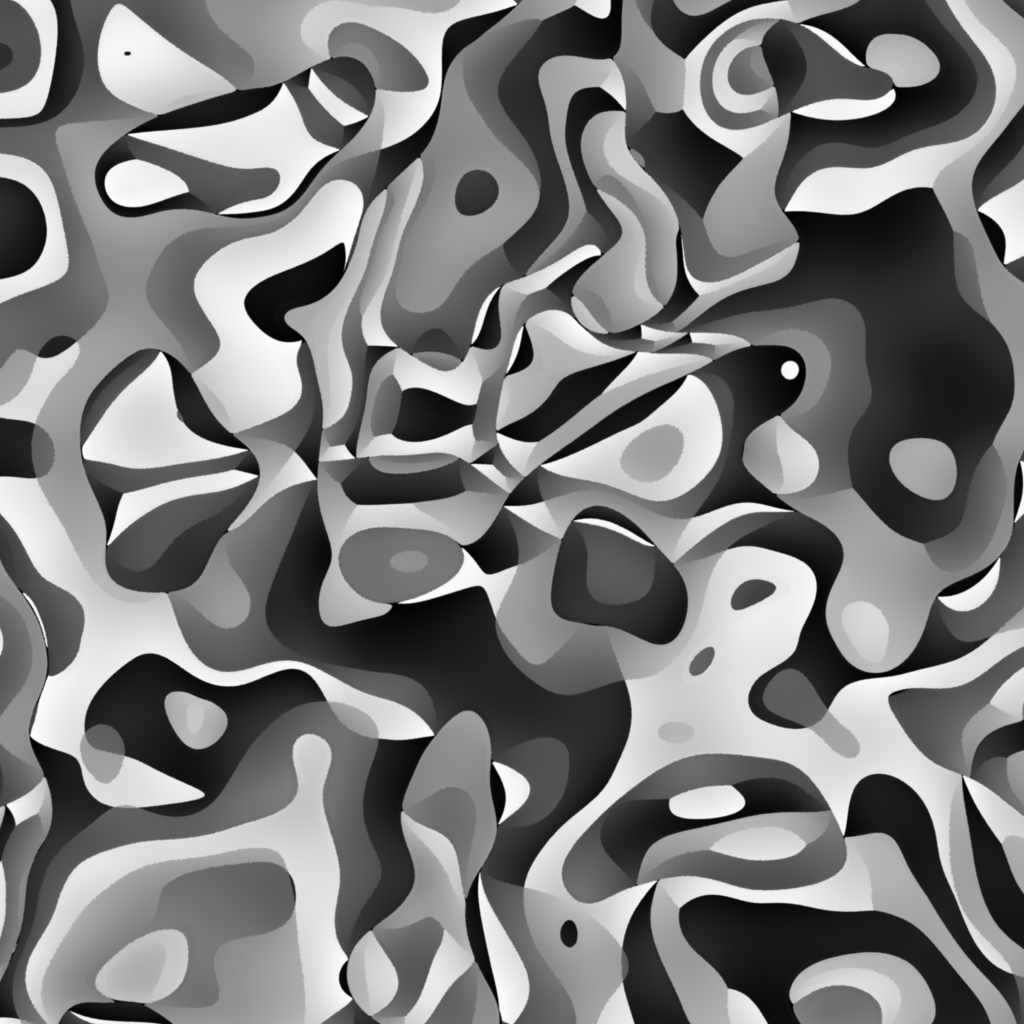}
	\caption*{phase}
\end{subfigure}

	\caption{\label{fig:variousresults} Various outputs using stochastic walls}
\end{figure}

Also, by varying input of the shader (uv offset or noise), we can obtain various network of lines and colored area, gradations that may be used for magma, mosaic or stained glass window visuals (see Figure \ref{fig:variousresults}). 

\begin{figure}[htb]
	\centering
	\includegraphics[width=0.3\linewidth]{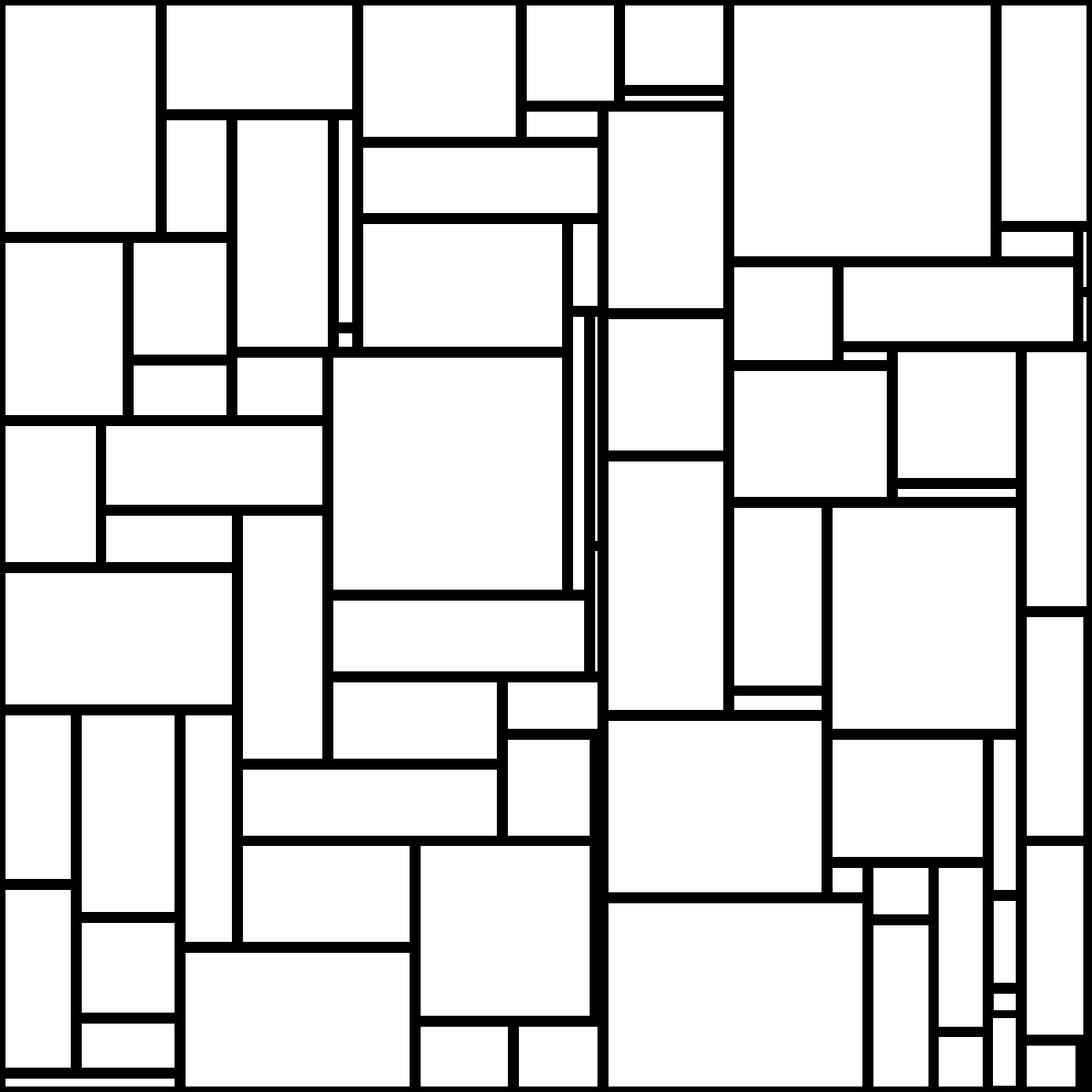}	
	\caption{\label{fig:boxpacking} Box packing method}
\end{figure}

Existing methods failed to reproduce our wall pattern structure.
We tried the box packing method from \cite{Miyata1990} (see Figure \ref{fig:boxpacking}).  
It produces elongated bricks, does not give control over cross patterns or line length. 
We have access to the structure of the bricks, but the method is not procedural.

\begin{figure}[htb]
	\centering
	\begin{subfigure}{0.31\linewidth}
		\includegraphics[width=\linewidth]{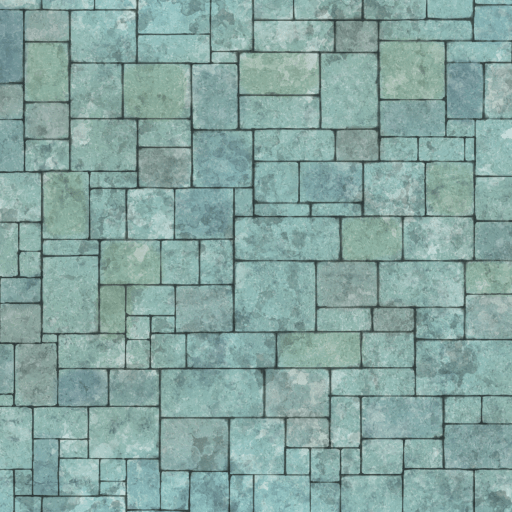}	
		\caption*{}
	\end{subfigure}
	\begin{subfigure}{0.31\linewidth}
		\includegraphics[width=\linewidth]{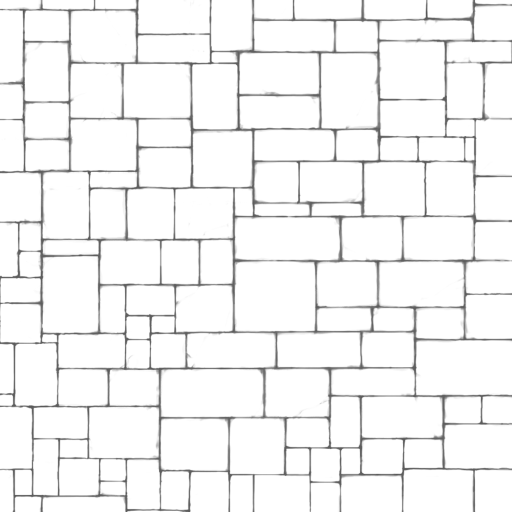}	
		\caption*{input}
	\end{subfigure}
	\begin{subfigure}{0.31\linewidth}
	\includegraphics[width=\linewidth]{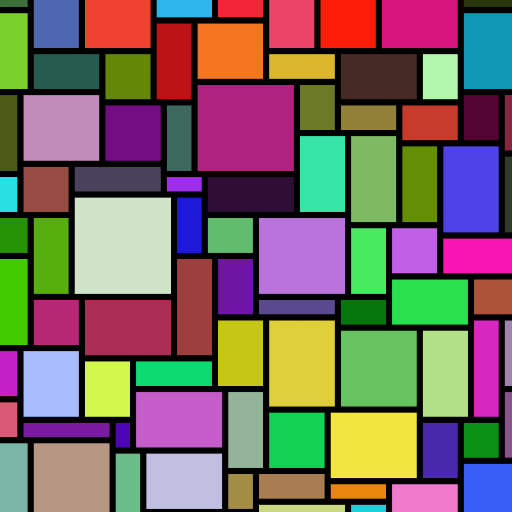}	
		\caption*{}
	\end{subfigure}

	\begin{subfigure}{0.31\linewidth}
	\includegraphics[width=\linewidth]{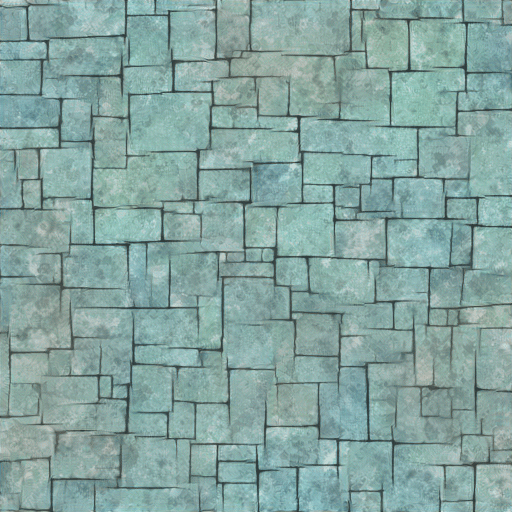}	
		\caption*{}
	\end{subfigure}
	\begin{subfigure}{0.31\linewidth}
	\includegraphics[width=\linewidth]{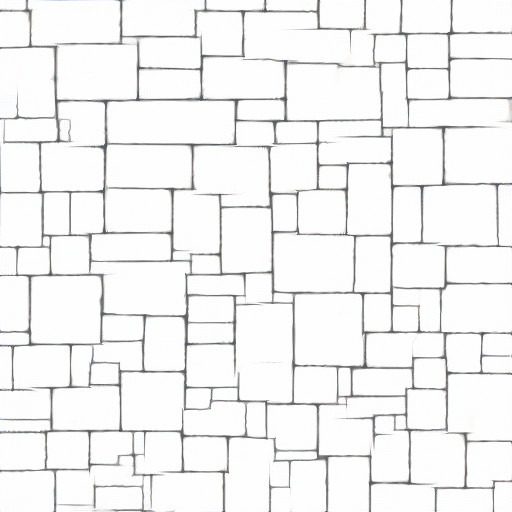}
	\caption*{cvn}	
	\end{subfigure}
	\begin{subfigure}{0.31\linewidth}
	\includegraphics[width=\linewidth]{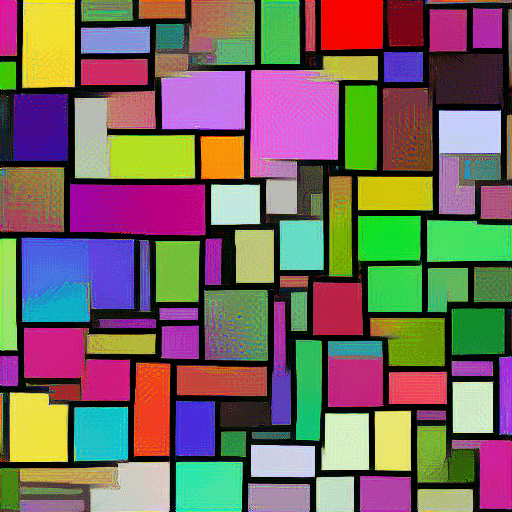}	
	\caption*{}
	\end{subfigure}

	\begin{subfigure}{0.31\linewidth}
	\includegraphics[width=\linewidth]{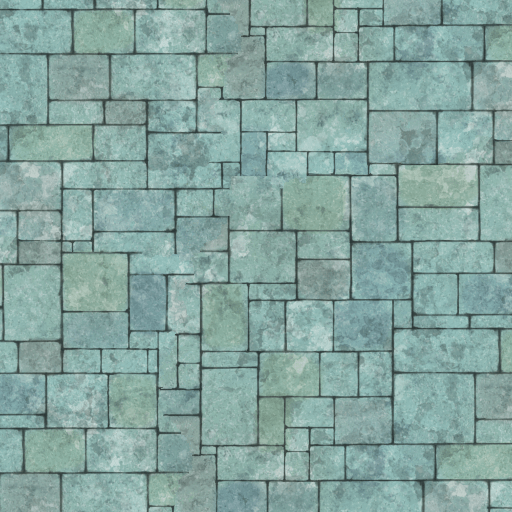}	
	\caption*{}
	\end{subfigure}
	\begin{subfigure}{0.31\linewidth}
	\includegraphics[width=\linewidth]{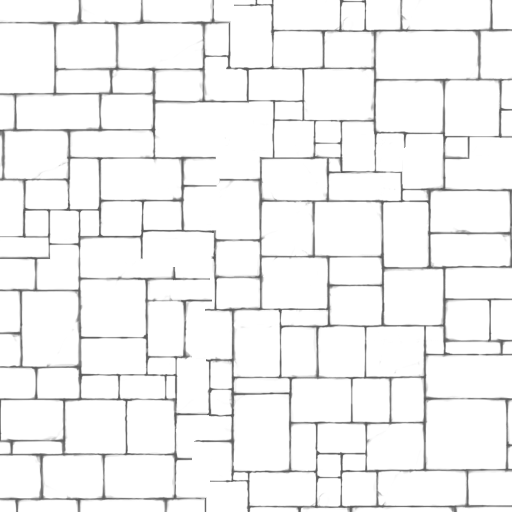}
	\caption*{graphcut}	
	\end{subfigure}
	\begin{subfigure}{0.31\linewidth}
	\includegraphics[width=\linewidth]{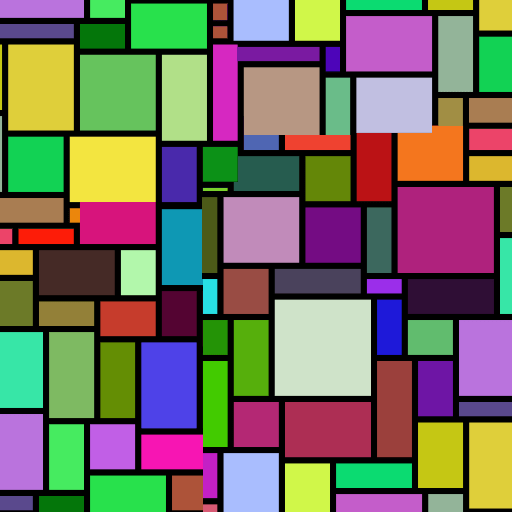}	
	\caption*{}
	\end{subfigure}

	\begin{subfigure}{0.31\linewidth}
		\includegraphics[width=\linewidth]{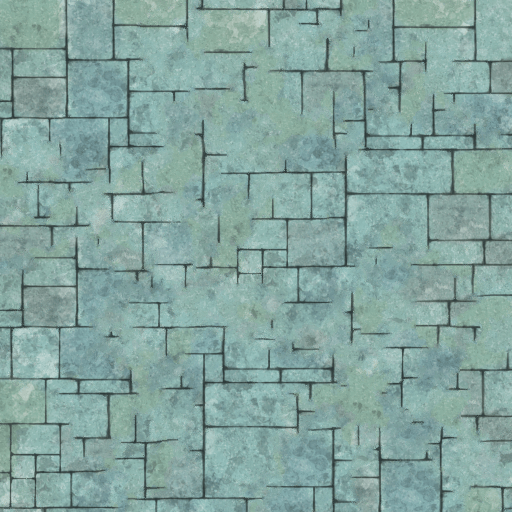}	
		\caption*{}
	\end{subfigure}
	\begin{subfigure}{0.31\linewidth}
		\includegraphics[width=\linewidth]{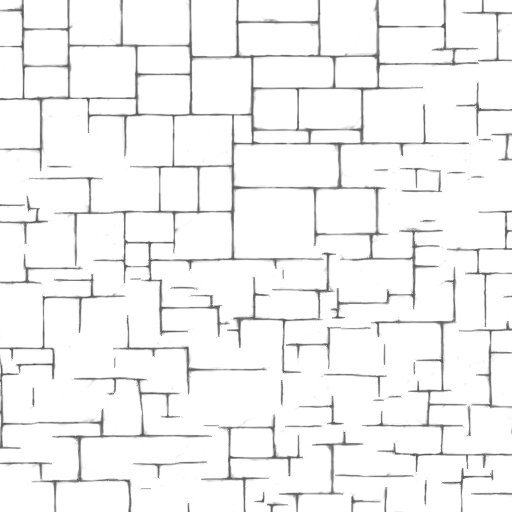}
		\caption*{patch based}	
	\end{subfigure}
	\begin{subfigure}{0.31\linewidth}
		\includegraphics[width=\linewidth]{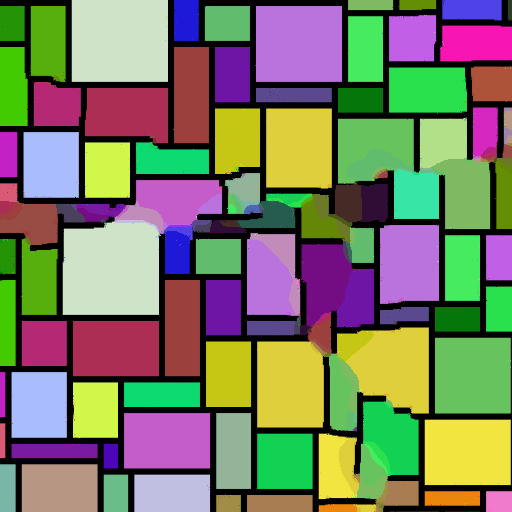}	
		\caption*{}
	\end{subfigure}

	\caption{\label{fig:otherMethods} Other methods }
\end{figure}

Texture synthesis methods are slow (minutes or hours of computation) and can only reproduce approximative look,  introducing holes in the lines and noises in the texture. The final look is hard to fine-tune due to the lack of access to the structure (see Figure \ref{fig:otherMethods}).

\subsection{GPU implementation}
\label{section:GPU}
The GPU implementation of the procedural approach was straightforward in OpenGL3 and  WebGL (to provide a ShaderToy implementation).
We use integer based hash functions in our production implementation.
However, the integer support seems to be limited in OpenGL, and we then switched to different, float based hash functions.
We plan to provide the shader source code online for anyone to test our method (both WebGL and OpenGL3).

We reach real time performance even with high number of samples as shown in Table~\ref{table:resultGPU}.
\begin{table}[htb]
	\centering	
	\begin{tabular}{|c|c|c|c|c|c|}
		\hline
		Nb Samples& 1& 2 & 8 & 16 & 32\\
		\hline
		VGA(1024x768) & 520& 515 & 450 & 180 & 148\\
		\hline
		HD(1920x1080) & 200& 220& 187 & 70 & 60\\
		\hline
	\end{tabular}
	\caption{\label{table:resultGPU} GPU algorithm performance (in frame per seconds). GPU results have been obtained using a Nvidia Quadro K620.}
\end{table}

\subsection{Computation time and discussion}
We measure and compare the performance of our algorithm against the general non-procedural algorithm described in Appendix \ref{appendix:nonProcedural}.
Both algorithms give the same kind of results, and computation time are given in Table~\ref{table:resultCPU}.

\begin{table}[htb]
	\centering
	\begin{tabular}{|c|c|c|c|c|c|}
		\hline
		& Initialization & Time for $10^7$ computation (x NP)\\
		\hline
		NP/NPD & 2.1/2.5& 790 (1x) \\
		\hline
		P/PD & 0&  3320 (4.2x)/3726 (4.7x) \\
		\hline
	\end{tabular}
	\caption{\label{table:resultCPU} CPU algorithms performance measured in $ms$. Time for the retrieval of a brick, using a 100x100 grid. Non procedural without/with dappling (NP/NPD $n=2$). Procedural without/with dappling(P/PD $n=2$)}
\end{table}
Although the procedural algorithms are 4 to 5 times slower than the non-procedural version, in production rendering context, this timings remain negligible compared to the full rendering times. 
It represents 2\% of  our wall rendering time (roughly 2 minutes per frame).

\begin{figure}[htb]
	\centering
	\includegraphics[width=0.47\linewidth]{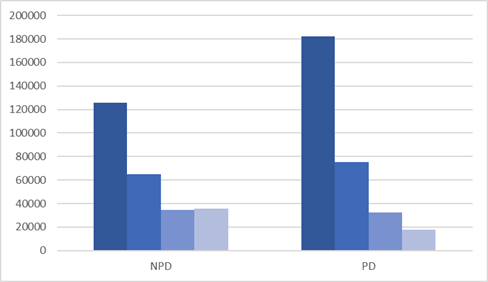}	
	\includegraphics[width=0.47\linewidth]{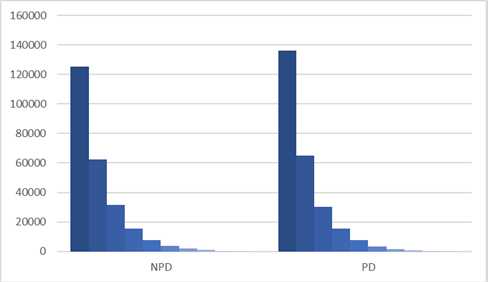}	
	\caption{\label{fig:algocomparison} Histogram of the vertical line length occurrence. Comparison between non procedural (NPD) and procedural (PD) version. Left: with maximum line length $n=4$. Right: with maximum line length $n=10$.}
\end{figure}

We evaluate the quality of the dappling results by computing histograms of the number of consecutive tiles of same orientation in rows (see Figure~\ref{fig:algocomparison}). 
We use dappling method \cite{Kaji2016} for the non-procedural algorithm, because it is the most general solution as far as we know.
Figure~\ref{fig:algocomparison} shows that the proportion of lines with the same length are similar with either procedural and non-procedural algorithms.
However we found a flaw in the algorithm \cite{Kaji2016}. 
In the case $n=4$, the number of lines of length 4 is oddly equivalent to line of length 3.
Our assumption is that every lines over 4 of length will be clamped to 4 and then artificially increase their occurrence, which insert a bias in the result.
Our procedural algorithm produces a better line length distribution than previous works.

The non-procedural approach is faster than our algorithm. It uses memory proportional to the number of bricks. 
The algorithm also produces repeatable patterns, that reduce memory usage, to the expense of visual artifacts.
It is also customizable and allows the caching of per brick information (color, randomization values...).
Procedural methods would need to recompute those information constantly, which explain the computation time difference.
However procedural approach enable to generate unbounded textures without any repetitions, and without using memory.

\section{Conclusion and future work}
By designing custom hash functions for our specific problem, we succeed to provide a simple yet general solution to the generation of stochastic wall patterns.
Our algorithm is fully procedural, avoids common visual artifacts and gives control to the user over the maximum line length. 
The computation overhead is low and the GPU implementation enables preview of the result, making it ready to integrate into movie production pipeline, extending artists' creation palette.

We are now considering the inclusion of multi resolution Wang tiles (\cite{Kopf2006}) or the combination of various sets of tiles to enable more variations in the brick sizes and patterns.
We are also thinking about using border constrained 2D Wang tiling solutions in the context of texture synthesis. 
We are also working on the 3D procedural texture generation of stochastic wall patterns, and we think about extending those results to general voxelization problems.
Our intuition is that the stochastic structure of the underlying grid may improve the quality of volume rendering and collision detections.

\bibliographystyle{ACM-Reference-Format}
\bibliography{bibliography}

\begin{appendices}
	\section{Non-procedural solution}
	\label{appendix:nonProcedural}	
\begin{figure}[htb]
		\begin{minipage}{0.49\linewidth}
			\centering
			\resizebox{0.99\linewidth}{!}{\input{border_constraint}}
		\end{minipage}
		\begin{minipage}{0.49\linewidth}
			\centering
			\resizebox{0.99\linewidth}{!}{\input{row_last_2_vert}}
		\end{minipage}
		\caption{\label{fig:WangTileborderConstraint} Left: Wang tile border constraint: left and right colors are equal, top and bottom colors are equal. Right: Solve the border constraint with dappling. }
	\end{figure}
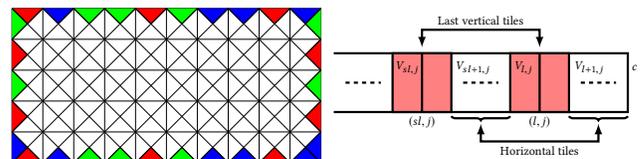

	This appendix presents the non-procedural version of the wall pattern generation algorithm.
	To generate an unbounded texture, we need to generate a texture that is repeatable.
	It translates into a Wang tiling problem with border constraints~\cite{Derouet-Jourdan2015} where the top border matches the bottom one and the left border matches the right one (see Figure~\ref{fig:WangTileborderConstraint} left).
	To avoid long lines as discussed in Section~\ref{subsection:existingAlgorithms}, we solve the border constrained Wang tiling problem on top of a cyclic dappling solution $D$ computed with~\cite{Kaji2016}. 
	Our algorithm solves the tiling problem in two phases.
	First it computes the colors of the vertical edges for each row $R$ of the tiling, using $D$. 
	Then we solve the colors of the horizontal edges in a similar fashion.
	
	Let's consider the length $n_R$ of row $R$, and a border constraint color $V_{n_R,j} = V_{0,j}=c$.
	We assign values to the vertical edge colors according to the model in Section~\ref{subsection:WangTileModel} until the last two vertical tiles ($V_{i,j} =V_{i+1,j}$ if it's horizontal, $V_{i+1,j} \neq V_{i,j}$ if it's vertical).
	
	Let $(sl, j)$ be the second to last vertical tile and $(l, j)$ be the last vertical tile (see Figure~\ref{fig:WangTileborderConstraint} right).
	Between the last two vertical tiles, there are only horizontal tiles.
	So we have the following constraints: 
	\begin{equation*}
		\begin{array}{ll}
			V_{sl,j} \neq V_{sl+1,j} & V_{sl+1,j}= V_{l,j}\\
			V_{l,j} \neq V_{l+1,j} & V_{l+1,j} = V_{n_R,j} =c\\
		\end{array}
	\end{equation*}
	So we just need to find a color for $CS = V_{sl+1,j}= V_{l,j}$, so that  $CS \neq V_{sl,j}, CS \neq  c$, which is always possible as long as we have at least 3 colors.
We give the pseudo-code for this algorithm on the rows, and it applies the same way to the columns.
\begin{algorithm}
	\caption{\label{algo:RowColoring}Color the vertical edges of a row.}
	\begin{algorithmic}
		\REQUIRE Row index $j$ to color, Dappling $D$, border color $c$
		\STATE $\text{last} \leftarrow -1$, $\text{secondToLast} \leftarrow -1$
		\FOR {$i \leftarrow n_R-1$ \TO $0$}
		\IF {$\text{last} = -1$ \AND $D[i] = \text{VERTICAL}$}
		\STATE $\text{last} \leftarrow i$ 
		\ELSIF {$D[i] = \text{VERTICAL}$}
		\STATE $\text{secondToLast} \leftarrow i$
		\STATE break
		\ENDIF
		\ENDFOR
		\STATE $\text{currentColor} \leftarrow c$, $V[0,j] \leftarrow c$
		\FOR {$i \leftarrow 0$ \TO $n_R-2$}
		\IF {$D[i] = \text{VERTICAL}$}
		\IF {$i = \text{secondToLast}$}
		\STATE $\text{currentColor} \leftarrow \text{diff} (\text{currentColor}, c)$ 
		\ELSIF {$i = \text{last}$}
		\STATE $\text{currentColor} \leftarrow c$
		\ELSE
		\STATE $\text{currentColor} \leftarrow  \text{diff}(c)$
		\ENDIF
		\ENDIF
		\STATE $V[i+1,j] \leftarrow \text{currentColor}$
		\ENDFOR
	\end{algorithmic}
	\label{alg:algononproc}
\end{algorithm}

\section{Base case solver}
\label{appendix:2x2solutions}
We separate the 16 cases for the equalities $H_{i, j+2}=H_{i, j}$ $H_{i+1, j}=H_{i+1, j+2}$ $V_{i, j+1}=V_{i+2, j+1}$ $V_{i, j}=V_{i+2, j}$. The solutions are given as 4 values for respectively $V_{i+1,j+1}$, $H_{i, j+1}$, $V_{i+1, j}$ and $H_{i+1,j+1}$.

\begin{enumerate}[leftmargin=1.5em]
\item solver 0000
\begin{itemize}[leftmargin=0em]
    \item $V_{i, j+1}, H_{i, j}, V_{i+2, j}, H_{i+1, j+2}$
	\item  $V_{i+2, j+1}, H_{i, j+2}, V_{i, j}, H_{i+1, j}$
\end{itemize}
\item solver 0001
\begin{itemize}[leftmargin=0em]
    \item $v_{d2}(i+1,j+1,V_{i, j+1}, V_{i+2, j+1}), H_{i, j+2}, V_{i, j}, H_{i+1, j+2}$
    \item $V_{i+2, j+1}, H_{i, j+2}, V_{i, j}, h_{d2}(i+1,j+1,H_{i+1, j}, H_{i+1, j+2})$
    \item $V_{i, j+1}, h_{d2}(i,j+1,H_{i, j+2}, H_{i, j}), V_{i, j}, H_{i+1, j+2}$
\end{itemize}

\item solver 0010
\begin{itemize}[leftmargin=0em]
    \item $V_{i, j+1}, H_{i, j}, v_{d2}(i+1,j,V_{i, j}, V_{i+2, j}), H_{i+1, j}$
    \item $V_{i, j+1}, H_{i, j}, V_{i+2, j}, h_{d2}(i+1,j+1,H_{i+1, j}, H_{i+1, j+2})$
   \item $V_{i, j+1}, h_{d2}(i,j+1,H_{i, j+2}, H_{i, j}), V_{i, j}, H_{i+1, j}$
\end{itemize}

\item solver 0011
\begin{itemize}[leftmargin=0em]
    \item $V_{i, j+1}, h_{d2}(i,j+1,H_{i, j+2}, H_{i, j}), V_{i, j}, h_{d2}(i+1,j+1,H_{i+1, j}, H_{i+1, j+2})$
    \item $V_{i, j+1}, H_{i, j}, v_{d2}(i+1,j,V_{i, j}, V_{i+2, j}), H_{i+1, j}$
   \item $v_{d2}(i+1,j+1,V_{i, j+1}, V_{i+2, j+1}), H_{i, j+2}, V_{i, j}, H_{i+1, j+2}$
\end{itemize}

\item solver 0100
\begin{itemize}[leftmargin=0em]
    \item $V_{i, j+1}, h_{d2}(i,j+1,H_{i, j+2}, H_{i, j}), V_{i, j}, H_{i+1, j}$
    \item $v_{d2}(i+1,j+1,V_{i, j+1}, V_{i+2, j+1}), H_{i, j+2}, V_{i, j}, H_{i+1, j}$
   \item $V_{i, j+1}, H_{i, j}, v_{d2}(i+1,j,V_{i, j}, V_{i+2, j}), H_{i+1, j}$
\end{itemize}

\item solver 0101
\begin{itemize}[leftmargin=0em]
   \item $V_{i, j+1}, H_{i, j}, v_{d2}(i+1,j,V_{i, j}, V_{i+2, j}), H_{i+1, j}$
   \item $V_{i+2, j+1}, H_{i, j+2}, V_{i, j}, h_{d2}(i+1,j+1,H_{i+1, j}, H_{i+1, j+2})$
\end{itemize}

\item solver 0110
\begin{itemize}[leftmargin=0em]
   \item $v_{d2}(i+1,j+1,V_{i, j+1}, V_{i+2, j+1}), H_{i, j+2}, V_{i, j}, H_{i+1, j}$
   \item $V_{i+2, j+1}, H_{i, j}, V_{i+2, j}, h_{d2}(i+1,j+1,H_{i+1, j}, H_{i+1, j+2})$
\end{itemize}

\item solver 0111
\begin{itemize}[leftmargin=0em]
    \item $V_{i, j+1}, h_{d2}(i,j+1,H_{i, j+2}, H_{i, j}), V_{i, j}, h_{d2}(i+1,j+1,H_{i+1, j}, H_{i+1, j+2})$
\end{itemize}
\item solver 1000
\begin{itemize}[leftmargin=0em]
    \item $V_{i+2, j+1}, H_{i, j+2}, V_{i+2, j}, h_{d2}(i+1,j+1,H_{i+1, j}, H_{i+1, j+2})$
    \item $v_{d2}(i+1,j+1,V_{i, j+1}, V_{i+2, j+1}), H_{i, j+2}, V_{i+2, j}, H_{i+1, j+2}$
   \item $V_{i+2, j+1}, H_{i, j+2}, v_{d2}(i+1,j,V_{i, j}, V_{i+2, j}), H_{i+1, j}$
\end{itemize}

\item solver 1001
\begin{itemize}[leftmargin=0em]
   \item $V_{i+2, j+1}, H_{i, j+2}, v_{d2}(i+1,j,V_{i, j}, V_{i+2, j}), H_{i+1, j}$
   \item $V_{i, j+1}, h_{d2}(i,j+1,H_{i, j+2}, H_{i, j}), V_{i, j}, H_{i+1, j+2}$
\end{itemize}

\item solver 1010
\begin{itemize}[leftmargin=0em]
   \item $v_{d2}(i+1,j+1,V_{i, j+1}, V_{i+2, j+1}), H_{i, j+2}, V_{i+2, j}, H_{i+1, j+2}$
   \item $V_{i+2, j+1}, h_{d2}(i,j+1,H_{i, j+2}, H_{i, j}), V_{i, j}, H_{i+1, j}$
\end{itemize}

\item solver 1011
\begin{itemize}[leftmargin=0em]
    \item $V_{i, j+1}, h_{d2}(i,j+1,H_{i, j+2}, H_{i, j}), V_{i, j}, h_{d2}(i+1,j+1,H_{i+1, j}, H_{i+1, j+2})$
\end{itemize}
\item solver 1100
\begin{itemize}[leftmargin=0em]
    \item $v_{d2}(i+1,j+1,V_{i, j+1}, V_{i+2, j+1}), H_{i, j+2}, v_{d2}(i+1,j,V_{i, j}, V_{i+2, j}), H_{i+1, j}$
    \item $V_{i, j+1}, h_{d2}(i,j+1,H_{i, j+2}, H_{i, j}), V_{i, j}, H_{i+1, j}$
   \item $V_{i+2, j+1}, H_{i, j+2}, V_{i+2, j}, h_{d2}(i+1,j+1,H_{i+1, j}, H_{i+1, j+2})$
\end{itemize}

\item solver 1101
\begin{itemize}[leftmargin=0em]
    \item $v_{d2}(i+1,j+1,V_{i, j+1}, V_{i+2, j+1}), H_{i, j+2}, v_{d2}(i+1,j,V_{i, j}, V_{i+2, j}), H_{i+1, j}$ 
\end{itemize}
\item solver 1110
\begin{itemize}[leftmargin=0em]
    \item $v_{d2}(i+1,j+1,V_{i, j+1}, V_{i+2, j+1}), H_{i, j+2}, v_{d2}(i+1,j,V_{i, j}, V_{i+2, j}), H_{i+1, j}$
\end{itemize}
\item solver 1111
\begin{itemize}[leftmargin=0em]
   \item $v_{d2}(i+1,j+1,V_{i, j+1}, V_{i+2, j+1}), H_{i, j+2}, v_{d2}(i+1,j,V_{i, j}, V_{i+2, j}), H_{i+1, j}$
   \item $V_{i, j+1}, h_{d2}(i,j+1,H_{i, j+2}, H_{i, j}), V_{i, j}, h_{d2}(i+1,j+1,H_{i+1, j}, H_{i+1, j+2})$ 
\end{itemize}
\end{enumerate}

with $v_{d1}$ and $v_{d2}$ (resp. $h_{d1}$ and $h_{d2}$) to compute random colors different from its input:
\begin{align*}
	v_{d1}(i,j,c) = & \left\{
	\begin{array}{ll}
		m & \text{if } m < c\\
		m+1 & \text{if } m \geq c
	\end{array}
	\right.\\
	& \text{where }m = \mathcal{V}(i,j)\% (n_c-1)\\
	v_{d2}(i,j,c_1, c_2) = & \left\{
	\begin{array}{ll}
		m' & \text{if } m' < \min(c_1,c_2) \\
		m'+1 & \text{if } \min(c_1,c_2) \leq m' < \max(c_1,c_2)-1 \\
		m'+2 & \text{if } m' \geq \max(c_1,c_2)-1 \\
	\end{array}
	\right.\\
	& \text{where }m' = \mathcal{V}(i,j)\% (n_c-2).\\
\end{align*}
\end{appendices}
\end{document}

%% file: tile.tex
\begin{tikzpicture}[set style={{help lines}+=[dashed]},scale=1.8,font=\boldmath]
		
		\draw[fill=blue!30] (0,0) -- (0.5,0.5) -- (1,0);
		\draw[fill=red!30] (0,1) -- (0.5,0.5) -- (1,1);
		\draw[fill=yellow!30] (0,0) -- (0.5,0.5) -- (0,1);
		\draw[fill=green!30] (1,0) -- (0.5,0.5) -- (1,1);
		\draw[fill=red!30] (0,1) -- (0.5,1.5) -- (1,1);
		\draw[fill=yellow!30] (0,0) -- (-0.5,0.5) -- (0,1);
		\draw[fill=green!30] (1,0) -- (1.5,0.5) -- (1,1);
		\draw[fill=blue!30] (0,0) -- (0.5,-0.5) -- (1,0);
		\draw[dashed,gray,very thick] (-0.5,-0.5) grid (1.5,1.5);
		
		\node[below] at (0.5,0) {$H_{i,j}$};
		\node[above] at (0.5,1) {$H_{i,j+1}$};
		\node[left] at (0,0.5) {$V_{i,j}$};
		\node[right, xshift=-3pt] at (1,0.5) {$V_{i+1,j}$};
		\node[] at (0.5, 0.7) {$(i,j)$};
	\end{tikzpicture}

%% file: tile_in_wall.tex
\begin{tikzpicture}[set style={{help lines}+=[dashed]},scale=1.6,font=\boldmath]
	\draw[dashed,gray,very thick] (-0.4,-0.4) grid (1.4,1.4);
	\draw[gray,thick] (-0.4,-0.4) rectangle (1.4,1.4);
	\draw[fill=blue!30] (0,0) -- (0.5,0.5) -- (1,0);
	\draw[fill=red!30] (0,1) -- (0.5,0.5) -- (1,1);
	\draw[fill=green!30] (0,0) -- (0.5,0.5) -- (0,1);
	\draw[fill=green!30] (1,0) -- (0.5,0.5) -- (1,1);
	
	\draw[very thick, black!30] (0,0) -- (1,1);
	\draw[very thick, black!30] (0,1) -- (1,0);
	
	
	\draw[very thick] (-0.4,0.25) -- (1.4,0.25);
	\draw[very thick] (0.25,0.25) -- (0.25,1.4);
	\draw[very thick] (0.75,-0.25) -- (0.75,0.25);
	
	\draw[very thick] (-0.25, -0.25) -- (1.4,-0.25);
	\draw[very thick] (0.5, -0.4) -- (0.5, -0.25);
	\draw[very thick] (1.25, -0.4) -- (1.25, -0.25);
	\draw[very thick] (-0.25, -0.4) -- (-0.25, 0.25);
	\draw[very thick] (-0.4, 1.25) -- (0.25,1.25);
	
	\draw [fill=black] (0, 0.25) circle (0.05) node[above, xshift=-7.5pt] {$V_{i,j}$};
	\draw [fill=black] (1, 0.25) circle (0.05) node[above, xshift=1.3pt] {$V_{i+1,j}$};
	\draw [fill=black] (0.75, 0) circle (0.05) node[left, yshift=-6pt] {$H_{i,j}$};
	\draw [fill=black] (0.25, 1) circle (0.05) node[right, yshift=7pt] {$H_{i,j+1}$};
	
	\end{tikzpicture}

%% file: invalid_tile_cross.tex
\begin{tikzpicture}[set style={{help lines}+=[dashed]},scale=3]
\draw[gray,very thick] (0,0) rectangle (1,1);
\draw[very thick, black] (0.25,0) -- (0.25,1);
\draw[very thick, black] (0,0.5) -- (1,0.5);
\end{tikzpicture}

%% file: invalid_tile_small_brick.tex
\begin{tikzpicture}[set style={{help lines}+=[dashed]},scale=3]
\draw[gray,very thick] (0,0) rectangle (1,1);
\draw[very thick, black] (0,0.75) -- (0.75,0.75);
\draw[very thick, black] (0.75, 1) -- (0.75, 0.25);
\draw[very thick, black] (1, 0.25) -- (0.25, 0.25);
\draw[very thick, black] (0.25, 0) -- (0.25, 0.75);
\end{tikzpicture}

%% file: invalid_tile_bottom.tex
\begin{tikzpicture}[set style={{help lines}+=[dashed]},scale=3]
\draw[gray,very thick] (0,0) rectangle (1,1);
\draw[very thick, black] (0,0.75) -- (0.25,0.75);
\draw[very thick, black] (0.25, 1) -- (0.25, 0.25);
\draw[very thick, black] (1, 0.25) -- (0.25, 0.25);
\draw[very thick, black] (0.5, 0) -- (0.5, 0.25);
\end{tikzpicture}

%% file: invalid_tile_top.tex
\begin{tikzpicture}[set style={{help lines}+=[dashed]},scale=3]
\draw[gray,very thick] (0,0) rectangle (1,1);
\draw[very thick, black] (0,0.75) -- (0.5,0.75);
\draw[very thick, black] (0.25, 1) -- (0.25, 0.75);
\draw[very thick, black] (1, 0.25) -- (0.5, 0.25);
\draw[very thick, black] (0.5, 0) -- (0.5, 0.75);
\end{tikzpicture}

%% file: invalid_tile_diagonal.tex
\begin{tikzpicture}[set style={{help lines}+=[dashed]},scale=3]
\draw[gray,very thick] (0,0) rectangle (1,1);
\draw[very thick, black] (0,0.75) -- (0.25,0.75);
\draw[very thick, black] (0.25, 1) -- (0.25, 0.75);
\draw[very thick, black] (1, 0.25) -- (0.5, 0.25);
\draw[very thick, black] (0.5, 0) -- (0.5, 0.25);
\draw[very thick, black] (0.25,0.75) -- (0.5,0.25);
\end{tikzpicture}

%% file: orientation_vertical.tex
\begin{tikzpicture}[set style={{help lines}+=[dashed]},scale=3]
\draw[gray,very thick] (0,0) rectangle (1,1);

\draw[very thick, black] (0.25,0.5) -- (0.25,1);
\draw[very thick, black] (0,0.5) -- (1,0.5);
\draw[very thick, black] (0.75, 0) -- (0.75,0.50);
\begin{scope}[xshift=2cm];
\draw[fill=blue] (0,0) -- (0.5,0.5) -- (1,0);
\draw[fill=red] (0,1) -- (0.5,0.5) -- (1,1);
\draw[fill=green] (0,0) -- (0.5,0.5) -- (0,1);
\draw[fill=green] (1,0) -- (0.5,0.5) -- (1,1);
\draw[very thick, black] (0,0) -- (1,1);
\draw[very thick, black] (0,1) -- (1,0);
\draw[very thick] (0,0) rectangle (1,1);
\end{scope}
\begin{scope}[xshift=4cm];
\draw[very thick,fill=yellow!30] (0,0) rectangle (1,1);
\draw[very thick] (0,0.5) -- (1,0.5);	
\end{scope}

\draw[ultra thick,arrows={triangle 45-triangle 45}] (1.1, 0.5) -- (1.9, 0.5);
\draw[ultra thick,arrows={triangle 45-triangle 45}] (3.1, 0.5) -- (3.9, 0.5);
\end{tikzpicture}

%% file: orientation+horizontal.tex
\begin{tikzpicture}[set style={{help lines}+=[dashed]},scale=3]
\draw[gray,very thick] (0,0) rectangle (1,1);

\draw[very thick, black] (0.5,0) -- (0.5,1);
\draw[very thick, black] (0,0.25) -- (0.5,0.25);
\draw[very thick, black] (0.5, 0.75) -- (1, 0.75);
\begin{scope}[xshift=2cm];
\draw[fill=green] (0,0) -- (0.5,0.5) -- (1,0);
\draw[fill=green] (0,1) -- (0.5,0.5) -- (1,1);
\draw[fill=red] (0,0) -- (0.5,0.5) -- (0,1);
\draw[fill=blue] (1,0) -- (0.5,0.5) -- (1,1);
\draw[very thick, black] (0,0) -- (1,1);
\draw[very thick, black] (0,1) -- (1,0);
\draw[very thick] (0,0) rectangle (1,1);
\end{scope}
\begin{scope}[xshift=4cm];
\draw[very thick,fill=red!50] (0,0) rectangle (1,1);
\draw[very thick] (0.5,0) -- (0.5,1);	
\end{scope}
\draw[ultra thick,arrows={triangle 45-triangle 45}] (1.1, 0.5) -- (1.9, 0.5);
\draw[ultra thick,arrows={triangle 45-triangle 45}] (3.1, 0.5) -- (3.9, 0.5);
\end{tikzpicture}

%% file: tile_set.tex
\begin{tikzpicture}[set style={{help lines}+=[dashed]},scale=1.5]
\begin{scope}[xshift=0.000000cm,yshift=0.000000cm];
\draw[fill=red] (0,0) -- (0.5,0.5) -- (1,0);
\draw[fill=red] (0,1) -- (0.5,0.5) -- (1,1);
\draw[fill=red] (0,0) -- (0.5,0.5) -- (0,1);
\draw[fill=green] (1,0) -- (0.5,0.5) -- (1,1);
\draw[very thick, black] (0,0) -- (1,1);
\draw[very thick, black] (0,1) -- (1,0);
\draw[very thick] (0,0) rectangle (1,1);
\end{scope}
\begin{scope}[xshift=1.100000cm,yshift=0.000000cm];
\draw[fill=red] (0,0) -- (0.5,0.5) -- (1,0);
\draw[fill=red] (0,1) -- (0.5,0.5) -- (1,1);
\draw[fill=red] (0,0) -- (0.5,0.5) -- (0,1);
\draw[fill=blue] (1,0) -- (0.5,0.5) -- (1,1);
\draw[very thick, black] (0,0) -- (1,1);
\draw[very thick, black] (0,1) -- (1,0);
\draw[very thick] (0,0) rectangle (1,1);
\end{scope}
\begin{scope}[xshift=2.200000cm,yshift=0.000000cm];
\draw[fill=red] (0,0) -- (0.5,0.5) -- (1,0);
\draw[fill=red] (0,1) -- (0.5,0.5) -- (1,1);
\draw[fill=green] (0,0) -- (0.5,0.5) -- (0,1);
\draw[fill=red] (1,0) -- (0.5,0.5) -- (1,1);
\draw[very thick, black] (0,0) -- (1,1);
\draw[very thick, black] (0,1) -- (1,0);
\draw[very thick] (0,0) rectangle (1,1);
\end{scope}
\begin{scope}[xshift=3.300000cm,yshift=0.000000cm];
\draw[fill=red] (0,0) -- (0.5,0.5) -- (1,0);
\draw[fill=red] (0,1) -- (0.5,0.5) -- (1,1);
\draw[fill=green] (0,0) -- (0.5,0.5) -- (0,1);
\draw[fill=blue] (1,0) -- (0.5,0.5) -- (1,1);
\draw[very thick, black] (0,0) -- (1,1);
\draw[very thick, black] (0,1) -- (1,0);
\draw[very thick] (0,0) rectangle (1,1);
\end{scope}
\begin{scope}[xshift=4.400000cm,yshift=0.000000cm];
\draw[fill=red] (0,0) -- (0.5,0.5) -- (1,0);
\draw[fill=red] (0,1) -- (0.5,0.5) -- (1,1);
\draw[fill=blue] (0,0) -- (0.5,0.5) -- (0,1);
\draw[fill=red] (1,0) -- (0.5,0.5) -- (1,1);
\draw[very thick, black] (0,0) -- (1,1);
\draw[very thick, black] (0,1) -- (1,0);
\draw[very thick] (0,0) rectangle (1,1);
\end{scope}
\begin{scope}[xshift=5.500000cm,yshift=0.000000cm];
\draw[fill=red] (0,0) -- (0.5,0.5) -- (1,0);
\draw[fill=red] (0,1) -- (0.5,0.5) -- (1,1);
\draw[fill=blue] (0,0) -- (0.5,0.5) -- (0,1);
\draw[fill=green] (1,0) -- (0.5,0.5) -- (1,1);
\draw[very thick, black] (0,0) -- (1,1);
\draw[very thick, black] (0,1) -- (1,0);
\draw[very thick] (0,0) rectangle (1,1);
\end{scope}
\begin{scope}[xshift=6.600000cm,yshift=0.000000cm];
\draw[fill=green] (0,0) -- (0.5,0.5) -- (1,0);
\draw[fill=red] (0,1) -- (0.5,0.5) -- (1,1);
\draw[fill=red] (0,0) -- (0.5,0.5) -- (0,1);
\draw[fill=red] (1,0) -- (0.5,0.5) -- (1,1);
\draw[very thick, black] (0,0) -- (1,1);
\draw[very thick, black] (0,1) -- (1,0);
\draw[very thick] (0,0) rectangle (1,1);
\end{scope}
\begin{scope}[xshift=7.700000cm,yshift=0.000000cm];
\draw[fill=green] (0,0) -- (0.5,0.5) -- (1,0);
\draw[fill=red] (0,1) -- (0.5,0.5) -- (1,1);
\draw[fill=green] (0,0) -- (0.5,0.5) -- (0,1);
\draw[fill=green] (1,0) -- (0.5,0.5) -- (1,1);
\draw[very thick, black] (0,0) -- (1,1);
\draw[very thick, black] (0,1) -- (1,0);
\draw[very thick] (0,0) rectangle (1,1);
\end{scope}
\begin{scope}[xshift=8.800000cm,yshift=0.000000cm];
\draw[fill=green] (0,0) -- (0.5,0.5) -- (1,0);
\draw[fill=red] (0,1) -- (0.5,0.5) -- (1,1);
\draw[fill=blue] (0,0) -- (0.5,0.5) -- (0,1);
\draw[fill=blue] (1,0) -- (0.5,0.5) -- (1,1);
\draw[very thick, black] (0,0) -- (1,1);
\draw[very thick, black] (0,1) -- (1,0);
\draw[very thick] (0,0) rectangle (1,1);
\end{scope}
\begin{scope}[xshift=0.000000cm,yshift=1.100000cm];
\draw[fill=blue] (0,0) -- (0.5,0.5) -- (1,0);
\draw[fill=red] (0,1) -- (0.5,0.5) -- (1,1);
\draw[fill=red] (0,0) -- (0.5,0.5) -- (0,1);
\draw[fill=red] (1,0) -- (0.5,0.5) -- (1,1);
\draw[very thick, black] (0,0) -- (1,1);
\draw[very thick, black] (0,1) -- (1,0);
\draw[very thick] (0,0) rectangle (1,1);
\end{scope}
\begin{scope}[xshift=1.100000cm,yshift=1.100000cm];
\draw[fill=blue] (0,0) -- (0.5,0.5) -- (1,0);
\draw[fill=red] (0,1) -- (0.5,0.5) -- (1,1);
\draw[fill=green] (0,0) -- (0.5,0.5) -- (0,1);
\draw[fill=green] (1,0) -- (0.5,0.5) -- (1,1);
\draw[very thick, black] (0,0) -- (1,1);
\draw[very thick, black] (0,1) -- (1,0);
\draw[very thick] (0,0) rectangle (1,1);
\end{scope}
\begin{scope}[xshift=2.200000cm,yshift=1.100000cm];
\draw[fill=blue] (0,0) -- (0.5,0.5) -- (1,0);
\draw[fill=red] (0,1) -- (0.5,0.5) -- (1,1);
\draw[fill=blue] (0,0) -- (0.5,0.5) -- (0,1);
\draw[fill=blue] (1,0) -- (0.5,0.5) -- (1,1);
\draw[very thick, black] (0,0) -- (1,1);
\draw[very thick, black] (0,1) -- (1,0);
\draw[very thick] (0,0) rectangle (1,1);
\end{scope}
\begin{scope}[xshift=3.300000cm,yshift=1.100000cm];
\draw[fill=red] (0,0) -- (0.5,0.5) -- (1,0);
\draw[fill=green] (0,1) -- (0.5,0.5) -- (1,1);
\draw[fill=red] (0,0) -- (0.5,0.5) -- (0,1);
\draw[fill=red] (1,0) -- (0.5,0.5) -- (1,1);
\draw[very thick, black] (0,0) -- (1,1);
\draw[very thick, black] (0,1) -- (1,0);
\draw[very thick] (0,0) rectangle (1,1);
\end{scope}
\begin{scope}[xshift=4.400000cm,yshift=1.100000cm];
\draw[fill=red] (0,0) -- (0.5,0.5) -- (1,0);
\draw[fill=green] (0,1) -- (0.5,0.5) -- (1,1);
\draw[fill=green] (0,0) -- (0.5,0.5) -- (0,1);
\draw[fill=green] (1,0) -- (0.5,0.5) -- (1,1);
\draw[very thick, black] (0,0) -- (1,1);
\draw[very thick, black] (0,1) -- (1,0);
\draw[very thick] (0,0) rectangle (1,1);
\end{scope}
\begin{scope}[xshift=5.500000cm,yshift=1.100000cm];
\draw[fill=red] (0,0) -- (0.5,0.5) -- (1,0);
\draw[fill=green] (0,1) -- (0.5,0.5) -- (1,1);
\draw[fill=blue] (0,0) -- (0.5,0.5) -- (0,1);
\draw[fill=blue] (1,0) -- (0.5,0.5) -- (1,1);
\draw[very thick, black] (0,0) -- (1,1);
\draw[very thick, black] (0,1) -- (1,0);
\draw[very thick] (0,0) rectangle (1,1);
\end{scope}
\begin{scope}[xshift=6.600000cm,yshift=1.100000cm];
\draw[fill=green] (0,0) -- (0.5,0.5) -- (1,0);
\draw[fill=green] (0,1) -- (0.5,0.5) -- (1,1);
\draw[fill=red] (0,0) -- (0.5,0.5) -- (0,1);
\draw[fill=green] (1,0) -- (0.5,0.5) -- (1,1);
\draw[very thick, black] (0,0) -- (1,1);
\draw[very thick, black] (0,1) -- (1,0);
\draw[very thick] (0,0) rectangle (1,1);
\end{scope}
\begin{scope}[xshift=7.700000cm,yshift=1.100000cm];
\draw[fill=green] (0,0) -- (0.5,0.5) -- (1,0);
\draw[fill=green] (0,1) -- (0.5,0.5) -- (1,1);
\draw[fill=red] (0,0) -- (0.5,0.5) -- (0,1);
\draw[fill=blue] (1,0) -- (0.5,0.5) -- (1,1);
\draw[very thick, black] (0,0) -- (1,1);
\draw[very thick, black] (0,1) -- (1,0);
\draw[very thick] (0,0) rectangle (1,1);
\end{scope}
\begin{scope}[xshift=8.800000cm,yshift=1.100000cm];
\draw[fill=green] (0,0) -- (0.5,0.5) -- (1,0);
\draw[fill=green] (0,1) -- (0.5,0.5) -- (1,1);
\draw[fill=green] (0,0) -- (0.5,0.5) -- (0,1);
\draw[fill=red] (1,0) -- (0.5,0.5) -- (1,1);
\draw[very thick, black] (0,0) -- (1,1);
\draw[very thick, black] (0,1) -- (1,0);
\draw[very thick] (0,0) rectangle (1,1);
\end{scope}
\begin{scope}[xshift=0.000000cm,yshift=2.200000cm];
\draw[fill=green] (0,0) -- (0.5,0.5) -- (1,0);
\draw[fill=green] (0,1) -- (0.5,0.5) -- (1,1);
\draw[fill=green] (0,0) -- (0.5,0.5) -- (0,1);
\draw[fill=blue] (1,0) -- (0.5,0.5) -- (1,1);
\draw[very thick, black] (0,0) -- (1,1);
\draw[very thick, black] (0,1) -- (1,0);
\draw[very thick] (0,0) rectangle (1,1);
\end{scope}
\begin{scope}[xshift=1.100000cm,yshift=2.200000cm];
\draw[fill=green] (0,0) -- (0.5,0.5) -- (1,0);
\draw[fill=green] (0,1) -- (0.5,0.5) -- (1,1);
\draw[fill=blue] (0,0) -- (0.5,0.5) -- (0,1);
\draw[fill=red] (1,0) -- (0.5,0.5) -- (1,1);
\draw[very thick, black] (0,0) -- (1,1);
\draw[very thick, black] (0,1) -- (1,0);
\draw[very thick] (0,0) rectangle (1,1);
\end{scope}
\begin{scope}[xshift=2.200000cm,yshift=2.200000cm];
\draw[fill=green] (0,0) -- (0.5,0.5) -- (1,0);
\draw[fill=green] (0,1) -- (0.5,0.5) -- (1,1);
\draw[fill=blue] (0,0) -- (0.5,0.5) -- (0,1);
\draw[fill=green] (1,0) -- (0.5,0.5) -- (1,1);
\draw[very thick, black] (0,0) -- (1,1);
\draw[very thick, black] (0,1) -- (1,0);
\draw[very thick] (0,0) rectangle (1,1);
\end{scope}
\begin{scope}[xshift=3.300000cm,yshift=2.200000cm];
\draw[fill=blue] (0,0) -- (0.5,0.5) -- (1,0);
\draw[fill=green] (0,1) -- (0.5,0.5) -- (1,1);
\draw[fill=red] (0,0) -- (0.5,0.5) -- (0,1);
\draw[fill=red] (1,0) -- (0.5,0.5) -- (1,1);
\draw[very thick, black] (0,0) -- (1,1);
\draw[very thick, black] (0,1) -- (1,0);
\draw[very thick] (0,0) rectangle (1,1);
\end{scope}
\begin{scope}[xshift=4.400000cm,yshift=2.200000cm];
\draw[fill=blue] (0,0) -- (0.5,0.5) -- (1,0);
\draw[fill=green] (0,1) -- (0.5,0.5) -- (1,1);
\draw[fill=green] (0,0) -- (0.5,0.5) -- (0,1);
\draw[fill=green] (1,0) -- (0.5,0.5) -- (1,1);
\draw[very thick, black] (0,0) -- (1,1);
\draw[very thick, black] (0,1) -- (1,0);
\draw[very thick] (0,0) rectangle (1,1);
\end{scope}
\begin{scope}[xshift=5.500000cm,yshift=2.200000cm];
\draw[fill=blue] (0,0) -- (0.5,0.5) -- (1,0);
\draw[fill=green] (0,1) -- (0.5,0.5) -- (1,1);
\draw[fill=blue] (0,0) -- (0.5,0.5) -- (0,1);
\draw[fill=blue] (1,0) -- (0.5,0.5) -- (1,1);
\draw[very thick, black] (0,0) -- (1,1);
\draw[very thick, black] (0,1) -- (1,0);
\draw[very thick] (0,0) rectangle (1,1);
\end{scope}
\begin{scope}[xshift=6.600000cm,yshift=2.200000cm];
\draw[fill=red] (0,0) -- (0.5,0.5) -- (1,0);
\draw[fill=blue] (0,1) -- (0.5,0.5) -- (1,1);
\draw[fill=red] (0,0) -- (0.5,0.5) -- (0,1);
\draw[fill=red] (1,0) -- (0.5,0.5) -- (1,1);
\draw[very thick, black] (0,0) -- (1,1);
\draw[very thick, black] (0,1) -- (1,0);
\draw[very thick] (0,0) rectangle (1,1);
\end{scope}
\begin{scope}[xshift=7.700000cm,yshift=2.200000cm];
\draw[fill=red] (0,0) -- (0.5,0.5) -- (1,0);
\draw[fill=blue] (0,1) -- (0.5,0.5) -- (1,1);
\draw[fill=green] (0,0) -- (0.5,0.5) -- (0,1);
\draw[fill=green] (1,0) -- (0.5,0.5) -- (1,1);
\draw[very thick, black] (0,0) -- (1,1);
\draw[very thick, black] (0,1) -- (1,0);
\draw[very thick] (0,0) rectangle (1,1);
\end{scope}
\begin{scope}[xshift=8.800000cm,yshift=2.200000cm];
\draw[fill=red] (0,0) -- (0.5,0.5) -- (1,0);
\draw[fill=blue] (0,1) -- (0.5,0.5) -- (1,1);
\draw[fill=blue] (0,0) -- (0.5,0.5) -- (0,1);
\draw[fill=blue] (1,0) -- (0.5,0.5) -- (1,1);
\draw[very thick, black] (0,0) -- (1,1);
\draw[very thick, black] (0,1) -- (1,0);
\draw[very thick] (0,0) rectangle (1,1);
\end{scope}
\begin{scope}[xshift=0.000000cm,yshift=3.300000cm];
\draw[fill=green] (0,0) -- (0.5,0.5) -- (1,0);
\draw[fill=blue] (0,1) -- (0.5,0.5) -- (1,1);
\draw[fill=red] (0,0) -- (0.5,0.5) -- (0,1);
\draw[fill=red] (1,0) -- (0.5,0.5) -- (1,1);
\draw[very thick, black] (0,0) -- (1,1);
\draw[very thick, black] (0,1) -- (1,0);
\draw[very thick] (0,0) rectangle (1,1);
\end{scope}
\begin{scope}[xshift=1.100000cm,yshift=3.300000cm];
\draw[fill=green] (0,0) -- (0.5,0.5) -- (1,0);
\draw[fill=blue] (0,1) -- (0.5,0.5) -- (1,1);
\draw[fill=green] (0,0) -- (0.5,0.5) -- (0,1);
\draw[fill=green] (1,0) -- (0.5,0.5) -- (1,1);
\draw[very thick, black] (0,0) -- (1,1);
\draw[very thick, black] (0,1) -- (1,0);
\draw[very thick] (0,0) rectangle (1,1);
\end{scope}
\begin{scope}[xshift=2.200000cm,yshift=3.300000cm];
\draw[fill=green] (0,0) -- (0.5,0.5) -- (1,0);
\draw[fill=blue] (0,1) -- (0.5,0.5) -- (1,1);
\draw[fill=blue] (0,0) -- (0.5,0.5) -- (0,1);
\draw[fill=blue] (1,0) -- (0.5,0.5) -- (1,1);
\draw[very thick, black] (0,0) -- (1,1);
\draw[very thick, black] (0,1) -- (1,0);
\draw[very thick] (0,0) rectangle (1,1);
\end{scope}
\begin{scope}[xshift=3.300000cm,yshift=3.300000cm];
\draw[fill=blue] (0,0) -- (0.5,0.5) -- (1,0);
\draw[fill=blue] (0,1) -- (0.5,0.5) -- (1,1);
\draw[fill=red] (0,0) -- (0.5,0.5) -- (0,1);
\draw[fill=green] (1,0) -- (0.5,0.5) -- (1,1);
\draw[very thick, black] (0,0) -- (1,1);
\draw[very thick, black] (0,1) -- (1,0);
\draw[very thick] (0,0) rectangle (1,1);
\end{scope}
\begin{scope}[xshift=4.400000cm,yshift=3.300000cm];
\draw[fill=blue] (0,0) -- (0.5,0.5) -- (1,0);
\draw[fill=blue] (0,1) -- (0.5,0.5) -- (1,1);
\draw[fill=red] (0,0) -- (0.5,0.5) -- (0,1);
\draw[fill=blue] (1,0) -- (0.5,0.5) -- (1,1);
\draw[very thick, black] (0,0) -- (1,1);
\draw[very thick, black] (0,1) -- (1,0);
\draw[very thick] (0,0) rectangle (1,1);
\end{scope}
\begin{scope}[xshift=5.500000cm,yshift=3.300000cm];
\draw[fill=blue] (0,0) -- (0.5,0.5) -- (1,0);
\draw[fill=blue] (0,1) -- (0.5,0.5) -- (1,1);
\draw[fill=green] (0,0) -- (0.5,0.5) -- (0,1);
\draw[fill=red] (1,0) -- (0.5,0.5) -- (1,1);
\draw[very thick, black] (0,0) -- (1,1);
\draw[very thick, black] (0,1) -- (1,0);
\draw[very thick] (0,0) rectangle (1,1);
\end{scope}
\begin{scope}[xshift=6.600000cm,yshift=3.300000cm];
\draw[fill=blue] (0,0) -- (0.5,0.5) -- (1,0);
\draw[fill=blue] (0,1) -- (0.5,0.5) -- (1,1);
\draw[fill=green] (0,0) -- (0.5,0.5) -- (0,1);
\draw[fill=blue] (1,0) -- (0.5,0.5) -- (1,1);
\draw[very thick, black] (0,0) -- (1,1);
\draw[very thick, black] (0,1) -- (1,0);
\draw[very thick] (0,0) rectangle (1,1);
\end{scope}
\begin{scope}[xshift=7.700000cm,yshift=3.300000cm];
\draw[fill=blue] (0,0) -- (0.5,0.5) -- (1,0);
\draw[fill=blue] (0,1) -- (0.5,0.5) -- (1,1);
\draw[fill=blue] (0,0) -- (0.5,0.5) -- (0,1);
\draw[fill=red] (1,0) -- (0.5,0.5) -- (1,1);
\draw[very thick, black] (0,0) -- (1,1);
\draw[very thick, black] (0,1) -- (1,0);
\draw[very thick] (0,0) rectangle (1,1);
\end{scope}
\begin{scope}[xshift=8.800000cm,yshift=3.300000cm];
\draw[fill=blue] (0,0) -- (0.5,0.5) -- (1,0);
\draw[fill=blue] (0,1) -- (0.5,0.5) -- (1,1);
\draw[fill=blue] (0,0) -- (0.5,0.5) -- (0,1);
\draw[fill=green] (1,0) -- (0.5,0.5) -- (1,1);
\draw[very thick, black] (0,0) -- (1,1);
\draw[very thick, black] (0,1) -- (1,0);
\draw[very thick] (0,0) rectangle (1,1);
\end{scope}
\end{tikzpicture}

%% file: long_lines_border_3.tex

\begin{tikzpicture}[set style={{help lines}+=[dashed]},scale=1]
\draw[very thick, black] (0.000000,0.750000) -- (1.000000,0.750000);
\draw[very thick, black] (0.500000,1.000000) -- (0.500000,0.750000);
\draw[very thick, black] (0.750000,0.000000) -- (0.750000,0.750000);
\draw[very thick, black] (1.750000,0.000000) -- (1.750000,1.000000);
\draw[very thick, black] (1.000000,0.750000) -- (1.750000,0.750000);
\draw[very thick, black] (1.750000,0.250000) -- (2.000000,0.250000);
\draw[very thick, black] (2.000000,0.250000) -- (3.000000,0.250000);
\draw[very thick, black] (2.500000,1.000000) -- (2.500000,0.250000);
\draw[very thick, black] (2.250000,0.000000) -- (2.250000,0.250000);
\draw[very thick, black] (3.000000,0.250000) -- (4.000000,0.250000);
\draw[very thick, black] (3.250000,1.000000) -- (3.250000,0.250000);
\draw[very thick, black] (3.750000,0.000000) -- (3.750000,0.250000);
\draw[very thick, black] (4.000000,0.250000) -- (5.000000,0.250000);
\draw[very thick, black] (4.250000,1.000000) -- (4.250000,0.250000);
\draw[very thick, black] (4.500000,0.000000) -- (4.500000,0.250000);
\draw[very thick, black] (5.500000,0.000000) -- (5.500000,1.000000);
\draw[very thick, black] (5.000000,0.250000) -- (5.500000,0.250000);
\draw[very thick, black] (5.500000,0.500000) -- (6.000000,0.500000);
\draw[very thick, black] (6.750000,0.000000) -- (6.750000,1.000000);
\draw[very thick, black] (6.000000,0.500000) -- (6.750000,0.500000);
\draw[very thick, black] (6.750000,0.750000) -- (7.000000,0.750000);
\draw[very thick, black] (7.000000,0.750000) -- (8.000000,0.750000);
\draw[very thick, black] (7.250000,1.000000) -- (7.250000,0.750000);
\draw[very thick, black] (7.500000,0.000000) -- (7.500000,0.750000);
\draw[very thick, black] (8.250000,0.000000) -- (8.250000,1.000000);
\draw[very thick, black] (8.000000,0.750000) -- (8.250000,0.750000);
\draw[very thick, black] (8.250000,0.500000) -- (9.000000,0.500000);
\draw[very thick, black] (9.000000,0.500000) -- (10.000000,0.500000);
\draw[very thick, black] (9.750000,1.000000) -- (9.750000,0.500000);
\draw[very thick, black] (9.500000,0.000000) -- (9.500000,0.500000);
\draw[very thick, black] (10.000000,0.500000) -- (11.000000,0.500000);
\draw[very thick, black] (10.750000,1.000000) -- (10.750000,0.500000);
\draw[very thick, black] (10.500000,0.000000) -- (10.500000,0.500000);
\draw[very thick, black] (11.250000,0.000000) -- (11.250000,1.000000);
\draw[very thick, black] (11.000000,0.500000) -- (11.250000,0.500000);
\draw[very thick, black] (11.250000,0.750000) -- (12.000000,0.750000);
\draw[very thick, black] (0.500000,1.000000) -- (0.500000,2.000000);
\draw[very thick, black] (0.000000,1.250000) -- (0.500000,1.250000);
\draw[very thick, black] (0.500000,1.500000) -- (1.000000,1.500000);
\draw[very thick, black] (1.000000,1.500000) -- (2.000000,1.500000);
\draw[very thick, black] (1.500000,2.000000) -- (1.500000,1.500000);
\draw[very thick, black] (1.750000,1.000000) -- (1.750000,1.500000);
\draw[very thick, black] (2.500000,1.000000) -- (2.500000,2.000000);
\draw[very thick, black] (2.000000,1.500000) -- (2.500000,1.500000);
\draw[very thick, black] (2.500000,1.750000) -- (3.000000,1.750000);
\draw[very thick, black] (3.250000,1.000000) -- (3.250000,2.000000);
\draw[very thick, black] (3.000000,1.750000) -- (3.250000,1.750000);
\draw[very thick, black] (3.250000,1.500000) -- (4.000000,1.500000);
\draw[very thick, black] (4.250000,1.000000) -- (4.250000,2.000000);
\draw[very thick, black] (4.000000,1.500000) -- (4.250000,1.500000);
\draw[very thick, black] (4.250000,1.750000) -- (5.000000,1.750000);
\draw[very thick, black] (5.500000,1.000000) -- (5.500000,2.000000);
\draw[very thick, black] (5.000000,1.750000) -- (5.500000,1.750000);
\draw[very thick, black] (5.500000,1.250000) -- (6.000000,1.250000);
\draw[very thick, black] (6.750000,1.000000) -- (6.750000,2.000000);
\draw[very thick, black] (6.000000,1.250000) -- (6.750000,1.250000);
\draw[very thick, black] (6.750000,1.750000) -- (7.000000,1.750000);
\draw[very thick, black] (7.250000,1.000000) -- (7.250000,2.000000);
\draw[very thick, black] (7.000000,1.750000) -- (7.250000,1.750000);
\draw[very thick, black] (7.250000,1.250000) -- (8.000000,1.250000);
\draw[very thick, black] (8.000000,1.250000) -- (9.000000,1.250000);
\draw[very thick, black] (8.750000,2.000000) -- (8.750000,1.250000);
\draw[very thick, black] (8.250000,1.000000) -- (8.250000,1.250000);
\draw[very thick, black] (9.750000,1.000000) -- (9.750000,2.000000);
\draw[very thick, black] (9.000000,1.250000) -- (9.750000,1.250000);
\draw[very thick, black] (9.750000,1.500000) -- (10.000000,1.500000);
\draw[very thick, black] (10.750000,1.000000) -- (10.750000,2.000000);
\draw[very thick, black] (10.000000,1.500000) -- (10.750000,1.500000);
\draw[very thick, black] (10.750000,1.250000) -- (11.000000,1.250000);
\draw[very thick, black] (11.000000,1.250000) -- (12.000000,1.250000);
\draw[very thick, black] (11.750000,2.000000) -- (11.750000,1.250000);
\draw[very thick, black] (11.250000,1.000000) -- (11.250000,1.250000);
\draw[very thick, black] (0.500000,2.000000) -- (0.500000,3.000000);
\draw[very thick, black] (0.000000,2.750000) -- (0.500000,2.750000);
\draw[very thick, black] (0.500000,2.250000) -- (1.000000,2.250000);
\draw[very thick, black] (1.500000,2.000000) -- (1.500000,3.000000);
\draw[very thick, black] (1.000000,2.250000) -- (1.500000,2.250000);
\draw[very thick, black] (1.500000,2.500000) -- (2.000000,2.500000);
\draw[very thick, black] (2.000000,2.500000) -- (3.000000,2.500000);
\draw[very thick, black] (2.750000,3.000000) -- (2.750000,2.500000);
\draw[very thick, black] (2.500000,2.000000) -- (2.500000,2.500000);
\draw[very thick, black] (3.250000,2.000000) -- (3.250000,3.000000);
\draw[very thick, black] (3.000000,2.500000) -- (3.250000,2.500000);
\draw[very thick, black] (3.250000,2.750000) -- (4.000000,2.750000);
\draw[very thick, black] (4.000000,2.750000) -- (5.000000,2.750000);
\draw[very thick, black] (4.500000,3.000000) -- (4.500000,2.750000);
\draw[very thick, black] (4.250000,2.000000) -- (4.250000,2.750000);
\draw[very thick, black] (5.000000,2.750000) -- (6.000000,2.750000);
\draw[very thick, black] (5.250000,3.000000) -- (5.250000,2.750000);
\draw[very thick, black] (5.500000,2.000000) -- (5.500000,2.750000);
\draw[very thick, black] (6.000000,2.750000) -- (7.000000,2.750000);
\draw[very thick, black] (6.250000,3.000000) -- (6.250000,2.750000);
\draw[very thick, black] (6.750000,2.000000) -- (6.750000,2.750000);
\draw[very thick, black] (7.250000,2.000000) -- (7.250000,3.000000);
\draw[very thick, black] (7.000000,2.750000) -- (7.250000,2.750000);
\draw[very thick, black] (7.250000,2.250000) -- (8.000000,2.250000);
\draw[very thick, black] (8.750000,2.000000) -- (8.750000,3.000000);
\draw[very thick, black] (8.000000,2.250000) -- (8.750000,2.250000);
\draw[very thick, black] (8.750000,2.500000) -- (9.000000,2.500000);
\draw[very thick, black] (9.750000,2.000000) -- (9.750000,3.000000);
\draw[very thick, black] (9.000000,2.500000) -- (9.750000,2.500000);
\draw[very thick, black] (9.750000,2.250000) -- (10.000000,2.250000);
\draw[very thick, black] (10.000000,2.250000) -- (11.000000,2.250000);
\draw[very thick, black] (10.500000,3.000000) -- (10.500000,2.250000);
\draw[very thick, black] (10.750000,2.000000) -- (10.750000,2.250000);
\draw[very thick, black] (11.750000,2.000000) -- (11.750000,3.000000);
\draw[very thick, black] (11.000000,2.250000) -- (11.750000,2.250000);
\draw[very thick, black] (11.750000,2.750000) -- (12.000000,2.750000);
\draw[very thick, black] (0.000000,3.250000) -- (1.000000,3.250000);
\draw[very thick, black] (0.250000,4.000000) -- (0.250000,3.250000);
\draw[very thick, black] (0.500000,3.000000) -- (0.500000,3.250000);
\draw[very thick, black] (1.500000,3.000000) -- (1.500000,4.000000);
\draw[very thick, black] (1.000000,3.250000) -- (1.500000,3.250000);
\draw[very thick, black] (1.500000,3.750000) -- (2.000000,3.750000);
\draw[very thick, black] (2.750000,3.000000) -- (2.750000,4.000000);
\draw[very thick, black] (2.000000,3.750000) -- (2.750000,3.750000);
\draw[very thick, black] (2.750000,3.500000) -- (3.000000,3.500000);
\draw[very thick, black] (3.000000,3.500000) -- (4.000000,3.500000);
\draw[very thick, black] (3.500000,4.000000) -- (3.500000,3.500000);
\draw[very thick, black] (3.250000,3.000000) -- (3.250000,3.500000);
\draw[very thick, black] (4.500000,3.000000) -- (4.500000,4.000000);
\draw[very thick, black] (4.000000,3.500000) -- (4.500000,3.500000);
\draw[very thick, black] (4.500000,3.750000) -- (5.000000,3.750000);
\draw[very thick, black] (5.000000,3.750000) -- (6.000000,3.750000);
\draw[very thick, black] (5.500000,4.000000) -- (5.500000,3.750000);
\draw[very thick, black] (5.250000,3.000000) -- (5.250000,3.750000);
\draw[very thick, black] (6.250000,3.000000) -- (6.250000,4.000000);
\draw[very thick, black] (6.000000,3.750000) -- (6.250000,3.750000);
\draw[very thick, black] (6.250000,3.250000) -- (7.000000,3.250000);
\draw[very thick, black] (7.250000,3.000000) -- (7.250000,4.000000);
\draw[very thick, black] (7.000000,3.250000) -- (7.250000,3.250000);
\draw[very thick, black] (7.250000,3.500000) -- (8.000000,3.500000);
\draw[very thick, black] (8.750000,3.000000) -- (8.750000,4.000000);
\draw[very thick, black] (8.000000,3.500000) -- (8.750000,3.500000);
\draw[very thick, black] (8.750000,3.250000) -- (9.000000,3.250000);
\draw[very thick, black] (9.750000,3.000000) -- (9.750000,4.000000);
\draw[very thick, black] (9.000000,3.250000) -- (9.750000,3.250000);
\draw[very thick, black] (9.750000,3.500000) -- (10.000000,3.500000);
\draw[very thick, black] (10.500000,3.000000) -- (10.500000,4.000000);
\draw[very thick, black] (10.000000,3.500000) -- (10.500000,3.500000);
\draw[very thick, black] (10.500000,3.250000) -- (11.000000,3.250000);
\draw[very thick, black] (11.000000,3.250000) -- (12.000000,3.250000);
\draw[very thick, black] (11.500000,4.000000) -- (11.500000,3.250000);
\draw[very thick, black] (11.750000,3.000000) -- (11.750000,3.250000);
\draw[very thick, black] (0.000000,4.500000) -- (1.000000,4.500000);
\draw[very thick, black] (0.750000,5.000000) -- (0.750000,4.500000);
\draw[very thick, black] (0.250000,4.000000) -- (0.250000,4.500000);
\draw[very thick, black] (1.000000,4.500000) -- (2.000000,4.500000);
\draw[very thick, black] (1.750000,5.000000) -- (1.750000,4.500000);
\draw[very thick, black] (1.500000,4.000000) -- (1.500000,4.500000);
\draw[very thick, black] (2.750000,4.000000) -- (2.750000,5.000000);
\draw[very thick, black] (2.000000,4.500000) -- (2.750000,4.500000);
\draw[very thick, black] (2.750000,4.250000) -- (3.000000,4.250000);
\draw[very thick, black] (3.000000,4.250000) -- (4.000000,4.250000);
\draw[very thick, black] (3.250000,5.000000) -- (3.250000,4.250000);
\draw[very thick, black] (3.500000,4.000000) -- (3.500000,4.250000);
\draw[very thick, black] (4.000000,4.250000) -- (5.000000,4.250000);
\draw[very thick, black] (4.250000,5.000000) -- (4.250000,4.250000);
\draw[very thick, black] (4.500000,4.000000) -- (4.500000,4.250000);
\draw[very thick, black] (5.000000,4.250000) -- (6.000000,4.250000);
\draw[very thick, black] (5.750000,5.000000) -- (5.750000,4.250000);
\draw[very thick, black] (5.500000,4.000000) -- (5.500000,4.250000);
\draw[very thick, black] (6.000000,4.250000) -- (7.000000,4.250000);
\draw[very thick, black] (6.750000,5.000000) -- (6.750000,4.250000);
\draw[very thick, black] (6.250000,4.000000) -- (6.250000,4.250000);
\draw[very thick, black] (7.250000,4.000000) -- (7.250000,5.000000);
\draw[very thick, black] (7.000000,4.250000) -- (7.250000,4.250000);
\draw[very thick, black] (7.250000,4.500000) -- (8.000000,4.500000);
\draw[very thick, black] (8.750000,4.000000) -- (8.750000,5.000000);
\draw[very thick, black] (8.000000,4.500000) -- (8.750000,4.500000);
\draw[very thick, black] (8.750000,4.250000) -- (9.000000,4.250000);
\draw[very thick, black] (9.000000,4.250000) -- (10.000000,4.250000);
\draw[very thick, black] (9.250000,5.000000) -- (9.250000,4.250000);
\draw[very thick, black] (9.750000,4.000000) -- (9.750000,4.250000);
\draw[very thick, black] (10.000000,4.250000) -- (11.000000,4.250000);
\draw[very thick, black] (10.250000,5.000000) -- (10.250000,4.250000);
\draw[very thick, black] (10.500000,4.000000) -- (10.500000,4.250000);
\draw[very thick, black] (11.500000,4.000000) -- (11.500000,5.000000);
\draw[very thick, black] (11.000000,4.250000) -- (11.500000,4.250000);
\draw[very thick, black] (11.500000,4.500000) -- (12.000000,4.500000);
\draw[very thick, black] (0.750000,5.000000) -- (0.750000,6.000000);
\draw[very thick, black] (0.000000,5.750000) -- (0.750000,5.750000);
\draw[very thick, black] (0.750000,5.500000) -- (1.000000,5.500000);
\draw[very thick, black] (1.750000,5.000000) -- (1.750000,6.000000);
\draw[very thick, black] (1.000000,5.500000) -- (1.750000,5.500000);
\draw[very thick, black] (1.750000,5.250000) -- (2.000000,5.250000);
\draw[very thick, black] (2.000000,5.250000) -- (3.000000,5.250000);
\draw[very thick, black] (2.250000,6.000000) -- (2.250000,5.250000);
\draw[very thick, black] (2.750000,5.000000) -- (2.750000,5.250000);
\draw[very thick, black] (3.000000,5.250000) -- (4.000000,5.250000);
\draw[very thick, black] (3.750000,6.000000) -- (3.750000,5.250000);
\draw[very thick, black] (3.250000,5.000000) -- (3.250000,5.250000);
\draw[very thick, black] (4.250000,5.000000) -- (4.250000,6.000000);
\draw[very thick, black] (4.000000,5.250000) -- (4.250000,5.250000);
\draw[very thick, black] (4.250000,5.750000) -- (5.000000,5.750000);
\draw[very thick, black] (5.750000,5.000000) -- (5.750000,6.000000);
\draw[very thick, black] (5.000000,5.750000) -- (5.750000,5.750000);
\draw[very thick, black] (5.750000,5.250000) -- (6.000000,5.250000);
\draw[very thick, black] (6.000000,5.250000) -- (7.000000,5.250000);
\draw[very thick, black] (6.500000,6.000000) -- (6.500000,5.250000);
\draw[very thick, black] (6.750000,5.000000) -- (6.750000,5.250000);
\draw[very thick, black] (7.250000,5.000000) -- (7.250000,6.000000);
\draw[very thick, black] (7.000000,5.250000) -- (7.250000,5.250000);
\draw[very thick, black] (7.250000,5.500000) -- (8.000000,5.500000);
\draw[very thick, black] (8.000000,5.500000) -- (9.000000,5.500000);
\draw[very thick, black] (8.500000,6.000000) -- (8.500000,5.500000);
\draw[very thick, black] (8.750000,5.000000) -- (8.750000,5.500000);
\draw[very thick, black] (9.000000,5.500000) -- (10.000000,5.500000);
\draw[very thick, black] (9.500000,6.000000) -- (9.500000,5.500000);
\draw[very thick, black] (9.250000,5.000000) -- (9.250000,5.500000);
\draw[very thick, black] (10.000000,5.500000) -- (11.000000,5.500000);
\draw[very thick, black] (10.500000,6.000000) -- (10.500000,5.500000);
\draw[very thick, black] (10.250000,5.000000) -- (10.250000,5.500000);
\draw[very thick, black] (11.500000,5.000000) -- (11.500000,6.000000);
\draw[very thick, black] (11.000000,5.500000) -- (11.500000,5.500000);
\draw[very thick, black] (11.500000,5.750000) -- (12.000000,5.750000);
\draw[very thick, black] (0.750000,6.000000) -- (0.750000,7.000000);
\draw[very thick, black] (0.000000,6.500000) -- (0.750000,6.500000);
\draw[very thick, black] (0.750000,6.750000) -- (1.000000,6.750000);
\draw[very thick, black] (1.000000,6.750000) -- (2.000000,6.750000);
\draw[very thick, black] (1.500000,7.000000) -- (1.500000,6.750000);
\draw[very thick, black] (1.750000,6.000000) -- (1.750000,6.750000);
\draw[very thick, black] (2.000000,6.750000) -- (3.000000,6.750000);
\draw[very thick, black] (2.750000,7.000000) -- (2.750000,6.750000);
\draw[very thick, black] (2.250000,6.000000) -- (2.250000,6.750000);
\draw[very thick, black] (3.000000,6.750000) -- (4.000000,6.750000);
\draw[very thick, black] (3.500000,7.000000) -- (3.500000,6.750000);
\draw[very thick, black] (3.750000,6.000000) -- (3.750000,6.750000);
\draw[very thick, black] (4.250000,6.000000) -- (4.250000,7.000000);
\draw[very thick, black] (4.000000,6.750000) -- (4.250000,6.750000);
\draw[very thick, black] (4.250000,6.500000) -- (5.000000,6.500000);
\draw[very thick, black] (5.750000,6.000000) -- (5.750000,7.000000);
\draw[very thick, black] (5.000000,6.500000) -- (5.750000,6.500000);
\draw[very thick, black] (5.750000,6.750000) -- (6.000000,6.750000);
\draw[very thick, black] (6.500000,6.000000) -- (6.500000,7.000000);
\draw[very thick, black] (6.000000,6.750000) -- (6.500000,6.750000);
\draw[very thick, black] (6.500000,6.500000) -- (7.000000,6.500000);
\draw[very thick, black] (7.250000,6.000000) -- (7.250000,7.000000);
\draw[very thick, black] (7.000000,6.500000) -- (7.250000,6.500000);
\draw[very thick, black] (7.250000,6.250000) -- (8.000000,6.250000);
\draw[very thick, black] (8.000000,6.250000) -- (9.000000,6.250000);
\draw[very thick, black] (8.250000,7.000000) -- (8.250000,6.250000);
\draw[very thick, black] (8.500000,6.000000) -- (8.500000,6.250000);
\draw[very thick, black] (9.000000,6.250000) -- (10.000000,6.250000);
\draw[very thick, black] (9.750000,7.000000) -- (9.750000,6.250000);
\draw[very thick, black] (9.500000,6.000000) -- (9.500000,6.250000);
\draw[very thick, black] (10.000000,6.250000) -- (11.000000,6.250000);
\draw[very thick, black] (10.750000,7.000000) -- (10.750000,6.250000);
\draw[very thick, black] (10.500000,6.000000) -- (10.500000,6.250000);
\draw[very thick, black] (11.500000,6.000000) -- (11.500000,7.000000);
\draw[very thick, black] (11.000000,6.250000) -- (11.500000,6.250000);
\draw[very thick, black] (11.500000,6.500000) -- (12.000000,6.500000);
\draw[very thick, black] (0.750000,7.000000) -- (0.750000,8.000000);
\draw[very thick, black] (0.000000,7.500000) -- (0.750000,7.500000);
\draw[very thick, black] (0.750000,7.250000) -- (1.000000,7.250000);
\draw[very thick, black] (1.000000,7.250000) -- (2.000000,7.250000);
\draw[very thick, black] (1.750000,8.000000) -- (1.750000,7.250000);
\draw[very thick, black] (1.500000,7.000000) -- (1.500000,7.250000);
\draw[very thick, black] (2.000000,7.250000) -- (3.000000,7.250000);
\draw[very thick, black] (2.250000,8.000000) -- (2.250000,7.250000);
\draw[very thick, black] (2.750000,7.000000) -- (2.750000,7.250000);
\draw[very thick, black] (3.000000,7.250000) -- (4.000000,7.250000);
\draw[very thick, black] (3.750000,8.000000) -- (3.750000,7.250000);
\draw[very thick, black] (3.500000,7.000000) -- (3.500000,7.250000);
\draw[very thick, black] (4.000000,7.250000) -- (5.000000,7.250000);
\draw[very thick, black] (4.750000,8.000000) -- (4.750000,7.250000);
\draw[very thick, black] (4.250000,7.000000) -- (4.250000,7.250000);
\draw[very thick, black] (5.750000,7.000000) -- (5.750000,8.000000);
\draw[very thick, black] (5.000000,7.250000) -- (5.750000,7.250000);
\draw[very thick, black] (5.750000,7.500000) -- (6.000000,7.500000);
\draw[very thick, black] (6.000000,7.500000) -- (7.000000,7.500000);
\draw[very thick, black] (6.750000,8.000000) -- (6.750000,7.500000);
\draw[very thick, black] (6.500000,7.000000) -- (6.500000,7.500000);
\draw[very thick, black] (7.250000,7.000000) -- (7.250000,8.000000);
\draw[very thick, black] (7.000000,7.500000) -- (7.250000,7.500000);
\draw[very thick, black] (7.250000,7.750000) -- (8.000000,7.750000);
\draw[very thick, black] (8.000000,7.750000) -- (9.000000,7.750000);
\draw[very thick, black] (8.750000,8.000000) -- (8.750000,7.750000);
\draw[very thick, black] (8.250000,7.000000) -- (8.250000,7.750000);
\draw[very thick, black] (9.000000,7.750000) -- (10.000000,7.750000);
\draw[very thick, black] (9.250000,8.000000) -- (9.250000,7.750000);
\draw[very thick, black] (9.750000,7.000000) -- (9.750000,7.750000);
\draw[very thick, black] (10.000000,7.750000) -- (11.000000,7.750000);
\draw[very thick, black] (10.250000,8.000000) -- (10.250000,7.750000);
\draw[very thick, black] (10.750000,7.000000) -- (10.750000,7.750000);
\draw[very thick, black] (11.500000,7.000000) -- (11.500000,8.000000);
\draw[very thick, black] (11.000000,7.750000) -- (11.500000,7.750000);
\draw[very thick, black] (11.500000,7.500000) -- (12.000000,7.500000);
\draw[very thick, black] (0.750000,8.000000) -- (0.750000,9.000000);
\draw[very thick, black] (0.000000,8.750000) -- (0.750000,8.750000);
\draw[very thick, black] (0.750000,8.250000) -- (1.000000,8.250000);
\draw[very thick, black] (1.000000,8.250000) -- (2.000000,8.250000);
\draw[very thick, black] (1.500000,9.000000) -- (1.500000,8.250000);
\draw[very thick, black] (1.750000,8.000000) -- (1.750000,8.250000);
\draw[very thick, black] (2.000000,8.250000) -- (3.000000,8.250000);
\draw[very thick, black] (2.500000,9.000000) -- (2.500000,8.250000);
\draw[very thick, black] (2.250000,8.000000) -- (2.250000,8.250000);
\draw[very thick, black] (3.000000,8.250000) -- (4.000000,8.250000);
\draw[very thick, black] (3.500000,9.000000) -- (3.500000,8.250000);
\draw[very thick, black] (3.750000,8.000000) -- (3.750000,8.250000);
\draw[very thick, black] (4.000000,8.250000) -- (5.000000,8.250000);
\draw[very thick, black] (4.500000,9.000000) -- (4.500000,8.250000);
\draw[very thick, black] (4.750000,8.000000) -- (4.750000,8.250000);
\draw[very thick, black] (5.000000,8.250000) -- (6.000000,8.250000);
\draw[very thick, black] (5.500000,9.000000) -- (5.500000,8.250000);
\draw[very thick, black] (5.750000,8.000000) -- (5.750000,8.250000);
\draw[very thick, black] (6.750000,8.000000) -- (6.750000,9.000000);
\draw[very thick, black] (6.000000,8.250000) -- (6.750000,8.250000);
\draw[very thick, black] (6.750000,8.500000) -- (7.000000,8.500000);
\draw[very thick, black] (7.000000,8.500000) -- (8.000000,8.500000);
\draw[very thick, black] (7.500000,9.000000) -- (7.500000,8.500000);
\draw[very thick, black] (7.250000,8.000000) -- (7.250000,8.500000);
\draw[very thick, black] (8.750000,8.000000) -- (8.750000,9.000000);
\draw[very thick, black] (8.000000,8.500000) -- (8.750000,8.500000);
\draw[very thick, black] (8.750000,8.250000) -- (9.000000,8.250000);
\draw[very thick, black] (9.000000,8.250000) -- (10.000000,8.250000);
\draw[very thick, black] (9.500000,9.000000) -- (9.500000,8.250000);
\draw[very thick, black] (9.250000,8.000000) -- (9.250000,8.250000);
\draw[very thick, black] (10.000000,8.250000) -- (11.000000,8.250000);
\draw[very thick, black] (10.750000,9.000000) -- (10.750000,8.250000);
\draw[very thick, black] (10.250000,8.000000) -- (10.250000,8.250000);
\draw[very thick, black] (11.500000,8.000000) -- (11.500000,9.000000);
\draw[very thick, black] (11.000000,8.250000) -- (11.500000,8.250000);
\draw[very thick, black] (11.500000,8.750000) -- (12.000000,8.750000);
\draw[very thick, black] (0.000000,9.500000) -- (1.000000,9.500000);
\draw[very thick, black] (0.500000,10.000000) -- (0.500000,9.500000);
\draw[very thick, black] (0.750000,9.000000) -- (0.750000,9.500000);
\draw[very thick, black] (1.000000,9.500000) -- (2.000000,9.500000);
\draw[very thick, black] (1.750000,10.000000) -- (1.750000,9.500000);
\draw[very thick, black] (1.500000,9.000000) -- (1.500000,9.500000);
\draw[very thick, black] (2.500000,9.000000) -- (2.500000,10.000000);
\draw[very thick, black] (2.000000,9.500000) -- (2.500000,9.500000);
\draw[very thick, black] (2.500000,9.750000) -- (3.000000,9.750000);
\draw[very thick, black] (3.000000,9.750000) -- (4.000000,9.750000);
\draw[very thick, black] (3.250000,10.000000) -- (3.250000,9.750000);
\draw[very thick, black] (3.500000,9.000000) -- (3.500000,9.750000);
\draw[very thick, black] (4.500000,9.000000) -- (4.500000,10.000000);
\draw[very thick, black] (4.000000,9.750000) -- (4.500000,9.750000);
\draw[very thick, black] (4.500000,9.250000) -- (5.000000,9.250000);
\draw[very thick, black] (5.000000,9.250000) -- (6.000000,9.250000);
\draw[very thick, black] (5.750000,10.000000) -- (5.750000,9.250000);
\draw[very thick, black] (5.500000,9.000000) -- (5.500000,9.250000);
\draw[very thick, black] (6.000000,9.250000) -- (7.000000,9.250000);
\draw[very thick, black] (6.500000,10.000000) -- (6.500000,9.250000);
\draw[very thick, black] (6.750000,9.000000) -- (6.750000,9.250000);
\draw[very thick, black] (7.000000,9.250000) -- (8.000000,9.250000);
\draw[very thick, black] (7.750000,10.000000) -- (7.750000,9.250000);
\draw[very thick, black] (7.500000,9.000000) -- (7.500000,9.250000);
\draw[very thick, black] (8.000000,9.250000) -- (9.000000,9.250000);
\draw[very thick, black] (8.250000,10.000000) -- (8.250000,9.250000);
\draw[very thick, black] (8.750000,9.000000) -- (8.750000,9.250000);
\draw[very thick, black] (9.500000,9.000000) -- (9.500000,10.000000);
\draw[very thick, black] (9.000000,9.250000) -- (9.500000,9.250000);
\draw[very thick, black] (9.500000,9.500000) -- (10.000000,9.500000);
\draw[very thick, black] (10.750000,9.000000) -- (10.750000,10.000000);
\draw[very thick, black] (10.000000,9.500000) -- (10.750000,9.500000);
\draw[very thick, black] (10.750000,9.750000) -- (11.000000,9.750000);
\draw[very thick, black] (11.500000,9.000000) -- (11.500000,10.000000);
\draw[very thick, black] (11.000000,9.750000) -- (11.500000,9.750000);
\draw[very thick, black] (11.500000,9.500000) -- (12.000000,9.500000);
\draw[very thick, black] (0.500000,10.000000) -- (0.500000,11.000000);
\draw[very thick, black] (0.000000,10.250000) -- (0.500000,10.250000);
\draw[very thick, black] (0.500000,10.750000) -- (1.000000,10.750000);
\draw[very thick, black] (1.000000,10.750000) -- (2.000000,10.750000);
\draw[very thick, black] (1.250000,11.000000) -- (1.250000,10.750000);
\draw[very thick, black] (1.750000,10.000000) -- (1.750000,10.750000);
\draw[very thick, black] (2.500000,10.000000) -- (2.500000,11.000000);
\draw[very thick, black] (2.000000,10.750000) -- (2.500000,10.750000);
\draw[very thick, black] (2.500000,10.500000) -- (3.000000,10.500000);
\draw[very thick, black] (3.000000,10.500000) -- (4.000000,10.500000);
\draw[very thick, black] (3.500000,11.000000) -- (3.500000,10.500000);
\draw[very thick, black] (3.250000,10.000000) -- (3.250000,10.500000);
\draw[very thick, black] (4.000000,10.500000) -- (5.000000,10.500000);
\draw[very thick, black] (4.250000,11.000000) -- (4.250000,10.500000);
\draw[very thick, black] (4.500000,10.000000) -- (4.500000,10.500000);
\draw[very thick, black] (5.750000,10.000000) -- (5.750000,11.000000);
\draw[very thick, black] (5.000000,10.500000) -- (5.750000,10.500000);
\draw[very thick, black] (5.750000,10.750000) -- (6.000000,10.750000);
\draw[very thick, black] (6.500000,10.000000) -- (6.500000,11.000000);
\draw[very thick, black] (6.000000,10.750000) -- (6.500000,10.750000);
\draw[very thick, black] (6.500000,10.500000) -- (7.000000,10.500000);
\draw[very thick, black] (7.000000,10.500000) -- (8.000000,10.500000);
\draw[very thick, black] (7.250000,11.000000) -- (7.250000,10.500000);
\draw[very thick, black] (7.750000,10.000000) -- (7.750000,10.500000);
\draw[very thick, black] (8.250000,10.000000) -- (8.250000,11.000000);
\draw[very thick, black] (8.000000,10.500000) -- (8.250000,10.500000);
\draw[very thick, black] (8.250000,10.250000) -- (9.000000,10.250000);
\draw[very thick, black] (9.500000,10.000000) -- (9.500000,11.000000);
\draw[very thick, black] (9.000000,10.250000) -- (9.500000,10.250000);
\draw[very thick, black] (9.500000,10.750000) -- (10.000000,10.750000);
\draw[very thick, black] (10.750000,10.000000) -- (10.750000,11.000000);
\draw[very thick, black] (10.000000,10.750000) -- (10.750000,10.750000);
\draw[very thick, black] (10.750000,10.250000) -- (11.000000,10.250000);
\draw[very thick, black] (11.000000,10.250000) -- (12.000000,10.250000);
\draw[very thick, black] (11.750000,11.000000) -- (11.750000,10.250000);
\draw[very thick, black] (11.500000,10.000000) -- (11.500000,10.250000);
\draw[very thick, black] (0.000000,11.250000) -- (1.000000,11.250000);
\draw[very thick, black] (0.750000,12.000000) -- (0.750000,11.250000);
\draw[very thick, black] (0.500000,11.000000) -- (0.500000,11.250000);
\draw[very thick, black] (1.000000,11.250000) -- (2.000000,11.250000);
\draw[very thick, black] (1.750000,12.000000) -- (1.750000,11.250000);
\draw[very thick, black] (1.250000,11.000000) -- (1.250000,11.250000);
\draw[very thick, black] (2.000000,11.250000) -- (3.000000,11.250000);
\draw[very thick, black] (2.250000,12.000000) -- (2.250000,11.250000);
\draw[very thick, black] (2.500000,11.000000) -- (2.500000,11.250000);
\draw[very thick, black] (3.000000,11.250000) -- (4.000000,11.250000);
\draw[very thick, black] (3.750000,12.000000) -- (3.750000,11.250000);
\draw[very thick, black] (3.500000,11.000000) -- (3.500000,11.250000);
\draw[very thick, black] (4.000000,11.250000) -- (5.000000,11.250000);
\draw[very thick, black] (4.500000,12.000000) -- (4.500000,11.250000);
\draw[very thick, black] (4.250000,11.000000) -- (4.250000,11.250000);
\draw[very thick, black] (5.000000,11.250000) -- (6.000000,11.250000);
\draw[very thick, black] (5.500000,12.000000) -- (5.500000,11.250000);
\draw[very thick, black] (5.750000,11.000000) -- (5.750000,11.250000);
\draw[very thick, black] (6.000000,11.250000) -- (7.000000,11.250000);
\draw[very thick, black] (6.750000,12.000000) -- (6.750000,11.250000);
\draw[very thick, black] (6.500000,11.000000) -- (6.500000,11.250000);
\draw[very thick, black] (7.000000,11.250000) -- (8.000000,11.250000);
\draw[very thick, black] (7.500000,12.000000) -- (7.500000,11.250000);
\draw[very thick, black] (7.250000,11.000000) -- (7.250000,11.250000);
\draw[very thick, black] (8.250000,11.000000) -- (8.250000,12.000000);
\draw[very thick, black] (8.000000,11.250000) -- (8.250000,11.250000);
\draw[very thick, black] (8.250000,11.500000) -- (9.000000,11.500000);
\draw[very thick, black] (9.500000,11.000000) -- (9.500000,12.000000);
\draw[very thick, black] (9.000000,11.500000) -- (9.500000,11.500000);
\draw[very thick, black] (9.500000,11.250000) -- (10.000000,11.250000);
\draw[very thick, black] (10.000000,11.250000) -- (11.000000,11.250000);
\draw[very thick, black] (10.500000,12.000000) -- (10.500000,11.250000);
\draw[very thick, black] (10.750000,11.000000) -- (10.750000,11.250000);
\draw[very thick, black] (11.000000,11.250000) -- (12.000000,11.250000);
\draw[very thick, black] (11.250000,12.000000) -- (11.250000,11.250000);
\draw[very thick, black] (11.750000,11.000000) -- (11.750000,11.250000);
\draw[dashed, gray] (0,0) grid (12,12);
\draw[line width=0.15cm, red] (0,11.25) -- (8.25,11.25);
\draw[line width=0.15cm, red] (11.5,3.25) -- (11.5,10.25);
\end{tikzpicture}

%% file: long_lines_border_5.tex

\begin{tikzpicture}[set style={{help lines}+=[dashed]},scale=1]
\draw[very thick, black] (0.666667,0.000000) -- (0.666667,1.000000);
\draw[very thick, black] (0.000000,0.166667) -- (0.666667,0.166667);
\draw[very thick, black] (0.666667,0.833333) -- (1.000000,0.833333);
\draw[very thick, black] (1.000000,0.833333) -- (2.000000,0.833333);
\draw[very thick, black] (1.833333,1.000000) -- (1.833333,0.833333);
\draw[very thick, black] (1.666667,0.000000) -- (1.666667,0.833333);
\draw[very thick, black] (2.000000,0.833333) -- (3.000000,0.833333);
\draw[very thick, black] (2.333333,1.000000) -- (2.333333,0.833333);
\draw[very thick, black] (2.500000,0.000000) -- (2.500000,0.833333);
\draw[very thick, black] (3.000000,0.833333) -- (4.000000,0.833333);
\draw[very thick, black] (3.833333,1.000000) -- (3.833333,0.833333);
\draw[very thick, black] (3.333333,0.000000) -- (3.333333,0.833333);
\draw[very thick, black] (4.166667,0.000000) -- (4.166667,1.000000);
\draw[very thick, black] (4.000000,0.833333) -- (4.166667,0.833333);
\draw[very thick, black] (4.166667,0.166667) -- (5.000000,0.166667);
\draw[very thick, black] (5.666667,0.000000) -- (5.666667,1.000000);
\draw[very thick, black] (5.000000,0.166667) -- (5.666667,0.166667);
\draw[very thick, black] (5.666667,0.500000) -- (6.000000,0.500000);
\draw[very thick, black] (6.500000,0.000000) -- (6.500000,1.000000);
\draw[very thick, black] (6.000000,0.500000) -- (6.500000,0.500000);
\draw[very thick, black] (6.500000,0.666667) -- (7.000000,0.666667);
\draw[very thick, black] (7.500000,0.000000) -- (7.500000,1.000000);
\draw[very thick, black] (7.000000,0.666667) -- (7.500000,0.666667);
\draw[very thick, black] (7.500000,0.333333) -- (8.000000,0.333333);
\draw[very thick, black] (8.666667,0.000000) -- (8.666667,1.000000);
\draw[very thick, black] (8.000000,0.333333) -- (8.666667,0.333333);
\draw[very thick, black] (8.666667,0.166667) -- (9.000000,0.166667);
\draw[very thick, black] (9.500000,0.000000) -- (9.500000,1.000000);
\draw[very thick, black] (9.000000,0.166667) -- (9.500000,0.166667);
\draw[very thick, black] (9.500000,0.500000) -- (10.000000,0.500000);
\draw[very thick, black] (10.000000,0.500000) -- (11.000000,0.500000);
\draw[very thick, black] (10.500000,1.000000) -- (10.500000,0.500000);
\draw[very thick, black] (10.666667,0.000000) -- (10.666667,0.500000);
\draw[very thick, black] (11.500000,0.000000) -- (11.500000,1.000000);
\draw[very thick, black] (11.000000,0.500000) -- (11.500000,0.500000);
\draw[very thick, black] (11.500000,0.166667) -- (12.000000,0.166667);
\draw[very thick, black] (0.000000,1.833333) -- (1.000000,1.833333);
\draw[very thick, black] (0.833333,2.000000) -- (0.833333,1.833333);
\draw[very thick, black] (0.666667,1.000000) -- (0.666667,1.833333);
\draw[very thick, black] (1.833333,1.000000) -- (1.833333,2.000000);
\draw[very thick, black] (1.000000,1.833333) -- (1.833333,1.833333);
\draw[very thick, black] (1.833333,1.333333) -- (2.000000,1.333333);
\draw[very thick, black] (2.000000,1.333333) -- (3.000000,1.333333);
\draw[very thick, black] (2.666667,2.000000) -- (2.666667,1.333333);
\draw[very thick, black] (2.333333,1.000000) -- (2.333333,1.333333);
\draw[very thick, black] (3.000000,1.333333) -- (4.000000,1.333333);
\draw[very thick, black] (3.333333,2.000000) -- (3.333333,1.333333);
\draw[very thick, black] (3.833333,1.000000) -- (3.833333,1.333333);
\draw[very thick, black] (4.000000,1.333333) -- (5.000000,1.333333);
\draw[very thick, black] (4.333333,2.000000) -- (4.333333,1.333333);
\draw[very thick, black] (4.166667,1.000000) -- (4.166667,1.333333);
\draw[very thick, black] (5.000000,1.333333) -- (6.000000,1.333333);
\draw[very thick, black] (5.333333,2.000000) -- (5.333333,1.333333);
\draw[very thick, black] (5.666667,1.000000) -- (5.666667,1.333333);
\draw[very thick, black] (6.000000,1.333333) -- (7.000000,1.333333);
\draw[very thick, black] (6.166667,2.000000) -- (6.166667,1.333333);
\draw[very thick, black] (6.500000,1.000000) -- (6.500000,1.333333);
\draw[very thick, black] (7.500000,1.000000) -- (7.500000,2.000000);
\draw[very thick, black] (7.000000,1.333333) -- (7.500000,1.333333);
\draw[very thick, black] (7.500000,1.500000) -- (8.000000,1.500000);
\draw[very thick, black] (8.666667,1.000000) -- (8.666667,2.000000);
\draw[very thick, black] (8.000000,1.500000) -- (8.666667,1.500000);
\draw[very thick, black] (8.666667,1.666667) -- (9.000000,1.666667);
\draw[very thick, black] (9.000000,1.666667) -- (10.000000,1.666667);
\draw[very thick, black] (9.333333,2.000000) -- (9.333333,1.666667);
\draw[very thick, black] (9.500000,1.000000) -- (9.500000,1.666667);
\draw[very thick, black] (10.500000,1.000000) -- (10.500000,2.000000);
\draw[very thick, black] (10.000000,1.666667) -- (10.500000,1.666667);
\draw[very thick, black] (10.500000,1.333333) -- (11.000000,1.333333);
\draw[very thick, black] (11.500000,1.000000) -- (11.500000,2.000000);
\draw[very thick, black] (11.000000,1.333333) -- (11.500000,1.333333);
\draw[very thick, black] (11.500000,1.833333) -- (12.000000,1.833333);
\draw[very thick, black] (0.833333,2.000000) -- (0.833333,3.000000);
\draw[very thick, black] (0.000000,2.833333) -- (0.833333,2.833333);
\draw[very thick, black] (0.833333,2.166667) -- (1.000000,2.166667);
\draw[very thick, black] (1.000000,2.166667) -- (2.000000,2.166667);
\draw[very thick, black] (1.333333,3.000000) -- (1.333333,2.166667);
\draw[very thick, black] (1.833333,2.000000) -- (1.833333,2.166667);
\draw[very thick, black] (2.000000,2.166667) -- (3.000000,2.166667);
\draw[very thick, black] (2.333333,3.000000) -- (2.333333,2.166667);
\draw[very thick, black] (2.666667,2.000000) -- (2.666667,2.166667);
\draw[very thick, black] (3.333333,2.000000) -- (3.333333,3.000000);
\draw[very thick, black] (3.000000,2.166667) -- (3.333333,2.166667);
\draw[very thick, black] (3.333333,2.500000) -- (4.000000,2.500000);
\draw[very thick, black] (4.000000,2.500000) -- (5.000000,2.500000);
\draw[very thick, black] (4.666667,3.000000) -- (4.666667,2.500000);
\draw[very thick, black] (4.333333,2.000000) -- (4.333333,2.500000);
\draw[very thick, black] (5.333333,2.000000) -- (5.333333,3.000000);
\draw[very thick, black] (5.000000,2.500000) -- (5.333333,2.500000);
\draw[very thick, black] (5.333333,2.333333) -- (6.000000,2.333333);
\draw[very thick, black] (6.166667,2.000000) -- (6.166667,3.000000);
\draw[very thick, black] (6.000000,2.333333) -- (6.166667,2.333333);
\draw[very thick, black] (6.166667,2.166667) -- (7.000000,2.166667);
\draw[very thick, black] (7.500000,2.000000) -- (7.500000,3.000000);
\draw[very thick, black] (7.000000,2.166667) -- (7.500000,2.166667);
\draw[very thick, black] (7.500000,2.333333) -- (8.000000,2.333333);
\draw[very thick, black] (8.000000,2.333333) -- (9.000000,2.333333);
\draw[very thick, black] (8.833333,3.000000) -- (8.833333,2.333333);
\draw[very thick, black] (8.666667,2.000000) -- (8.666667,2.333333);
\draw[very thick, black] (9.000000,2.333333) -- (10.000000,2.333333);
\draw[very thick, black] (9.500000,3.000000) -- (9.500000,2.333333);
\draw[very thick, black] (9.333333,2.000000) -- (9.333333,2.333333);
\draw[very thick, black] (10.500000,2.000000) -- (10.500000,3.000000);
\draw[very thick, black] (10.000000,2.333333) -- (10.500000,2.333333);
\draw[very thick, black] (10.500000,2.166667) -- (11.000000,2.166667);
\draw[very thick, black] (11.500000,2.000000) -- (11.500000,3.000000);
\draw[very thick, black] (11.000000,2.166667) -- (11.500000,2.166667);
\draw[very thick, black] (11.500000,2.833333) -- (12.000000,2.833333);
\draw[very thick, black] (0.000000,3.166667) -- (1.000000,3.166667);
\draw[very thick, black] (0.333333,4.000000) -- (0.333333,3.166667);
\draw[very thick, black] (0.833333,3.000000) -- (0.833333,3.166667);
\draw[very thick, black] (1.000000,3.166667) -- (2.000000,3.166667);
\draw[very thick, black] (1.666667,4.000000) -- (1.666667,3.166667);
\draw[very thick, black] (1.333333,3.000000) -- (1.333333,3.166667);
\draw[very thick, black] (2.000000,3.166667) -- (3.000000,3.166667);
\draw[very thick, black] (2.166667,4.000000) -- (2.166667,3.166667);
\draw[very thick, black] (2.333333,3.000000) -- (2.333333,3.166667);
\draw[very thick, black] (3.333333,3.000000) -- (3.333333,4.000000);
\draw[very thick, black] (3.000000,3.166667) -- (3.333333,3.166667);
\draw[very thick, black] (3.333333,3.333333) -- (4.000000,3.333333);
\draw[very thick, black] (4.000000,3.333333) -- (5.000000,3.333333);
\draw[very thick, black] (4.333333,4.000000) -- (4.333333,3.333333);
\draw[very thick, black] (4.666667,3.000000) -- (4.666667,3.333333);
\draw[very thick, black] (5.000000,3.333333) -- (6.000000,3.333333);
\draw[very thick, black] (5.500000,4.000000) -- (5.500000,3.333333);
\draw[very thick, black] (5.333333,3.000000) -- (5.333333,3.333333);
\draw[very thick, black] (6.000000,3.333333) -- (7.000000,3.333333);
\draw[very thick, black] (6.666667,4.000000) -- (6.666667,3.333333);
\draw[very thick, black] (6.166667,3.000000) -- (6.166667,3.333333);
\draw[very thick, black] (7.000000,3.333333) -- (8.000000,3.333333);
\draw[very thick, black] (7.833333,4.000000) -- (7.833333,3.333333);
\draw[very thick, black] (7.500000,3.000000) -- (7.500000,3.333333);
\draw[very thick, black] (8.000000,3.333333) -- (9.000000,3.333333);
\draw[very thick, black] (8.666667,4.000000) -- (8.666667,3.333333);
\draw[very thick, black] (8.833333,3.000000) -- (8.833333,3.333333);
\draw[very thick, black] (9.500000,3.000000) -- (9.500000,4.000000);
\draw[very thick, black] (9.000000,3.333333) -- (9.500000,3.333333);
\draw[very thick, black] (9.500000,3.833333) -- (10.000000,3.833333);
\draw[very thick, black] (10.000000,3.833333) -- (11.000000,3.833333);
\draw[very thick, black] (10.333333,4.000000) -- (10.333333,3.833333);
\draw[very thick, black] (10.500000,3.000000) -- (10.500000,3.833333);
\draw[very thick, black] (11.500000,3.000000) -- (11.500000,4.000000);
\draw[very thick, black] (11.000000,3.833333) -- (11.500000,3.833333);
\draw[very thick, black] (11.500000,3.166667) -- (12.000000,3.166667);
\draw[very thick, black] (0.333333,4.000000) -- (0.333333,5.000000);
\draw[very thick, black] (0.000000,4.500000) -- (0.333333,4.500000);
\draw[very thick, black] (0.333333,4.333333) -- (1.000000,4.333333);
\draw[very thick, black] (1.000000,4.333333) -- (2.000000,4.333333);
\draw[very thick, black] (1.333333,5.000000) -- (1.333333,4.333333);
\draw[very thick, black] (1.666667,4.000000) -- (1.666667,4.333333);
\draw[very thick, black] (2.000000,4.333333) -- (3.000000,4.333333);
\draw[very thick, black] (2.666667,5.000000) -- (2.666667,4.333333);
\draw[very thick, black] (2.166667,4.000000) -- (2.166667,4.333333);
\draw[very thick, black] (3.000000,4.333333) -- (4.000000,4.333333);
\draw[very thick, black] (3.500000,5.000000) -- (3.500000,4.333333);
\draw[very thick, black] (3.333333,4.000000) -- (3.333333,4.333333);
\draw[very thick, black] (4.333333,4.000000) -- (4.333333,5.000000);
\draw[very thick, black] (4.000000,4.333333) -- (4.333333,4.333333);
\draw[very thick, black] (4.333333,4.500000) -- (5.000000,4.500000);
\draw[very thick, black] (5.000000,4.500000) -- (6.000000,4.500000);
\draw[very thick, black] (5.166667,5.000000) -- (5.166667,4.500000);
\draw[very thick, black] (5.500000,4.000000) -- (5.500000,4.500000);
\draw[very thick, black] (6.000000,4.500000) -- (7.000000,4.500000);
\draw[very thick, black] (6.833333,5.000000) -- (6.833333,4.500000);
\draw[very thick, black] (6.666667,4.000000) -- (6.666667,4.500000);
\draw[very thick, black] (7.833333,4.000000) -- (7.833333,5.000000);
\draw[very thick, black] (7.000000,4.500000) -- (7.833333,4.500000);
\draw[very thick, black] (7.833333,4.333333) -- (8.000000,4.333333);
\draw[very thick, black] (8.000000,4.333333) -- (9.000000,4.333333);
\draw[very thick, black] (8.333333,5.000000) -- (8.333333,4.333333);
\draw[very thick, black] (8.666667,4.000000) -- (8.666667,4.333333);
\draw[very thick, black] (9.000000,4.333333) -- (10.000000,4.333333);
\draw[very thick, black] (9.333333,5.000000) -- (9.333333,4.333333);
\draw[very thick, black] (9.500000,4.000000) -- (9.500000,4.333333);
\draw[very thick, black] (10.333333,4.000000) -- (10.333333,5.000000);
\draw[very thick, black] (10.000000,4.333333) -- (10.333333,4.333333);
\draw[very thick, black] (10.333333,4.666667) -- (11.000000,4.666667);
\draw[very thick, black] (11.500000,4.000000) -- (11.500000,5.000000);
\draw[very thick, black] (11.000000,4.666667) -- (11.500000,4.666667);
\draw[very thick, black] (11.500000,4.500000) -- (12.000000,4.500000);
\draw[very thick, black] (0.000000,5.166667) -- (1.000000,5.166667);
\draw[very thick, black] (0.833333,6.000000) -- (0.833333,5.166667);
\draw[very thick, black] (0.333333,5.000000) -- (0.333333,5.166667);
\draw[very thick, black] (1.000000,5.166667) -- (2.000000,5.166667);
\draw[very thick, black] (1.666667,6.000000) -- (1.666667,5.166667);
\draw[very thick, black] (1.333333,5.000000) -- (1.333333,5.166667);
\draw[very thick, black] (2.000000,5.166667) -- (3.000000,5.166667);
\draw[very thick, black] (2.833333,6.000000) -- (2.833333,5.166667);
\draw[very thick, black] (2.666667,5.000000) -- (2.666667,5.166667);
\draw[very thick, black] (3.000000,5.166667) -- (4.000000,5.166667);
\draw[very thick, black] (3.666667,6.000000) -- (3.666667,5.166667);
\draw[very thick, black] (3.500000,5.000000) -- (3.500000,5.166667);
\draw[very thick, black] (4.333333,5.000000) -- (4.333333,6.000000);
\draw[very thick, black] (4.000000,5.166667) -- (4.333333,5.166667);
\draw[very thick, black] (4.333333,5.833333) -- (5.000000,5.833333);
\draw[very thick, black] (5.000000,5.833333) -- (6.000000,5.833333);
\draw[very thick, black] (5.333333,6.000000) -- (5.333333,5.833333);
\draw[very thick, black] (5.166667,5.000000) -- (5.166667,5.833333);
\draw[very thick, black] (6.833333,5.000000) -- (6.833333,6.000000);
\draw[very thick, black] (6.000000,5.833333) -- (6.833333,5.833333);
\draw[very thick, black] (6.833333,5.666667) -- (7.000000,5.666667);
\draw[very thick, black] (7.000000,5.666667) -- (8.000000,5.666667);
\draw[very thick, black] (7.166667,6.000000) -- (7.166667,5.666667);
\draw[very thick, black] (7.833333,5.000000) -- (7.833333,5.666667);
\draw[very thick, black] (8.000000,5.666667) -- (9.000000,5.666667);
\draw[very thick, black] (8.500000,6.000000) -- (8.500000,5.666667);
\draw[very thick, black] (8.333333,5.000000) -- (8.333333,5.666667);
\draw[very thick, black] (9.000000,5.666667) -- (10.000000,5.666667);
\draw[very thick, black] (9.833333,6.000000) -- (9.833333,5.666667);
\draw[very thick, black] (9.333333,5.000000) -- (9.333333,5.666667);
\draw[very thick, black] (10.000000,5.666667) -- (11.000000,5.666667);
\draw[very thick, black] (10.666667,6.000000) -- (10.666667,5.666667);
\draw[very thick, black] (10.333333,5.000000) -- (10.333333,5.666667);
\draw[very thick, black] (11.500000,5.000000) -- (11.500000,6.000000);
\draw[very thick, black] (11.000000,5.666667) -- (11.500000,5.666667);
\draw[very thick, black] (11.500000,5.166667) -- (12.000000,5.166667);
\draw[very thick, black] (0.833333,6.000000) -- (0.833333,7.000000);
\draw[very thick, black] (0.000000,6.500000) -- (0.833333,6.500000);
\draw[very thick, black] (0.833333,6.833333) -- (1.000000,6.833333);
\draw[very thick, black] (1.666667,6.000000) -- (1.666667,7.000000);
\draw[very thick, black] (1.000000,6.833333) -- (1.666667,6.833333);
\draw[very thick, black] (1.666667,6.166667) -- (2.000000,6.166667);
\draw[very thick, black] (2.833333,6.000000) -- (2.833333,7.000000);
\draw[very thick, black] (2.000000,6.166667) -- (2.833333,6.166667);
\draw[very thick, black] (2.833333,6.833333) -- (3.000000,6.833333);
\draw[very thick, black] (3.666667,6.000000) -- (3.666667,7.000000);
\draw[very thick, black] (3.000000,6.833333) -- (3.666667,6.833333);
\draw[very thick, black] (3.666667,6.166667) -- (4.000000,6.166667);
\draw[very thick, black] (4.333333,6.000000) -- (4.333333,7.000000);
\draw[very thick, black] (4.000000,6.166667) -- (4.333333,6.166667);
\draw[very thick, black] (4.333333,6.666667) -- (5.000000,6.666667);
\draw[very thick, black] (5.333333,6.000000) -- (5.333333,7.000000);
\draw[very thick, black] (5.000000,6.666667) -- (5.333333,6.666667);
\draw[very thick, black] (5.333333,6.833333) -- (6.000000,6.833333);
\draw[very thick, black] (6.833333,6.000000) -- (6.833333,7.000000);
\draw[very thick, black] (6.000000,6.833333) -- (6.833333,6.833333);
\draw[very thick, black] (6.833333,6.500000) -- (7.000000,6.500000);
\draw[very thick, black] (7.166667,6.000000) -- (7.166667,7.000000);
\draw[very thick, black] (7.000000,6.500000) -- (7.166667,6.500000);
\draw[very thick, black] (7.166667,6.333333) -- (8.000000,6.333333);
\draw[very thick, black] (8.000000,6.333333) -- (9.000000,6.333333);
\draw[very thick, black] (8.666667,7.000000) -- (8.666667,6.333333);
\draw[very thick, black] (8.500000,6.000000) -- (8.500000,6.333333);
\draw[very thick, black] (9.000000,6.333333) -- (10.000000,6.333333);
\draw[very thick, black] (9.500000,7.000000) -- (9.500000,6.333333);
\draw[very thick, black] (9.833333,6.000000) -- (9.833333,6.333333);
\draw[very thick, black] (10.000000,6.333333) -- (11.000000,6.333333);
\draw[very thick, black] (10.166667,7.000000) -- (10.166667,6.333333);
\draw[very thick, black] (10.666667,6.000000) -- (10.666667,6.333333);
\draw[very thick, black] (11.500000,6.000000) -- (11.500000,7.000000);
\draw[very thick, black] (11.000000,6.333333) -- (11.500000,6.333333);
\draw[very thick, black] (11.500000,6.500000) -- (12.000000,6.500000);
\draw[very thick, black] (0.000000,7.833333) -- (1.000000,7.833333);
\draw[very thick, black] (0.666667,8.000000) -- (0.666667,7.833333);
\draw[very thick, black] (0.833333,7.000000) -- (0.833333,7.833333);
\draw[very thick, black] (1.666667,7.000000) -- (1.666667,8.000000);
\draw[very thick, black] (1.000000,7.833333) -- (1.666667,7.833333);
\draw[very thick, black] (1.666667,7.666667) -- (2.000000,7.666667);
\draw[very thick, black] (2.000000,7.666667) -- (3.000000,7.666667);
\draw[very thick, black] (2.166667,8.000000) -- (2.166667,7.666667);
\draw[very thick, black] (2.833333,7.000000) -- (2.833333,7.666667);
\draw[very thick, black] (3.000000,7.666667) -- (4.000000,7.666667);
\draw[very thick, black] (3.333333,8.000000) -- (3.333333,7.666667);
\draw[very thick, black] (3.666667,7.000000) -- (3.666667,7.666667);
\draw[very thick, black] (4.000000,7.666667) -- (5.000000,7.666667);
\draw[very thick, black] (4.833333,8.000000) -- (4.833333,7.666667);
\draw[very thick, black] (4.333333,7.000000) -- (4.333333,7.666667);
\draw[very thick, black] (5.000000,7.666667) -- (6.000000,7.666667);
\draw[very thick, black] (5.500000,8.000000) -- (5.500000,7.666667);
\draw[very thick, black] (5.333333,7.000000) -- (5.333333,7.666667);
\draw[very thick, black] (6.000000,7.666667) -- (7.000000,7.666667);
\draw[very thick, black] (6.166667,8.000000) -- (6.166667,7.666667);
\draw[very thick, black] (6.833333,7.000000) -- (6.833333,7.666667);
\draw[very thick, black] (7.166667,7.000000) -- (7.166667,8.000000);
\draw[very thick, black] (7.000000,7.666667) -- (7.166667,7.666667);
\draw[very thick, black] (7.166667,7.500000) -- (8.000000,7.500000);
\draw[very thick, black] (8.666667,7.000000) -- (8.666667,8.000000);
\draw[very thick, black] (8.000000,7.500000) -- (8.666667,7.500000);
\draw[very thick, black] (8.666667,7.833333) -- (9.000000,7.833333);
\draw[very thick, black] (9.500000,7.000000) -- (9.500000,8.000000);
\draw[very thick, black] (9.000000,7.833333) -- (9.500000,7.833333);
\draw[very thick, black] (9.500000,7.666667) -- (10.000000,7.666667);
\draw[very thick, black] (10.000000,7.666667) -- (11.000000,7.666667);
\draw[very thick, black] (10.500000,8.000000) -- (10.500000,7.666667);
\draw[very thick, black] (10.166667,7.000000) -- (10.166667,7.666667);
\draw[very thick, black] (11.500000,7.000000) -- (11.500000,8.000000);
\draw[very thick, black] (11.000000,7.666667) -- (11.500000,7.666667);
\draw[very thick, black] (11.500000,7.833333) -- (12.000000,7.833333);
\draw[very thick, black] (0.666667,8.000000) -- (0.666667,9.000000);
\draw[very thick, black] (0.000000,8.166667) -- (0.666667,8.166667);
\draw[very thick, black] (0.666667,8.500000) -- (1.000000,8.500000);
\draw[very thick, black] (1.000000,8.500000) -- (2.000000,8.500000);
\draw[very thick, black] (1.500000,9.000000) -- (1.500000,8.500000);
\draw[very thick, black] (1.666667,8.000000) -- (1.666667,8.500000);
\draw[very thick, black] (2.166667,8.000000) -- (2.166667,9.000000);
\draw[very thick, black] (2.000000,8.500000) -- (2.166667,8.500000);
\draw[very thick, black] (2.166667,8.333333) -- (3.000000,8.333333);
\draw[very thick, black] (3.000000,8.333333) -- (4.000000,8.333333);
\draw[very thick, black] (3.166667,9.000000) -- (3.166667,8.333333);
\draw[very thick, black] (3.333333,8.000000) -- (3.333333,8.333333);
\draw[very thick, black] (4.000000,8.333333) -- (5.000000,8.333333);
\draw[very thick, black] (4.666667,9.000000) -- (4.666667,8.333333);
\draw[very thick, black] (4.833333,8.000000) -- (4.833333,8.333333);
\draw[very thick, black] (5.000000,8.333333) -- (6.000000,8.333333);
\draw[very thick, black] (5.666667,9.000000) -- (5.666667,8.333333);
\draw[very thick, black] (5.500000,8.000000) -- (5.500000,8.333333);
\draw[very thick, black] (6.166667,8.000000) -- (6.166667,9.000000);
\draw[very thick, black] (6.000000,8.333333) -- (6.166667,8.333333);
\draw[very thick, black] (6.166667,8.500000) -- (7.000000,8.500000);
\draw[very thick, black] (7.166667,8.000000) -- (7.166667,9.000000);
\draw[very thick, black] (7.000000,8.500000) -- (7.166667,8.500000);
\draw[very thick, black] (7.166667,8.833333) -- (8.000000,8.833333);
\draw[very thick, black] (8.000000,8.833333) -- (9.000000,8.833333);
\draw[very thick, black] (8.500000,9.000000) -- (8.500000,8.833333);
\draw[very thick, black] (8.666667,8.000000) -- (8.666667,8.833333);
\draw[very thick, black] (9.000000,8.833333) -- (10.000000,8.833333);
\draw[very thick, black] (9.833333,9.000000) -- (9.833333,8.833333);
\draw[very thick, black] (9.500000,8.000000) -- (9.500000,8.833333);
\draw[very thick, black] (10.500000,8.000000) -- (10.500000,9.000000);
\draw[very thick, black] (10.000000,8.833333) -- (10.500000,8.833333);
\draw[very thick, black] (10.500000,8.500000) -- (11.000000,8.500000);
\draw[very thick, black] (11.500000,8.000000) -- (11.500000,9.000000);
\draw[very thick, black] (11.000000,8.500000) -- (11.500000,8.500000);
\draw[very thick, black] (11.500000,8.166667) -- (12.000000,8.166667);
\draw[very thick, black] (0.000000,9.166667) -- (1.000000,9.166667);
\draw[very thick, black] (0.166667,10.000000) -- (0.166667,9.166667);
\draw[very thick, black] (0.666667,9.000000) -- (0.666667,9.166667);
\draw[very thick, black] (1.500000,9.000000) -- (1.500000,10.000000);
\draw[very thick, black] (1.000000,9.166667) -- (1.500000,9.166667);
\draw[very thick, black] (1.500000,9.500000) -- (2.000000,9.500000);
\draw[very thick, black] (2.000000,9.500000) -- (3.000000,9.500000);
\draw[very thick, black] (2.833333,10.000000) -- (2.833333,9.500000);
\draw[very thick, black] (2.166667,9.000000) -- (2.166667,9.500000);
\draw[very thick, black] (3.166667,9.000000) -- (3.166667,10.000000);
\draw[very thick, black] (3.000000,9.500000) -- (3.166667,9.500000);
\draw[very thick, black] (3.166667,9.666667) -- (4.000000,9.666667);
\draw[very thick, black] (4.000000,9.666667) -- (5.000000,9.666667);
\draw[very thick, black] (4.500000,10.000000) -- (4.500000,9.666667);
\draw[very thick, black] (4.666667,9.000000) -- (4.666667,9.666667);
\draw[very thick, black] (5.666667,9.000000) -- (5.666667,10.000000);
\draw[very thick, black] (5.000000,9.666667) -- (5.666667,9.666667);
\draw[very thick, black] (5.666667,9.500000) -- (6.000000,9.500000);
\draw[very thick, black] (6.000000,9.500000) -- (7.000000,9.500000);
\draw[very thick, black] (6.833333,10.000000) -- (6.833333,9.500000);
\draw[very thick, black] (6.166667,9.000000) -- (6.166667,9.500000);
\draw[very thick, black] (7.000000,9.500000) -- (8.000000,9.500000);
\draw[very thick, black] (7.500000,10.000000) -- (7.500000,9.500000);
\draw[very thick, black] (7.166667,9.000000) -- (7.166667,9.500000);
\draw[very thick, black] (8.000000,9.500000) -- (9.000000,9.500000);
\draw[very thick, black] (8.166667,10.000000) -- (8.166667,9.500000);
\draw[very thick, black] (8.500000,9.000000) -- (8.500000,9.500000);
\draw[very thick, black] (9.000000,9.500000) -- (10.000000,9.500000);
\draw[very thick, black] (9.333333,10.000000) -- (9.333333,9.500000);
\draw[very thick, black] (9.833333,9.000000) -- (9.833333,9.500000);
\draw[very thick, black] (10.000000,9.500000) -- (11.000000,9.500000);
\draw[very thick, black] (10.666667,10.000000) -- (10.666667,9.500000);
\draw[very thick, black] (10.500000,9.000000) -- (10.500000,9.500000);
\draw[very thick, black] (11.500000,9.000000) -- (11.500000,10.000000);
\draw[very thick, black] (11.000000,9.500000) -- (11.500000,9.500000);
\draw[very thick, black] (11.500000,9.166667) -- (12.000000,9.166667);
\draw[very thick, black] (0.166667,10.000000) -- (0.166667,11.000000);
\draw[very thick, black] (0.000000,10.166667) -- (0.166667,10.166667);
\draw[very thick, black] (0.166667,10.500000) -- (1.000000,10.500000);
\draw[very thick, black] (1.500000,10.000000) -- (1.500000,11.000000);
\draw[very thick, black] (1.000000,10.500000) -- (1.500000,10.500000);
\draw[very thick, black] (1.500000,10.166667) -- (2.000000,10.166667);
\draw[very thick, black] (2.000000,10.166667) -- (3.000000,10.166667);
\draw[very thick, black] (2.166667,11.000000) -- (2.166667,10.166667);
\draw[very thick, black] (2.833333,10.000000) -- (2.833333,10.166667);
\draw[very thick, black] (3.166667,10.000000) -- (3.166667,11.000000);
\draw[very thick, black] (3.000000,10.166667) -- (3.166667,10.166667);
\draw[very thick, black] (3.166667,10.500000) -- (4.000000,10.500000);
\draw[very thick, black] (4.000000,10.500000) -- (5.000000,10.500000);
\draw[very thick, black] (4.666667,11.000000) -- (4.666667,10.500000);
\draw[very thick, black] (4.500000,10.000000) -- (4.500000,10.500000);
\draw[very thick, black] (5.666667,10.000000) -- (5.666667,11.000000);
\draw[very thick, black] (5.000000,10.500000) -- (5.666667,10.500000);
\draw[very thick, black] (5.666667,10.333333) -- (6.000000,10.333333);
\draw[very thick, black] (6.833333,10.000000) -- (6.833333,11.000000);
\draw[very thick, black] (6.000000,10.333333) -- (6.833333,10.333333);
\draw[very thick, black] (6.833333,10.166667) -- (7.000000,10.166667);
\draw[very thick, black] (7.000000,10.166667) -- (8.000000,10.166667);
\draw[very thick, black] (7.333333,11.000000) -- (7.333333,10.166667);
\draw[very thick, black] (7.500000,10.000000) -- (7.500000,10.166667);
\draw[very thick, black] (8.000000,10.166667) -- (9.000000,10.166667);
\draw[very thick, black] (8.666667,11.000000) -- (8.666667,10.166667);
\draw[very thick, black] (8.166667,10.000000) -- (8.166667,10.166667);
\draw[very thick, black] (9.333333,10.000000) -- (9.333333,11.000000);
\draw[very thick, black] (9.000000,10.166667) -- (9.333333,10.166667);
\draw[very thick, black] (9.333333,10.666667) -- (10.000000,10.666667);
\draw[very thick, black] (10.666667,10.000000) -- (10.666667,11.000000);
\draw[very thick, black] (10.000000,10.666667) -- (10.666667,10.666667);
\draw[very thick, black] (10.666667,10.500000) -- (11.000000,10.500000);
\draw[very thick, black] (11.500000,10.000000) -- (11.500000,11.000000);
\draw[very thick, black] (11.000000,10.500000) -- (11.500000,10.500000);
\draw[very thick, black] (11.500000,10.166667) -- (12.000000,10.166667);
\draw[very thick, black] (0.000000,11.333333) -- (1.000000,11.333333);
\draw[very thick, black] (0.666667,12.000000) -- (0.666667,11.333333);
\draw[very thick, black] (0.166667,11.000000) -- (0.166667,11.333333);
\draw[very thick, black] (1.000000,11.333333) -- (2.000000,11.333333);
\draw[very thick, black] (1.666667,12.000000) -- (1.666667,11.333333);
\draw[very thick, black] (1.500000,11.000000) -- (1.500000,11.333333);
\draw[very thick, black] (2.000000,11.333333) -- (3.000000,11.333333);
\draw[very thick, black] (2.500000,12.000000) -- (2.500000,11.333333);
\draw[very thick, black] (2.166667,11.000000) -- (2.166667,11.333333);
\draw[very thick, black] (3.000000,11.333333) -- (4.000000,11.333333);
\draw[very thick, black] (3.333333,12.000000) -- (3.333333,11.333333);
\draw[very thick, black] (3.166667,11.000000) -- (3.166667,11.333333);
\draw[very thick, black] (4.000000,11.333333) -- (5.000000,11.333333);
\draw[very thick, black] (4.166667,12.000000) -- (4.166667,11.333333);
\draw[very thick, black] (4.666667,11.000000) -- (4.666667,11.333333);
\draw[very thick, black] (5.666667,11.000000) -- (5.666667,12.000000);
\draw[very thick, black] (5.000000,11.333333) -- (5.666667,11.333333);
\draw[very thick, black] (5.666667,11.500000) -- (6.000000,11.500000);
\draw[very thick, black] (6.000000,11.500000) -- (7.000000,11.500000);
\draw[very thick, black] (6.500000,12.000000) -- (6.500000,11.500000);
\draw[very thick, black] (6.833333,11.000000) -- (6.833333,11.500000);
\draw[very thick, black] (7.000000,11.500000) -- (8.000000,11.500000);
\draw[very thick, black] (7.500000,12.000000) -- (7.500000,11.500000);
\draw[very thick, black] (7.333333,11.000000) -- (7.333333,11.500000);
\draw[very thick, black] (8.666667,11.000000) -- (8.666667,12.000000);
\draw[very thick, black] (8.000000,11.500000) -- (8.666667,11.500000);
\draw[very thick, black] (8.666667,11.833333) -- (9.000000,11.833333);
\draw[very thick, black] (9.000000,11.833333) -- (10.000000,11.833333);
\draw[very thick, black] (9.500000,12.000000) -- (9.500000,11.833333);
\draw[very thick, black] (9.333333,11.000000) -- (9.333333,11.833333);
\draw[very thick, black] (10.666667,11.000000) -- (10.666667,12.000000);
\draw[very thick, black] (10.000000,11.833333) -- (10.666667,11.833333);
\draw[very thick, black] (10.666667,11.666667) -- (11.000000,11.666667);
\draw[very thick, black] (11.500000,11.000000) -- (11.500000,12.000000);
\draw[very thick, black] (11.000000,11.666667) -- (11.500000,11.666667);
\draw[very thick, black] (11.500000,11.333333) -- (12.000000,11.333333);
\draw[dashed, gray] (0,0) grid (12,12);

\draw[line width=0.15cm, red] (0,11.33) -- (5.66,11.33);
\draw[line width=0.15cm, red] (11.5,0) -- (11.5,12);

\end{tikzpicture}

%% file: long_lines_border_10.tex

\begin{tikzpicture}[set style={{help lines}+=[dashed]},scale=1]
\draw[very thick, black] (0.000000,0.181818) -- (1.000000,0.181818);
\draw[very thick, black] (0.545455,1.000000) -- (0.545455,0.181818);
\draw[very thick, black] (0.090909,0.000000) -- (0.090909,0.181818);
\draw[very thick, black] (1.000000,0.181818) -- (2.000000,0.181818);
\draw[very thick, black] (1.727273,1.000000) -- (1.727273,0.181818);
\draw[very thick, black] (1.636364,0.000000) -- (1.636364,0.181818);
\draw[very thick, black] (2.000000,0.181818) -- (3.000000,0.181818);
\draw[very thick, black] (2.909091,1.000000) -- (2.909091,0.181818);
\draw[very thick, black] (2.454545,0.000000) -- (2.454545,0.181818);
\draw[very thick, black] (3.454545,0.000000) -- (3.454545,1.000000);
\draw[very thick, black] (3.000000,0.181818) -- (3.454545,0.181818);
\draw[very thick, black] (3.454545,0.272727) -- (4.000000,0.272727);
\draw[very thick, black] (4.363636,0.000000) -- (4.363636,1.000000);
\draw[very thick, black] (4.000000,0.272727) -- (4.363636,0.272727);
\draw[very thick, black] (4.363636,0.545455) -- (5.000000,0.545455);
\draw[very thick, black] (5.909091,0.000000) -- (5.909091,1.000000);
\draw[very thick, black] (5.000000,0.545455) -- (5.909091,0.545455);
\draw[very thick, black] (5.909091,0.818182) -- (6.000000,0.818182);
\draw[very thick, black] (6.000000,0.818182) -- (7.000000,0.818182);
\draw[very thick, black] (6.727273,1.000000) -- (6.727273,0.818182);
\draw[very thick, black] (6.454545,0.000000) -- (6.454545,0.818182);
\draw[very thick, black] (7.727273,0.000000) -- (7.727273,1.000000);
\draw[very thick, black] (7.000000,0.818182) -- (7.727273,0.818182);
\draw[very thick, black] (7.727273,0.909091) -- (8.000000,0.909091);
\draw[very thick, black] (8.000000,0.909091) -- (9.000000,0.909091);
\draw[very thick, black] (8.909091,1.000000) -- (8.909091,0.909091);
\draw[very thick, black] (8.545455,0.000000) -- (8.545455,0.909091);
\draw[very thick, black] (9.909091,0.000000) -- (9.909091,1.000000);
\draw[very thick, black] (9.000000,0.909091) -- (9.909091,0.909091);
\draw[very thick, black] (9.909091,0.818182) -- (10.000000,0.818182);
\draw[very thick, black] (10.909091,0.000000) -- (10.909091,1.000000);
\draw[very thick, black] (10.000000,0.818182) -- (10.909091,0.818182);
\draw[very thick, black] (10.909091,0.363636) -- (11.000000,0.363636);
\draw[very thick, black] (11.272727,0.000000) -- (11.272727,1.000000);
\draw[very thick, black] (11.000000,0.363636) -- (11.272727,0.363636);
\draw[very thick, black] (11.272727,0.181818) -- (12.000000,0.181818);
\draw[very thick, black] (0.000000,1.454545) -- (1.000000,1.454545);
\draw[very thick, black] (0.272727,2.000000) -- (0.272727,1.454545);
\draw[very thick, black] (0.545455,1.000000) -- (0.545455,1.454545);
\draw[very thick, black] (1.727273,1.000000) -- (1.727273,2.000000);
\draw[very thick, black] (1.000000,1.454545) -- (1.727273,1.454545);
\draw[very thick, black] (1.727273,1.727273) -- (2.000000,1.727273);
\draw[very thick, black] (2.909091,1.000000) -- (2.909091,2.000000);
\draw[very thick, black] (2.000000,1.727273) -- (2.909091,1.727273);
\draw[very thick, black] (2.909091,1.545455) -- (3.000000,1.545455);
\draw[very thick, black] (3.000000,1.545455) -- (4.000000,1.545455);
\draw[very thick, black] (3.272727,2.000000) -- (3.272727,1.545455);
\draw[very thick, black] (3.454545,1.000000) -- (3.454545,1.545455);
\draw[very thick, black] (4.363636,1.000000) -- (4.363636,2.000000);
\draw[very thick, black] (4.000000,1.545455) -- (4.363636,1.545455);
\draw[very thick, black] (4.363636,1.363636) -- (5.000000,1.363636);
\draw[very thick, black] (5.909091,1.000000) -- (5.909091,2.000000);
\draw[very thick, black] (5.000000,1.363636) -- (5.909091,1.363636);
\draw[very thick, black] (5.909091,1.272727) -- (6.000000,1.272727);
\draw[very thick, black] (6.727273,1.000000) -- (6.727273,2.000000);
\draw[very thick, black] (6.000000,1.272727) -- (6.727273,1.272727);
\draw[very thick, black] (6.727273,1.727273) -- (7.000000,1.727273);
\draw[very thick, black] (7.727273,1.000000) -- (7.727273,2.000000);
\draw[very thick, black] (7.000000,1.727273) -- (7.727273,1.727273);
\draw[very thick, black] (7.727273,1.909091) -- (8.000000,1.909091);
\draw[very thick, black] (8.000000,1.909091) -- (9.000000,1.909091);
\draw[very thick, black] (8.545455,2.000000) -- (8.545455,1.909091);
\draw[very thick, black] (8.909091,1.000000) -- (8.909091,1.909091);
\draw[very thick, black] (9.909091,1.000000) -- (9.909091,2.000000);
\draw[very thick, black] (9.000000,1.909091) -- (9.909091,1.909091);
\draw[very thick, black] (9.909091,1.545455) -- (10.000000,1.545455);
\draw[very thick, black] (10.909091,1.000000) -- (10.909091,2.000000);
\draw[very thick, black] (10.000000,1.545455) -- (10.909091,1.545455);
\draw[very thick, black] (10.909091,1.818182) -- (11.000000,1.818182);
\draw[very thick, black] (11.272727,1.000000) -- (11.272727,2.000000);
\draw[very thick, black] (11.000000,1.818182) -- (11.272727,1.818182);
\draw[very thick, black] (11.272727,1.454545) -- (12.000000,1.454545);
\draw[very thick, black] (0.272727,2.000000) -- (0.272727,3.000000);
\draw[very thick, black] (0.000000,2.454545) -- (0.272727,2.454545);
\draw[very thick, black] (0.272727,2.181818) -- (1.000000,2.181818);
\draw[very thick, black] (1.000000,2.181818) -- (2.000000,2.181818);
\draw[very thick, black] (1.636364,3.000000) -- (1.636364,2.181818);
\draw[very thick, black] (1.727273,2.000000) -- (1.727273,2.181818);
\draw[very thick, black] (2.000000,2.181818) -- (3.000000,2.181818);
\draw[very thick, black] (2.272727,3.000000) -- (2.272727,2.181818);
\draw[very thick, black] (2.909091,2.000000) -- (2.909091,2.181818);
\draw[very thick, black] (3.000000,2.181818) -- (4.000000,2.181818);
\draw[very thick, black] (3.727273,3.000000) -- (3.727273,2.181818);
\draw[very thick, black] (3.272727,2.000000) -- (3.272727,2.181818);
\draw[very thick, black] (4.363636,2.000000) -- (4.363636,3.000000);
\draw[very thick, black] (4.000000,2.181818) -- (4.363636,2.181818);
\draw[very thick, black] (4.363636,2.272727) -- (5.000000,2.272727);
\draw[very thick, black] (5.909091,2.000000) -- (5.909091,3.000000);
\draw[very thick, black] (5.000000,2.272727) -- (5.909091,2.272727);
\draw[very thick, black] (5.909091,2.727273) -- (6.000000,2.727273);
\draw[very thick, black] (6.727273,2.000000) -- (6.727273,3.000000);
\draw[very thick, black] (6.000000,2.727273) -- (6.727273,2.727273);
\draw[very thick, black] (6.727273,2.636364) -- (7.000000,2.636364);
\draw[very thick, black] (7.000000,2.636364) -- (8.000000,2.636364);
\draw[very thick, black] (7.363636,3.000000) -- (7.363636,2.636364);
\draw[very thick, black] (7.727273,2.000000) -- (7.727273,2.636364);
\draw[very thick, black] (8.545455,2.000000) -- (8.545455,3.000000);
\draw[very thick, black] (8.000000,2.636364) -- (8.545455,2.636364);
\draw[very thick, black] (8.545455,2.090909) -- (9.000000,2.090909);
\draw[very thick, black] (9.000000,2.090909) -- (10.000000,2.090909);
\draw[very thick, black] (9.818182,3.000000) -- (9.818182,2.090909);
\draw[very thick, black] (9.909091,2.000000) -- (9.909091,2.090909);
\draw[very thick, black] (10.000000,2.090909) -- (11.000000,2.090909);
\draw[very thick, black] (10.363636,3.000000) -- (10.363636,2.090909);
\draw[very thick, black] (10.909091,2.000000) -- (10.909091,2.090909);
\draw[very thick, black] (11.272727,2.000000) -- (11.272727,3.000000);
\draw[very thick, black] (11.000000,2.090909) -- (11.272727,2.090909);
\draw[very thick, black] (11.272727,2.454545) -- (12.000000,2.454545);
\draw[very thick, black] (0.000000,3.363636) -- (1.000000,3.363636);
\draw[very thick, black] (0.636364,4.000000) -- (0.636364,3.363636);
\draw[very thick, black] (0.272727,3.000000) -- (0.272727,3.363636);
\draw[very thick, black] (1.636364,3.000000) -- (1.636364,4.000000);
\draw[very thick, black] (1.000000,3.363636) -- (1.636364,3.363636);
\draw[very thick, black] (1.636364,3.545455) -- (2.000000,3.545455);
\draw[very thick, black] (2.000000,3.545455) -- (3.000000,3.545455);
\draw[very thick, black] (2.818182,4.000000) -- (2.818182,3.545455);
\draw[very thick, black] (2.272727,3.000000) -- (2.272727,3.545455);
\draw[very thick, black] (3.727273,3.000000) -- (3.727273,4.000000);
\draw[very thick, black] (3.000000,3.545455) -- (3.727273,3.545455);
\draw[very thick, black] (3.727273,3.090909) -- (4.000000,3.090909);
\draw[very thick, black] (4.000000,3.090909) -- (5.000000,3.090909);
\draw[very thick, black] (4.818182,4.000000) -- (4.818182,3.090909);
\draw[very thick, black] (4.363636,3.000000) -- (4.363636,3.090909);
\draw[very thick, black] (5.000000,3.090909) -- (6.000000,3.090909);
\draw[very thick, black] (5.181818,4.000000) -- (5.181818,3.090909);
\draw[very thick, black] (5.909091,3.000000) -- (5.909091,3.090909);
\draw[very thick, black] (6.000000,3.090909) -- (7.000000,3.090909);
\draw[very thick, black] (6.909091,4.000000) -- (6.909091,3.090909);
\draw[very thick, black] (6.727273,3.000000) -- (6.727273,3.090909);
\draw[very thick, black] (7.363636,3.000000) -- (7.363636,4.000000);
\draw[very thick, black] (7.000000,3.090909) -- (7.363636,3.090909);
\draw[very thick, black] (7.363636,3.272727) -- (8.000000,3.272727);
\draw[very thick, black] (8.000000,3.272727) -- (9.000000,3.272727);
\draw[very thick, black] (8.454545,4.000000) -- (8.454545,3.272727);
\draw[very thick, black] (8.545455,3.000000) -- (8.545455,3.272727);
\draw[very thick, black] (9.000000,3.272727) -- (10.000000,3.272727);
\draw[very thick, black] (9.363636,4.000000) -- (9.363636,3.272727);
\draw[very thick, black] (9.818182,3.000000) -- (9.818182,3.272727);
\draw[very thick, black] (10.000000,3.272727) -- (11.000000,3.272727);
\draw[very thick, black] (10.818182,4.000000) -- (10.818182,3.272727);
\draw[very thick, black] (10.363636,3.000000) -- (10.363636,3.272727);
\draw[very thick, black] (11.272727,3.000000) -- (11.272727,4.000000);
\draw[very thick, black] (11.000000,3.272727) -- (11.272727,3.272727);
\draw[very thick, black] (11.272727,3.363636) -- (12.000000,3.363636);
\draw[very thick, black] (0.000000,4.545455) -- (1.000000,4.545455);
\draw[very thick, black] (0.909091,5.000000) -- (0.909091,4.545455);
\draw[very thick, black] (0.636364,4.000000) -- (0.636364,4.545455);
\draw[very thick, black] (1.636364,4.000000) -- (1.636364,5.000000);
\draw[very thick, black] (1.000000,4.545455) -- (1.636364,4.545455);
\draw[very thick, black] (1.636364,4.454545) -- (2.000000,4.454545);
\draw[very thick, black] (2.818182,4.000000) -- (2.818182,5.000000);
\draw[very thick, black] (2.000000,4.454545) -- (2.818182,4.454545);
\draw[very thick, black] (2.818182,4.727273) -- (3.000000,4.727273);
\draw[very thick, black] (3.727273,4.000000) -- (3.727273,5.000000);
\draw[very thick, black] (3.000000,4.727273) -- (3.727273,4.727273);
\draw[very thick, black] (3.727273,4.636364) -- (4.000000,4.636364);
\draw[very thick, black] (4.000000,4.636364) -- (5.000000,4.636364);
\draw[very thick, black] (4.454545,5.000000) -- (4.454545,4.636364);
\draw[very thick, black] (4.818182,4.000000) -- (4.818182,4.636364);
\draw[very thick, black] (5.181818,4.000000) -- (5.181818,5.000000);
\draw[very thick, black] (5.000000,4.636364) -- (5.181818,4.636364);
\draw[very thick, black] (5.181818,4.090909) -- (6.000000,4.090909);
\draw[very thick, black] (6.000000,4.090909) -- (7.000000,4.090909);
\draw[very thick, black] (6.272727,5.000000) -- (6.272727,4.090909);
\draw[very thick, black] (6.909091,4.000000) -- (6.909091,4.090909);
\draw[very thick, black] (7.000000,4.090909) -- (8.000000,4.090909);
\draw[very thick, black] (7.181818,5.000000) -- (7.181818,4.090909);
\draw[very thick, black] (7.363636,4.000000) -- (7.363636,4.090909);
\draw[very thick, black] (8.000000,4.090909) -- (9.000000,4.090909);
\draw[very thick, black] (8.090909,5.000000) -- (8.090909,4.090909);
\draw[very thick, black] (8.454545,4.000000) -- (8.454545,4.090909);
\draw[very thick, black] (9.363636,4.000000) -- (9.363636,5.000000);
\draw[very thick, black] (9.000000,4.090909) -- (9.363636,4.090909);
\draw[very thick, black] (9.363636,4.909091) -- (10.000000,4.909091);
\draw[very thick, black] (10.818182,4.000000) -- (10.818182,5.000000);
\draw[very thick, black] (10.000000,4.909091) -- (10.818182,4.909091);
\draw[very thick, black] (10.818182,4.636364) -- (11.000000,4.636364);
\draw[very thick, black] (11.272727,4.000000) -- (11.272727,5.000000);
\draw[very thick, black] (11.000000,4.636364) -- (11.272727,4.636364);
\draw[very thick, black] (11.272727,4.545455) -- (12.000000,4.545455);
\draw[very thick, black] (0.000000,5.090909) -- (1.000000,5.090909);
\draw[very thick, black] (0.545455,6.000000) -- (0.545455,5.090909);
\draw[very thick, black] (0.909091,5.000000) -- (0.909091,5.090909);
\draw[very thick, black] (1.000000,5.090909) -- (2.000000,5.090909);
\draw[very thick, black] (1.545455,6.000000) -- (1.545455,5.090909);
\draw[very thick, black] (1.636364,5.000000) -- (1.636364,5.090909);
\draw[very thick, black] (2.818182,5.000000) -- (2.818182,6.000000);
\draw[very thick, black] (2.000000,5.090909) -- (2.818182,5.090909);
\draw[very thick, black] (2.818182,5.454545) -- (3.000000,5.454545);
\draw[very thick, black] (3.000000,5.454545) -- (4.000000,5.454545);
\draw[very thick, black] (3.636364,6.000000) -- (3.636364,5.454545);
\draw[very thick, black] (3.727273,5.000000) -- (3.727273,5.454545);
\draw[very thick, black] (4.000000,5.454545) -- (5.000000,5.454545);
\draw[very thick, black] (4.727273,6.000000) -- (4.727273,5.454545);
\draw[very thick, black] (4.454545,5.000000) -- (4.454545,5.454545);
\draw[very thick, black] (5.181818,5.000000) -- (5.181818,6.000000);
\draw[very thick, black] (5.000000,5.454545) -- (5.181818,5.454545);
\draw[very thick, black] (5.181818,5.909091) -- (6.000000,5.909091);
\draw[very thick, black] (6.000000,5.909091) -- (7.000000,5.909091);
\draw[very thick, black] (6.636364,6.000000) -- (6.636364,5.909091);
\draw[very thick, black] (6.272727,5.000000) -- (6.272727,5.909091);
\draw[very thick, black] (7.000000,5.909091) -- (8.000000,5.909091);
\draw[very thick, black] (7.454545,6.000000) -- (7.454545,5.909091);
\draw[very thick, black] (7.181818,5.000000) -- (7.181818,5.909091);
\draw[very thick, black] (8.000000,5.909091) -- (9.000000,5.909091);
\draw[very thick, black] (8.636364,6.000000) -- (8.636364,5.909091);
\draw[very thick, black] (8.090909,5.000000) -- (8.090909,5.909091);
\draw[very thick, black] (9.363636,5.000000) -- (9.363636,6.000000);
\draw[very thick, black] (9.000000,5.909091) -- (9.363636,5.909091);
\draw[very thick, black] (9.363636,5.545455) -- (10.000000,5.545455);
\draw[very thick, black] (10.818182,5.000000) -- (10.818182,6.000000);
\draw[very thick, black] (10.000000,5.545455) -- (10.818182,5.545455);
\draw[very thick, black] (10.818182,5.727273) -- (11.000000,5.727273);
\draw[very thick, black] (11.272727,5.000000) -- (11.272727,6.000000);
\draw[very thick, black] (11.000000,5.727273) -- (11.272727,5.727273);
\draw[very thick, black] (11.272727,5.090909) -- (12.000000,5.090909);
\draw[very thick, black] (0.545455,6.000000) -- (0.545455,7.000000);
\draw[very thick, black] (0.000000,6.181818) -- (0.545455,6.181818);
\draw[very thick, black] (0.545455,6.727273) -- (1.000000,6.727273);
\draw[very thick, black] (1.000000,6.727273) -- (2.000000,6.727273);
\draw[very thick, black] (1.727273,7.000000) -- (1.727273,6.727273);
\draw[very thick, black] (1.545455,6.000000) -- (1.545455,6.727273);
\draw[very thick, black] (2.818182,6.000000) -- (2.818182,7.000000);
\draw[very thick, black] (2.000000,6.727273) -- (2.818182,6.727273);
\draw[very thick, black] (2.818182,6.363636) -- (3.000000,6.363636);
\draw[very thick, black] (3.000000,6.363636) -- (4.000000,6.363636);
\draw[very thick, black] (3.545455,7.000000) -- (3.545455,6.363636);
\draw[very thick, black] (3.636364,6.000000) -- (3.636364,6.363636);
\draw[very thick, black] (4.000000,6.363636) -- (5.000000,6.363636);
\draw[very thick, black] (4.454545,7.000000) -- (4.454545,6.363636);
\draw[very thick, black] (4.727273,6.000000) -- (4.727273,6.363636);
\draw[very thick, black] (5.000000,6.363636) -- (6.000000,6.363636);
\draw[very thick, black] (5.454545,7.000000) -- (5.454545,6.363636);
\draw[very thick, black] (5.181818,6.000000) -- (5.181818,6.363636);
\draw[very thick, black] (6.636364,6.000000) -- (6.636364,7.000000);
\draw[very thick, black] (6.000000,6.363636) -- (6.636364,6.363636);
\draw[very thick, black] (6.636364,6.727273) -- (7.000000,6.727273);
\draw[very thick, black] (7.454545,6.000000) -- (7.454545,7.000000);
\draw[very thick, black] (7.000000,6.727273) -- (7.454545,6.727273);
\draw[very thick, black] (7.454545,6.454545) -- (8.000000,6.454545);
\draw[very thick, black] (8.636364,6.000000) -- (8.636364,7.000000);
\draw[very thick, black] (8.000000,6.454545) -- (8.636364,6.454545);
\draw[very thick, black] (8.636364,6.090909) -- (9.000000,6.090909);
\draw[very thick, black] (9.000000,6.090909) -- (10.000000,6.090909);
\draw[very thick, black] (9.272727,7.000000) -- (9.272727,6.090909);
\draw[very thick, black] (9.363636,6.000000) -- (9.363636,6.090909);
\draw[very thick, black] (10.818182,6.000000) -- (10.818182,7.000000);
\draw[very thick, black] (10.000000,6.090909) -- (10.818182,6.090909);
\draw[very thick, black] (10.818182,6.727273) -- (11.000000,6.727273);
\draw[very thick, black] (11.272727,6.000000) -- (11.272727,7.000000);
\draw[very thick, black] (11.000000,6.727273) -- (11.272727,6.727273);
\draw[very thick, black] (11.272727,6.181818) -- (12.000000,6.181818);
\draw[very thick, black] (0.000000,7.727273) -- (1.000000,7.727273);
\draw[very thick, black] (0.818182,8.000000) -- (0.818182,7.727273);
\draw[very thick, black] (0.545455,7.000000) -- (0.545455,7.727273);
\draw[very thick, black] (1.727273,7.000000) -- (1.727273,8.000000);
\draw[very thick, black] (1.000000,7.727273) -- (1.727273,7.727273);
\draw[very thick, black] (1.727273,7.363636) -- (2.000000,7.363636);
\draw[very thick, black] (2.000000,7.363636) -- (3.000000,7.363636);
\draw[very thick, black] (2.181818,8.000000) -- (2.181818,7.363636);
\draw[very thick, black] (2.818182,7.000000) -- (2.818182,7.363636);
\draw[very thick, black] (3.000000,7.363636) -- (4.000000,7.363636);
\draw[very thick, black] (3.818182,8.000000) -- (3.818182,7.363636);
\draw[very thick, black] (3.545455,7.000000) -- (3.545455,7.363636);
\draw[very thick, black] (4.454545,7.000000) -- (4.454545,8.000000);
\draw[very thick, black] (4.000000,7.363636) -- (4.454545,7.363636);
\draw[very thick, black] (4.454545,7.181818) -- (5.000000,7.181818);
\draw[very thick, black] (5.454545,7.000000) -- (5.454545,8.000000);
\draw[very thick, black] (5.000000,7.181818) -- (5.454545,7.181818);
\draw[very thick, black] (5.454545,7.454545) -- (6.000000,7.454545);
\draw[very thick, black] (6.636364,7.000000) -- (6.636364,8.000000);
\draw[very thick, black] (6.000000,7.454545) -- (6.636364,7.454545);
\draw[very thick, black] (6.636364,7.181818) -- (7.000000,7.181818);
\draw[very thick, black] (7.000000,7.181818) -- (8.000000,7.181818);
\draw[very thick, black] (7.636364,8.000000) -- (7.636364,7.181818);
\draw[very thick, black] (7.454545,7.000000) -- (7.454545,7.181818);
\draw[very thick, black] (8.000000,7.181818) -- (9.000000,7.181818);
\draw[very thick, black] (8.454545,8.000000) -- (8.454545,7.181818);
\draw[very thick, black] (8.636364,7.000000) -- (8.636364,7.181818);
\draw[very thick, black] (9.000000,7.181818) -- (10.000000,7.181818);
\draw[very thick, black] (9.363636,8.000000) -- (9.363636,7.181818);
\draw[very thick, black] (9.272727,7.000000) -- (9.272727,7.181818);
\draw[very thick, black] (10.818182,7.000000) -- (10.818182,8.000000);
\draw[very thick, black] (10.000000,7.181818) -- (10.818182,7.181818);
\draw[very thick, black] (10.818182,7.363636) -- (11.000000,7.363636);
\draw[very thick, black] (11.272727,7.000000) -- (11.272727,8.000000);
\draw[very thick, black] (11.000000,7.363636) -- (11.272727,7.363636);
\draw[very thick, black] (11.272727,7.727273) -- (12.000000,7.727273);
\draw[very thick, black] (0.000000,8.818182) -- (1.000000,8.818182);
\draw[very thick, black] (0.545455,9.000000) -- (0.545455,8.818182);
\draw[very thick, black] (0.818182,8.000000) -- (0.818182,8.818182);
\draw[very thick, black] (1.000000,8.818182) -- (2.000000,8.818182);
\draw[very thick, black] (1.909091,9.000000) -- (1.909091,8.818182);
\draw[very thick, black] (1.727273,8.000000) -- (1.727273,8.818182);
\draw[very thick, black] (2.000000,8.818182) -- (3.000000,8.818182);
\draw[very thick, black] (2.090909,9.000000) -- (2.090909,8.818182);
\draw[very thick, black] (2.181818,8.000000) -- (2.181818,8.818182);
\draw[very thick, black] (3.000000,8.818182) -- (4.000000,8.818182);
\draw[very thick, black] (3.363636,9.000000) -- (3.363636,8.818182);
\draw[very thick, black] (3.818182,8.000000) -- (3.818182,8.818182);
\draw[very thick, black] (4.000000,8.818182) -- (5.000000,8.818182);
\draw[very thick, black] (4.545455,9.000000) -- (4.545455,8.818182);
\draw[very thick, black] (4.454545,8.000000) -- (4.454545,8.818182);
\draw[very thick, black] (5.000000,8.818182) -- (6.000000,8.818182);
\draw[very thick, black] (5.363636,9.000000) -- (5.363636,8.818182);
\draw[very thick, black] (5.454545,8.000000) -- (5.454545,8.818182);
\draw[very thick, black] (6.636364,8.000000) -- (6.636364,9.000000);
\draw[very thick, black] (6.000000,8.818182) -- (6.636364,8.818182);
\draw[very thick, black] (6.636364,8.727273) -- (7.000000,8.727273);
\draw[very thick, black] (7.000000,8.727273) -- (8.000000,8.727273);
\draw[very thick, black] (7.090909,9.000000) -- (7.090909,8.727273);
\draw[very thick, black] (7.636364,8.000000) -- (7.636364,8.727273);
\draw[very thick, black] (8.454545,8.000000) -- (8.454545,9.000000);
\draw[very thick, black] (8.000000,8.727273) -- (8.454545,8.727273);
\draw[very thick, black] (8.454545,8.090909) -- (9.000000,8.090909);
\draw[very thick, black] (9.000000,8.090909) -- (10.000000,8.090909);
\draw[very thick, black] (9.545455,9.000000) -- (9.545455,8.090909);
\draw[very thick, black] (9.363636,8.000000) -- (9.363636,8.090909);
\draw[very thick, black] (10.818182,8.000000) -- (10.818182,9.000000);
\draw[very thick, black] (10.000000,8.090909) -- (10.818182,8.090909);
\draw[very thick, black] (10.818182,8.545455) -- (11.000000,8.545455);
\draw[very thick, black] (11.272727,8.000000) -- (11.272727,9.000000);
\draw[very thick, black] (11.000000,8.545455) -- (11.272727,8.545455);
\draw[very thick, black] (11.272727,8.818182) -- (12.000000,8.818182);
\draw[very thick, black] (0.000000,9.090909) -- (1.000000,9.090909);
\draw[very thick, black] (0.636364,10.000000) -- (0.636364,9.090909);
\draw[very thick, black] (0.545455,9.000000) -- (0.545455,9.090909);
\draw[very thick, black] (1.909091,9.000000) -- (1.909091,10.000000);
\draw[very thick, black] (1.000000,9.090909) -- (1.909091,9.090909);
\draw[very thick, black] (1.909091,9.909091) -- (2.000000,9.909091);
\draw[very thick, black] (2.000000,9.909091) -- (3.000000,9.909091);
\draw[very thick, black] (2.545455,10.000000) -- (2.545455,9.909091);
\draw[very thick, black] (2.090909,9.000000) -- (2.090909,9.909091);
\draw[very thick, black] (3.363636,9.000000) -- (3.363636,10.000000);
\draw[very thick, black] (3.000000,9.909091) -- (3.363636,9.909091);
\draw[very thick, black] (3.363636,9.363636) -- (4.000000,9.363636);
\draw[very thick, black] (4.000000,9.363636) -- (5.000000,9.363636);
\draw[very thick, black] (4.090909,10.000000) -- (4.090909,9.363636);
\draw[very thick, black] (4.545455,9.000000) -- (4.545455,9.363636);
\draw[very thick, black] (5.000000,9.363636) -- (6.000000,9.363636);
\draw[very thick, black] (5.818182,10.000000) -- (5.818182,9.363636);
\draw[very thick, black] (5.363636,9.000000) -- (5.363636,9.363636);
\draw[very thick, black] (6.636364,9.000000) -- (6.636364,10.000000);
\draw[very thick, black] (6.000000,9.363636) -- (6.636364,9.363636);
\draw[very thick, black] (6.636364,9.727273) -- (7.000000,9.727273);
\draw[very thick, black] (7.090909,9.000000) -- (7.090909,10.000000);
\draw[very thick, black] (7.000000,9.727273) -- (7.090909,9.727273);
\draw[very thick, black] (7.090909,9.272727) -- (8.000000,9.272727);
\draw[very thick, black] (8.454545,9.000000) -- (8.454545,10.000000);
\draw[very thick, black] (8.000000,9.272727) -- (8.454545,9.272727);
\draw[very thick, black] (8.454545,9.181818) -- (9.000000,9.181818);
\draw[very thick, black] (9.000000,9.181818) -- (10.000000,9.181818);
\draw[very thick, black] (9.090909,10.000000) -- (9.090909,9.181818);
\draw[very thick, black] (9.545455,9.000000) -- (9.545455,9.181818);
\draw[very thick, black] (10.000000,9.181818) -- (11.000000,9.181818);
\draw[very thick, black] (10.545455,10.000000) -- (10.545455,9.181818);
\draw[very thick, black] (10.818182,9.000000) -- (10.818182,9.181818);
\draw[very thick, black] (11.272727,9.000000) -- (11.272727,10.000000);
\draw[very thick, black] (11.000000,9.181818) -- (11.272727,9.181818);
\draw[very thick, black] (11.272727,9.090909) -- (12.000000,9.090909);
\draw[very thick, black] (0.000000,10.454545) -- (1.000000,10.454545);
\draw[very thick, black] (0.727273,11.000000) -- (0.727273,10.454545);
\draw[very thick, black] (0.636364,10.000000) -- (0.636364,10.454545);
\draw[very thick, black] (1.909091,10.000000) -- (1.909091,11.000000);
\draw[very thick, black] (1.000000,10.454545) -- (1.909091,10.454545);
\draw[very thick, black] (1.909091,10.909091) -- (2.000000,10.909091);
\draw[very thick, black] (2.545455,10.000000) -- (2.545455,11.000000);
\draw[very thick, black] (2.000000,10.909091) -- (2.545455,10.909091);
\draw[very thick, black] (2.545455,10.272727) -- (3.000000,10.272727);
\draw[very thick, black] (3.000000,10.272727) -- (4.000000,10.272727);
\draw[very thick, black] (3.727273,11.000000) -- (3.727273,10.272727);
\draw[very thick, black] (3.363636,10.000000) -- (3.363636,10.272727);
\draw[very thick, black] (4.090909,10.000000) -- (4.090909,11.000000);
\draw[very thick, black] (4.000000,10.272727) -- (4.090909,10.272727);
\draw[very thick, black] (4.090909,10.818182) -- (5.000000,10.818182);
\draw[very thick, black] (5.818182,10.000000) -- (5.818182,11.000000);
\draw[very thick, black] (5.000000,10.818182) -- (5.818182,10.818182);
\draw[very thick, black] (5.818182,10.272727) -- (6.000000,10.272727);
\draw[very thick, black] (6.636364,10.000000) -- (6.636364,11.000000);
\draw[very thick, black] (6.000000,10.272727) -- (6.636364,10.272727);
\draw[very thick, black] (6.636364,10.454545) -- (7.000000,10.454545);
\draw[very thick, black] (7.090909,10.000000) -- (7.090909,11.000000);
\draw[very thick, black] (7.000000,10.454545) -- (7.090909,10.454545);
\draw[very thick, black] (7.090909,10.545455) -- (8.000000,10.545455);
\draw[very thick, black] (8.454545,10.000000) -- (8.454545,11.000000);
\draw[very thick, black] (8.000000,10.545455) -- (8.454545,10.545455);
\draw[very thick, black] (8.454545,10.272727) -- (9.000000,10.272727);
\draw[very thick, black] (9.000000,10.272727) -- (10.000000,10.272727);
\draw[very thick, black] (9.545455,11.000000) -- (9.545455,10.272727);
\draw[very thick, black] (9.090909,10.000000) -- (9.090909,10.272727);
\draw[very thick, black] (10.000000,10.272727) -- (11.000000,10.272727);
\draw[very thick, black] (10.909091,11.000000) -- (10.909091,10.272727);
\draw[very thick, black] (10.545455,10.000000) -- (10.545455,10.272727);
\draw[very thick, black] (11.272727,10.000000) -- (11.272727,11.000000);
\draw[very thick, black] (11.000000,10.272727) -- (11.272727,10.272727);
\draw[very thick, black] (11.272727,10.454545) -- (12.000000,10.454545);
\draw[very thick, black] (0.000000,11.727273) -- (1.000000,11.727273);
\draw[very thick, black] (0.090909,12.000000) -- (0.090909,11.727273);
\draw[very thick, black] (0.727273,11.000000) -- (0.727273,11.727273);
\draw[very thick, black] (1.000000,11.727273) -- (2.000000,11.727273);
\draw[very thick, black] (1.636364,12.000000) -- (1.636364,11.727273);
\draw[very thick, black] (1.909091,11.000000) -- (1.909091,11.727273);
\draw[very thick, black] (2.000000,11.727273) -- (3.000000,11.727273);
\draw[very thick, black] (2.454545,12.000000) -- (2.454545,11.727273);
\draw[very thick, black] (2.545455,11.000000) -- (2.545455,11.727273);
\draw[very thick, black] (3.000000,11.727273) -- (4.000000,11.727273);
\draw[very thick, black] (3.454545,12.000000) -- (3.454545,11.727273);
\draw[very thick, black] (3.727273,11.000000) -- (3.727273,11.727273);
\draw[very thick, black] (4.000000,11.727273) -- (5.000000,11.727273);
\draw[very thick, black] (4.363636,12.000000) -- (4.363636,11.727273);
\draw[very thick, black] (4.090909,11.000000) -- (4.090909,11.727273);
\draw[very thick, black] (5.000000,11.727273) -- (6.000000,11.727273);
\draw[very thick, black] (5.909091,12.000000) -- (5.909091,11.727273);
\draw[very thick, black] (5.818182,11.000000) -- (5.818182,11.727273);
\draw[very thick, black] (6.000000,11.727273) -- (7.000000,11.727273);
\draw[very thick, black] (6.454545,12.000000) -- (6.454545,11.727273);
\draw[very thick, black] (6.636364,11.000000) -- (6.636364,11.727273);
\draw[very thick, black] (7.000000,11.727273) -- (8.000000,11.727273);
\draw[very thick, black] (7.727273,12.000000) -- (7.727273,11.727273);
\draw[very thick, black] (7.090909,11.000000) -- (7.090909,11.727273);
\draw[very thick, black] (8.000000,11.727273) -- (9.000000,11.727273);
\draw[very thick, black] (8.545455,12.000000) -- (8.545455,11.727273);
\draw[very thick, black] (8.454545,11.000000) -- (8.454545,11.727273);
\draw[very thick, black] (9.000000,11.727273) -- (10.000000,11.727273);
\draw[very thick, black] (9.909091,12.000000) -- (9.909091,11.727273);
\draw[very thick, black] (9.545455,11.000000) -- (9.545455,11.727273);
\draw[very thick, black] (10.909091,11.000000) -- (10.909091,12.000000);
\draw[very thick, black] (10.000000,11.727273) -- (10.909091,11.727273);
\draw[very thick, black] (10.909091,11.363636) -- (11.000000,11.363636);
\draw[very thick, black] (11.272727,11.000000) -- (11.272727,12.000000);
\draw[very thick, black] (11.000000,11.363636) -- (11.272727,11.363636);
\draw[very thick, black] (11.272727,11.727273) -- (12.000000,11.727273);
\draw[dashed, gray] (0,0) grid (12,12);

\draw[line width=0.15cm, red] (0,11.727273) -- (10.909091,11.727273);
\draw[line width=0.15cm, red] (11.272727,0) -- (11.272727,12);

\end{tikzpicture}

%% file: 2x2_problem.tex
\begin{tikzpicture}[set style={{help lines}+=[dashed]},scale=1.7]
\draw[style=help lines] (-0.1,-0.1) grid +(2.2,2.2);
\node at (0.6,0.6) {$(i',j')$};
\draw [fill=black] (0, 0.5) circle (0.05) node[below] {$V_{i',j'}$};
\draw [fill=black] (0, 1.5) circle (0.05) node[below, xshift=3pt] {$V_{i',j'+1}$};
\draw [fill=black] (0.75, 0) circle (0.05) node [below, xshift=-3pt] {$H_{i',j'}$};
\draw [fill=black] (1.25, 0) circle (0.05) node [below, xshift=5pt] {$H_{i'+1,j'}$};
\draw [fill=black] (0.75, 2) circle (0.05) node [above, xshift=-3pt]  {$H_{i',j'+2}$};
\draw [fill=black] (1.75, 2) circle (0.05) node [above]  {$H_{i'+1,j'+2}$};
\draw [fill=black] (2, 0.5) circle (0.05) node[below] {$V_{i'+2,j'}$};
\draw [fill=black] (2, 1.75) circle (0.05) node[below, xshift=-2pt] {$V_{i'+2,j'+1}$};

\node at (1, -0.5) {(a)};

\begin{scope}[xshift=2.6cm]
\draw[style=help lines] (-0.1,-0.1) grid +(2.2,2.2);
\draw [fill=black] (0, 0.5) circle (0.05);
\draw [fill=black] (0, 1.5) circle (0.05);
\draw [fill=black] (0.75, 0) circle (0.05);
\draw [fill=black] (1.25, 0) circle (0.05);
\draw [fill=black] (0.75, 2) circle (0.05);
\draw [fill=black] (1.75, 2) circle (0.05);
\draw [fill=black] (2, 0.5) circle (0.05);
\draw [fill=black] (2, 1.75) circle (0.05);
\draw [fill=black] (1, 0.25) circle (0.05) node[xshift=0.2pt,yshift=7pt] {$V_{i'+1,j'}$ };
\draw [fill=black] (1, 1.75) circle (0.05) node[xshift=17pt,yshift=6pt] {$V_{i'+1,j'+1}$ };
\draw [fill=black] (0.75, 1) circle (0.05) node[xshift=-13pt,yshift=-6.5pt] {$H_{i',j'+1}$ };
\draw [fill=black] (1.25, 1) circle (0.05) node[xshift=18pt,yshift=7pt] {$H_{i'+1,j'+1}$ };
\draw [] (0.75,0) -- (0.75,2);
\draw [] (0,0.5) -- (0.75,0.5);
\draw [] (0.75, 0.25) -- (1.25, 0.25);
\draw [] (1.25,0) -- (1.25,1.75);
\draw [] (0.75, 1.75) -- (2, 1.75);
\draw [] (1.25,0.5) -- (2, 0.5);
\draw [] (1.75,1.75) -- (1.75, 2);
\draw [] (0, 1.5) -- (0.75,1.5);

\node at (1, -0.5) {(b)};
\end{scope}

\end{tikzpicture}

%% file: widgets.tex
\begin{tikzpicture}[set style={{help lines}+=[dashed]},scale=1]
\widgetone{0}{0}
\begin{scope}[xshift=2.2cm]
\widgettwo{0}{0}	
\end{scope}
\end{tikzpicture}

%% file: dappling_even.tex
\begin{tikzpicture}[set style={{help lines}+=[dashed]},scale=0.7]
\draw[style=help lines] (0,0) grid +(10,10);

\widgetone{0}{0};
\widgetone{2}{2};
\widgettwo{4}{4};
\widgettwo{6}{6};
\widgetone{8}{8};

\widgetone{6}{0};
\widgettwo{8}{2};

\widgettwo{0}{6};
\widgettwo{2}{8};

\draw [decorate,decoration={brace,amplitude=3pt},xshift=-4pt,yshift=0pt]
(0,2) -- (0,6) node [black,midway,xshift=-0.9cm]  {$n-2$};

\draw [decorate,decoration={brace,amplitude=3pt,mirror},yshift=-4pt,xshift=0pt]
(2,0) -- (6,0) node [black,midway,below,yshift=-0.4cm]  {$n-2$};
\end{tikzpicture}

%% file: dappling_odd.tex
\begin{tikzpicture}[set style={{help lines}+=[dashed]},scale=0.7]
\draw[style=help lines] (0,0) grid +(10,10);

\widgetone{0}{0};
\widgetone{2}{2};
\widgetone{4}{4};
\widgetone{6}{6};
\widgetone{8}{8};

\widgetone{6}{0};
\widgetone{8}{2};

\widgetone{0}{6};
\widgetone{2}{8};

\draw [decorate,decoration={brace,amplitude=3pt},xshift=-4pt,yshift=0pt]
(0,2) -- (0,6) node [black,midway,xshift=-0.9cm]  {$n-1$};

\draw [decorate,decoration={brace,amplitude=3pt,mirror},yshift=-4pt,xshift=0pt]
(2,0) -- (6,0) node [black,midway,below,yshift=-0.4cm]  {$n-1$};
\end{tikzpicture}

%% file: result_n_1.tex
\begin{tikzpicture}[set style={{help lines}+=[dashed]}]
\draw[very thick, black] (0.250000,0.000000) -- (0.250000,1.000000);
\draw[very thick, black] (0.000000,0.500000) -- (0.250000,0.500000);
\draw[very thick, black] (0.250000,0.250000) -- (1.000000,0.250000);
\draw[very thick, black] (1.000000,0.250000) -- (2.000000,0.250000);
\draw[very thick, black] (1.500000,1.000000) -- (1.500000,0.250000);
\draw[very thick, black] (1.250000,0.000000) -- (1.250000,0.250000);
\draw[very thick, black] (2.750000,0.000000) -- (2.750000,1.000000);
\draw[very thick, black] (2.000000,0.250000) -- (2.750000,0.250000);
\draw[very thick, black] (2.750000,0.500000) -- (3.000000,0.500000);
\draw[very thick, black] (3.000000,0.500000) -- (4.000000,0.500000);
\draw[very thick, black] (3.750000,1.000000) -- (3.750000,0.500000);
\draw[very thick, black] (3.250000,0.000000) -- (3.250000,0.500000);
\draw[very thick, black] (4.250000,0.000000) -- (4.250000,1.000000);
\draw[very thick, black] (4.000000,0.500000) -- (4.250000,0.500000);
\draw[very thick, black] (4.250000,0.250000) -- (5.000000,0.250000);
\draw[very thick, black] (5.000000,0.250000) -- (6.000000,0.250000);
\draw[very thick, black] (5.750000,1.000000) -- (5.750000,0.250000);
\draw[very thick, black] (5.250000,0.000000) -- (5.250000,0.250000);
\draw[very thick, black] (6.500000,0.000000) -- (6.500000,1.000000);
\draw[very thick, black] (6.000000,0.250000) -- (6.500000,0.250000);
\draw[very thick, black] (6.500000,0.750000) -- (7.000000,0.750000);
\draw[very thick, black] (7.000000,0.750000) -- (8.000000,0.750000);
\draw[very thick, black] (7.750000,1.000000) -- (7.750000,0.750000);
\draw[very thick, black] (7.250000,0.000000) -- (7.250000,0.750000);
\draw[very thick, black] (8.750000,0.000000) -- (8.750000,1.000000);
\draw[very thick, black] (8.000000,0.750000) -- (8.750000,0.750000);
\draw[very thick, black] (8.750000,0.500000) -- (9.000000,0.500000);
\draw[very thick, black] (9.000000,0.500000) -- (10.000000,0.500000);
\draw[very thick, black] (9.750000,1.000000) -- (9.750000,0.500000);
\draw[very thick, black] (9.250000,0.000000) -- (9.250000,0.500000);
\draw[very thick, black] (10.250000,0.000000) -- (10.250000,1.000000);
\draw[very thick, black] (10.000000,0.500000) -- (10.250000,0.500000);
\draw[very thick, black] (10.250000,0.250000) -- (11.000000,0.250000);
\draw[very thick, black] (11.000000,0.250000) -- (12.000000,0.250000);
\draw[very thick, black] (11.250000,1.000000) -- (11.250000,0.250000);
\draw[very thick, black] (11.500000,0.000000) -- (11.500000,0.250000);
\draw[very thick, black] (0.000000,1.750000) -- (1.000000,1.750000);
\draw[very thick, black] (0.750000,2.000000) -- (0.750000,1.750000);
\draw[very thick, black] (0.250000,1.000000) -- (0.250000,1.750000);
\draw[very thick, black] (1.500000,1.000000) -- (1.500000,2.000000);
\draw[very thick, black] (1.000000,1.750000) -- (1.500000,1.750000);
\draw[very thick, black] (1.500000,1.500000) -- (2.000000,1.500000);
\draw[very thick, black] (2.000000,1.500000) -- (3.000000,1.500000);
\draw[very thick, black] (2.250000,2.000000) -- (2.250000,1.500000);
\draw[very thick, black] (2.750000,1.000000) -- (2.750000,1.500000);
\draw[very thick, black] (3.750000,1.000000) -- (3.750000,2.000000);
\draw[very thick, black] (3.000000,1.500000) -- (3.750000,1.500000);
\draw[very thick, black] (3.750000,1.750000) -- (4.000000,1.750000);
\draw[very thick, black] (4.000000,1.750000) -- (5.000000,1.750000);
\draw[very thick, black] (4.500000,2.000000) -- (4.500000,1.750000);
\draw[very thick, black] (4.250000,1.000000) -- (4.250000,1.750000);
\draw[very thick, black] (5.750000,1.000000) -- (5.750000,2.000000);
\draw[very thick, black] (5.000000,1.750000) -- (5.750000,1.750000);
\draw[very thick, black] (5.750000,1.500000) -- (6.000000,1.500000);
\draw[very thick, black] (6.000000,1.500000) -- (7.000000,1.500000);
\draw[very thick, black] (6.250000,2.000000) -- (6.250000,1.500000);
\draw[very thick, black] (6.500000,1.000000) -- (6.500000,1.500000);
\draw[very thick, black] (7.750000,1.000000) -- (7.750000,2.000000);
\draw[very thick, black] (7.000000,1.500000) -- (7.750000,1.500000);
\draw[very thick, black] (7.750000,1.250000) -- (8.000000,1.250000);
\draw[very thick, black] (8.000000,1.250000) -- (9.000000,1.250000);
\draw[very thick, black] (8.500000,2.000000) -- (8.500000,1.250000);
\draw[very thick, black] (8.750000,1.000000) -- (8.750000,1.250000);
\draw[very thick, black] (9.750000,1.000000) -- (9.750000,2.000000);
\draw[very thick, black] (9.000000,1.250000) -- (9.750000,1.250000);
\draw[very thick, black] (9.750000,1.750000) -- (10.000000,1.750000);
\draw[very thick, black] (10.000000,1.750000) -- (11.000000,1.750000);
\draw[very thick, black] (10.750000,2.000000) -- (10.750000,1.750000);
\draw[very thick, black] (10.250000,1.000000) -- (10.250000,1.750000);
\draw[very thick, black] (11.250000,1.000000) -- (11.250000,2.000000);
\draw[very thick, black] (11.000000,1.750000) -- (11.250000,1.750000);
\draw[very thick, black] (11.250000,1.500000) -- (12.000000,1.500000);
\draw[very thick, black] (0.750000,2.000000) -- (0.750000,3.000000);
\draw[very thick, black] (0.000000,2.250000) -- (0.750000,2.250000);
\draw[very thick, black] (0.750000,2.500000) -- (1.000000,2.500000);
\draw[very thick, black] (1.000000,2.500000) -- (2.000000,2.500000);
\draw[very thick, black] (1.750000,3.000000) -- (1.750000,2.500000);
\draw[very thick, black] (1.500000,2.000000) -- (1.500000,2.500000);
\draw[very thick, black] (2.250000,2.000000) -- (2.250000,3.000000);
\draw[very thick, black] (2.000000,2.500000) -- (2.250000,2.500000);
\draw[very thick, black] (2.250000,2.250000) -- (3.000000,2.250000);
\draw[very thick, black] (3.000000,2.250000) -- (4.000000,2.250000);
\draw[very thick, black] (3.500000,3.000000) -- (3.500000,2.250000);
\draw[very thick, black] (3.750000,2.000000) -- (3.750000,2.250000);
\draw[very thick, black] (4.500000,2.000000) -- (4.500000,3.000000);
\draw[very thick, black] (4.000000,2.250000) -- (4.500000,2.250000);
\draw[very thick, black] (4.500000,2.500000) -- (5.000000,2.500000);
\draw[very thick, black] (5.000000,2.500000) -- (6.000000,2.500000);
\draw[very thick, black] (5.250000,3.000000) -- (5.250000,2.500000);
\draw[very thick, black] (5.750000,2.000000) -- (5.750000,2.500000);
\draw[very thick, black] (6.250000,2.000000) -- (6.250000,3.000000);
\draw[very thick, black] (6.000000,2.500000) -- (6.250000,2.500000);
\draw[very thick, black] (6.250000,2.250000) -- (7.000000,2.250000);
\draw[very thick, black] (7.000000,2.250000) -- (8.000000,2.250000);
\draw[very thick, black] (7.500000,3.000000) -- (7.500000,2.250000);
\draw[very thick, black] (7.750000,2.000000) -- (7.750000,2.250000);
\draw[very thick, black] (8.500000,2.000000) -- (8.500000,3.000000);
\draw[very thick, black] (8.000000,2.250000) -- (8.500000,2.250000);
\draw[very thick, black] (8.500000,2.500000) -- (9.000000,2.500000);
\draw[very thick, black] (9.000000,2.500000) -- (10.000000,2.500000);
\draw[very thick, black] (9.250000,3.000000) -- (9.250000,2.500000);
\draw[very thick, black] (9.750000,2.000000) -- (9.750000,2.500000);
\draw[very thick, black] (10.750000,2.000000) -- (10.750000,3.000000);
\draw[very thick, black] (10.000000,2.500000) -- (10.750000,2.500000);
\draw[very thick, black] (10.750000,2.250000) -- (11.000000,2.250000);
\draw[very thick, black] (11.000000,2.250000) -- (12.000000,2.250000);
\draw[very thick, black] (11.750000,3.000000) -- (11.750000,2.250000);
\draw[very thick, black] (11.250000,2.000000) -- (11.250000,2.250000);
\draw[very thick, black] (0.000000,3.250000) -- (1.000000,3.250000);
\draw[very thick, black] (0.500000,4.000000) -- (0.500000,3.250000);
\draw[very thick, black] (0.750000,3.000000) -- (0.750000,3.250000);
\draw[very thick, black] (1.750000,3.000000) -- (1.750000,4.000000);
\draw[very thick, black] (1.000000,3.250000) -- (1.750000,3.250000);
\draw[very thick, black] (1.750000,3.500000) -- (2.000000,3.500000);
\draw[very thick, black] (2.000000,3.500000) -- (3.000000,3.500000);
\draw[very thick, black] (2.500000,4.000000) -- (2.500000,3.500000);
\draw[very thick, black] (2.250000,3.000000) -- (2.250000,3.500000);
\draw[very thick, black] (3.500000,3.000000) -- (3.500000,4.000000);
\draw[very thick, black] (3.000000,3.500000) -- (3.500000,3.500000);
\draw[very thick, black] (3.500000,3.750000) -- (4.000000,3.750000);
\draw[very thick, black] (4.000000,3.750000) -- (5.000000,3.750000);
\draw[very thick, black] (4.750000,4.000000) -- (4.750000,3.750000);
\draw[very thick, black] (4.500000,3.000000) -- (4.500000,3.750000);
\draw[very thick, black] (5.250000,3.000000) -- (5.250000,4.000000);
\draw[very thick, black] (5.000000,3.750000) -- (5.250000,3.750000);
\draw[very thick, black] (5.250000,3.500000) -- (6.000000,3.500000);
\draw[very thick, black] (6.000000,3.500000) -- (7.000000,3.500000);
\draw[very thick, black] (6.750000,4.000000) -- (6.750000,3.500000);
\draw[very thick, black] (6.250000,3.000000) -- (6.250000,3.500000);
\draw[very thick, black] (7.500000,3.000000) -- (7.500000,4.000000);
\draw[very thick, black] (7.000000,3.500000) -- (7.500000,3.500000);
\draw[very thick, black] (7.500000,3.250000) -- (8.000000,3.250000);
\draw[very thick, black] (8.000000,3.250000) -- (9.000000,3.250000);
\draw[very thick, black] (8.250000,4.000000) -- (8.250000,3.250000);
\draw[very thick, black] (8.500000,3.000000) -- (8.500000,3.250000);
\draw[very thick, black] (9.250000,3.000000) -- (9.250000,4.000000);
\draw[very thick, black] (9.000000,3.250000) -- (9.250000,3.250000);
\draw[very thick, black] (9.250000,3.750000) -- (10.000000,3.750000);
\draw[very thick, black] (10.000000,3.750000) -- (11.000000,3.750000);
\draw[very thick, black] (10.500000,4.000000) -- (10.500000,3.750000);
\draw[very thick, black] (10.750000,3.000000) -- (10.750000,3.750000);
\draw[very thick, black] (11.750000,3.000000) -- (11.750000,4.000000);
\draw[very thick, black] (11.000000,3.750000) -- (11.750000,3.750000);
\draw[very thick, black] (11.750000,3.500000) -- (12.000000,3.500000);
\draw[very thick, black] (0.500000,4.000000) -- (0.500000,5.000000);
\draw[very thick, black] (0.000000,4.250000) -- (0.500000,4.250000);
\draw[very thick, black] (0.500000,4.500000) -- (1.000000,4.500000);
\draw[very thick, black] (1.000000,4.500000) -- (2.000000,4.500000);
\draw[very thick, black] (1.250000,5.000000) -- (1.250000,4.500000);
\draw[very thick, black] (1.750000,4.000000) -- (1.750000,4.500000);
\draw[very thick, black] (2.500000,4.000000) -- (2.500000,5.000000);
\draw[very thick, black] (2.000000,4.500000) -- (2.500000,4.500000);
\draw[very thick, black] (2.500000,4.750000) -- (3.000000,4.750000);
\draw[very thick, black] (3.000000,4.750000) -- (4.000000,4.750000);
\draw[very thick, black] (3.750000,5.000000) -- (3.750000,4.750000);
\draw[very thick, black] (3.500000,4.000000) -- (3.500000,4.750000);
\draw[very thick, black] (4.750000,4.000000) -- (4.750000,5.000000);
\draw[very thick, black] (4.000000,4.750000) -- (4.750000,4.750000);
\draw[very thick, black] (4.750000,4.250000) -- (5.000000,4.250000);
\draw[very thick, black] (5.000000,4.250000) -- (6.000000,4.250000);
\draw[very thick, black] (5.500000,5.000000) -- (5.500000,4.250000);
\draw[very thick, black] (5.250000,4.000000) -- (5.250000,4.250000);
\draw[very thick, black] (6.750000,4.000000) -- (6.750000,5.000000);
\draw[very thick, black] (6.000000,4.250000) -- (6.750000,4.250000);
\draw[very thick, black] (6.750000,4.750000) -- (7.000000,4.750000);
\draw[very thick, black] (7.000000,4.750000) -- (8.000000,4.750000);
\draw[very thick, black] (7.750000,5.000000) -- (7.750000,4.750000);
\draw[very thick, black] (7.500000,4.000000) -- (7.500000,4.750000);
\draw[very thick, black] (8.250000,4.000000) -- (8.250000,5.000000);
\draw[very thick, black] (8.000000,4.750000) -- (8.250000,4.750000);
\draw[very thick, black] (8.250000,4.500000) -- (9.000000,4.500000);
\draw[very thick, black] (9.000000,4.500000) -- (10.000000,4.500000);
\draw[very thick, black] (9.500000,5.000000) -- (9.500000,4.500000);
\draw[very thick, black] (9.250000,4.000000) -- (9.250000,4.500000);
\draw[very thick, black] (10.500000,4.000000) -- (10.500000,5.000000);
\draw[very thick, black] (10.000000,4.500000) -- (10.500000,4.500000);
\draw[very thick, black] (10.500000,4.750000) -- (11.000000,4.750000);
\draw[very thick, black] (11.000000,4.750000) -- (12.000000,4.750000);
\draw[very thick, black] (11.250000,5.000000) -- (11.250000,4.750000);
\draw[very thick, black] (11.750000,4.000000) -- (11.750000,4.750000);
\draw[very thick, black] (0.000000,5.750000) -- (1.000000,5.750000);
\draw[very thick, black] (0.750000,6.000000) -- (0.750000,5.750000);
\draw[very thick, black] (0.500000,5.000000) -- (0.500000,5.750000);
\draw[very thick, black] (1.250000,5.000000) -- (1.250000,6.000000);
\draw[very thick, black] (1.000000,5.750000) -- (1.250000,5.750000);
\draw[very thick, black] (1.250000,5.250000) -- (2.000000,5.250000);
\draw[very thick, black] (2.000000,5.250000) -- (3.000000,5.250000);
\draw[very thick, black] (2.250000,6.000000) -- (2.250000,5.250000);
\draw[very thick, black] (2.500000,5.000000) -- (2.500000,5.250000);
\draw[very thick, black] (3.750000,5.000000) -- (3.750000,6.000000);
\draw[very thick, black] (3.000000,5.250000) -- (3.750000,5.250000);
\draw[very thick, black] (3.750000,5.750000) -- (4.000000,5.750000);
\draw[very thick, black] (4.000000,5.750000) -- (5.000000,5.750000);
\draw[very thick, black] (4.250000,6.000000) -- (4.250000,5.750000);
\draw[very thick, black] (4.750000,5.000000) -- (4.750000,5.750000);
\draw[very thick, black] (5.500000,5.000000) -- (5.500000,6.000000);
\draw[very thick, black] (5.000000,5.750000) -- (5.500000,5.750000);
\draw[very thick, black] (5.500000,5.250000) -- (6.000000,5.250000);
\draw[very thick, black] (6.000000,5.250000) -- (7.000000,5.250000);
\draw[very thick, black] (6.500000,6.000000) -- (6.500000,5.250000);
\draw[very thick, black] (6.750000,5.000000) -- (6.750000,5.250000);
\draw[very thick, black] (7.750000,5.000000) -- (7.750000,6.000000);
\draw[very thick, black] (7.000000,5.250000) -- (7.750000,5.250000);
\draw[very thick, black] (7.750000,5.500000) -- (8.000000,5.500000);
\draw[very thick, black] (8.000000,5.500000) -- (9.000000,5.500000);
\draw[very thick, black] (8.500000,6.000000) -- (8.500000,5.500000);
\draw[very thick, black] (8.250000,5.000000) -- (8.250000,5.500000);
\draw[very thick, black] (9.500000,5.000000) -- (9.500000,6.000000);
\draw[very thick, black] (9.000000,5.500000) -- (9.500000,5.500000);
\draw[very thick, black] (9.500000,5.250000) -- (10.000000,5.250000);
\draw[very thick, black] (10.000000,5.250000) -- (11.000000,5.250000);
\draw[very thick, black] (10.250000,6.000000) -- (10.250000,5.250000);
\draw[very thick, black] (10.500000,5.000000) -- (10.500000,5.250000);
\draw[very thick, black] (11.250000,5.000000) -- (11.250000,6.000000);
\draw[very thick, black] (11.000000,5.250000) -- (11.250000,5.250000);
\draw[very thick, black] (11.250000,5.750000) -- (12.000000,5.750000);
\draw[very thick, black] (0.750000,6.000000) -- (0.750000,7.000000);
\draw[very thick, black] (0.000000,6.500000) -- (0.750000,6.500000);
\draw[very thick, black] (0.750000,6.750000) -- (1.000000,6.750000);
\draw[very thick, black] (1.000000,6.750000) -- (2.000000,6.750000);
\draw[very thick, black] (1.500000,7.000000) -- (1.500000,6.750000);
\draw[very thick, black] (1.250000,6.000000) -- (1.250000,6.750000);
\draw[very thick, black] (2.250000,6.000000) -- (2.250000,7.000000);
\draw[very thick, black] (2.000000,6.750000) -- (2.250000,6.750000);
\draw[very thick, black] (2.250000,6.250000) -- (3.000000,6.250000);
\draw[very thick, black] (3.000000,6.250000) -- (4.000000,6.250000);
\draw[very thick, black] (3.500000,7.000000) -- (3.500000,6.250000);
\draw[very thick, black] (3.750000,6.000000) -- (3.750000,6.250000);
\draw[very thick, black] (4.250000,6.000000) -- (4.250000,7.000000);
\draw[very thick, black] (4.000000,6.250000) -- (4.250000,6.250000);
\draw[very thick, black] (4.250000,6.750000) -- (5.000000,6.750000);
\draw[very thick, black] (5.000000,6.750000) -- (6.000000,6.750000);
\draw[very thick, black] (5.250000,7.000000) -- (5.250000,6.750000);
\draw[very thick, black] (5.500000,6.000000) -- (5.500000,6.750000);
\draw[very thick, black] (6.500000,6.000000) -- (6.500000,7.000000);
\draw[very thick, black] (6.000000,6.750000) -- (6.500000,6.750000);
\draw[very thick, black] (6.500000,6.500000) -- (7.000000,6.500000);
\draw[very thick, black] (7.000000,6.500000) -- (8.000000,6.500000);
\draw[very thick, black] (7.250000,7.000000) -- (7.250000,6.500000);
\draw[very thick, black] (7.750000,6.000000) -- (7.750000,6.500000);
\draw[very thick, black] (8.500000,6.000000) -- (8.500000,7.000000);
\draw[very thick, black] (8.000000,6.500000) -- (8.500000,6.500000);
\draw[very thick, black] (8.500000,6.750000) -- (9.000000,6.750000);
\draw[very thick, black] (9.000000,6.750000) -- (10.000000,6.750000);
\draw[very thick, black] (9.250000,7.000000) -- (9.250000,6.750000);
\draw[very thick, black] (9.500000,6.000000) -- (9.500000,6.750000);
\draw[very thick, black] (10.250000,6.000000) -- (10.250000,7.000000);
\draw[very thick, black] (10.000000,6.750000) -- (10.250000,6.750000);
\draw[very thick, black] (10.250000,6.500000) -- (11.000000,6.500000);
\draw[very thick, black] (11.000000,6.500000) -- (12.000000,6.500000);
\draw[very thick, black] (11.500000,7.000000) -- (11.500000,6.500000);
\draw[very thick, black] (11.250000,6.000000) -- (11.250000,6.500000);
\draw[very thick, black] (0.000000,7.250000) -- (1.000000,7.250000);
\draw[very thick, black] (0.250000,8.000000) -- (0.250000,7.250000);
\draw[very thick, black] (0.750000,7.000000) -- (0.750000,7.250000);
\draw[very thick, black] (1.500000,7.000000) -- (1.500000,8.000000);
\draw[very thick, black] (1.000000,7.250000) -- (1.500000,7.250000);
\draw[very thick, black] (1.500000,7.500000) -- (2.000000,7.500000);
\draw[very thick, black] (2.000000,7.500000) -- (3.000000,7.500000);
\draw[very thick, black] (2.750000,8.000000) -- (2.750000,7.500000);
\draw[very thick, black] (2.250000,7.000000) -- (2.250000,7.500000);
\draw[very thick, black] (3.500000,7.000000) -- (3.500000,8.000000);
\draw[very thick, black] (3.000000,7.500000) -- (3.500000,7.500000);
\draw[very thick, black] (3.500000,7.750000) -- (4.000000,7.750000);
\draw[very thick, black] (4.000000,7.750000) -- (5.000000,7.750000);
\draw[very thick, black] (4.500000,8.000000) -- (4.500000,7.750000);
\draw[very thick, black] (4.250000,7.000000) -- (4.250000,7.750000);
\draw[very thick, black] (5.250000,7.000000) -- (5.250000,8.000000);
\draw[very thick, black] (5.000000,7.750000) -- (5.250000,7.750000);
\draw[very thick, black] (5.250000,7.500000) -- (6.000000,7.500000);
\draw[very thick, black] (6.000000,7.500000) -- (7.000000,7.500000);
\draw[very thick, black] (6.750000,8.000000) -- (6.750000,7.500000);
\draw[very thick, black] (6.500000,7.000000) -- (6.500000,7.500000);
\draw[very thick, black] (7.250000,7.000000) -- (7.250000,8.000000);
\draw[very thick, black] (7.000000,7.500000) -- (7.250000,7.500000);
\draw[very thick, black] (7.250000,7.250000) -- (8.000000,7.250000);
\draw[very thick, black] (8.000000,7.250000) -- (9.000000,7.250000);
\draw[very thick, black] (8.750000,8.000000) -- (8.750000,7.250000);
\draw[very thick, black] (8.500000,7.000000) -- (8.500000,7.250000);
\draw[very thick, black] (9.250000,7.000000) -- (9.250000,8.000000);
\draw[very thick, black] (9.000000,7.250000) -- (9.250000,7.250000);
\draw[very thick, black] (9.250000,7.750000) -- (10.000000,7.750000);
\draw[very thick, black] (10.000000,7.750000) -- (11.000000,7.750000);
\draw[very thick, black] (10.500000,8.000000) -- (10.500000,7.750000);
\draw[very thick, black] (10.250000,7.000000) -- (10.250000,7.750000);
\draw[very thick, black] (11.500000,7.000000) -- (11.500000,8.000000);
\draw[very thick, black] (11.000000,7.750000) -- (11.500000,7.750000);
\draw[very thick, black] (11.500000,7.250000) -- (12.000000,7.250000);
\draw[very thick, black] (0.250000,8.000000) -- (0.250000,9.000000);
\draw[very thick, black] (0.000000,8.750000) -- (0.250000,8.750000);
\draw[very thick, black] (0.250000,8.250000) -- (1.000000,8.250000);
\draw[very thick, black] (1.000000,8.250000) -- (2.000000,8.250000);
\draw[very thick, black] (1.250000,9.000000) -- (1.250000,8.250000);
\draw[very thick, black] (1.500000,8.000000) -- (1.500000,8.250000);
\draw[very thick, black] (2.750000,8.000000) -- (2.750000,9.000000);
\draw[very thick, black] (2.000000,8.250000) -- (2.750000,8.250000);
\draw[very thick, black] (2.750000,8.750000) -- (3.000000,8.750000);
\draw[very thick, black] (3.000000,8.750000) -- (4.000000,8.750000);
\draw[very thick, black] (3.750000,9.000000) -- (3.750000,8.750000);
\draw[very thick, black] (3.500000,8.000000) -- (3.500000,8.750000);
\draw[very thick, black] (4.500000,8.000000) -- (4.500000,9.000000);
\draw[very thick, black] (4.000000,8.750000) -- (4.500000,8.750000);
\draw[very thick, black] (4.500000,8.500000) -- (5.000000,8.500000);
\draw[very thick, black] (5.000000,8.500000) -- (6.000000,8.500000);
\draw[very thick, black] (5.750000,9.000000) -- (5.750000,8.500000);
\draw[very thick, black] (5.250000,8.000000) -- (5.250000,8.500000);
\draw[very thick, black] (6.750000,8.000000) -- (6.750000,9.000000);
\draw[very thick, black] (6.000000,8.500000) -- (6.750000,8.500000);
\draw[very thick, black] (6.750000,8.250000) -- (7.000000,8.250000);
\draw[very thick, black] (7.000000,8.250000) -- (8.000000,8.250000);
\draw[very thick, black] (7.500000,9.000000) -- (7.500000,8.250000);
\draw[very thick, black] (7.250000,8.000000) -- (7.250000,8.250000);
\draw[very thick, black] (8.750000,8.000000) -- (8.750000,9.000000);
\draw[very thick, black] (8.000000,8.250000) -- (8.750000,8.250000);
\draw[very thick, black] (8.750000,8.750000) -- (9.000000,8.750000);
\draw[very thick, black] (9.000000,8.750000) -- (10.000000,8.750000);
\draw[very thick, black] (9.750000,9.000000) -- (9.750000,8.750000);
\draw[very thick, black] (9.250000,8.000000) -- (9.250000,8.750000);
\draw[very thick, black] (10.500000,8.000000) -- (10.500000,9.000000);
\draw[very thick, black] (10.000000,8.750000) -- (10.500000,8.750000);
\draw[very thick, black] (10.500000,8.500000) -- (11.000000,8.500000);
\draw[very thick, black] (11.000000,8.500000) -- (12.000000,8.500000);
\draw[very thick, black] (11.750000,9.000000) -- (11.750000,8.500000);
\draw[very thick, black] (11.500000,8.000000) -- (11.500000,8.500000);
\draw[very thick, black] (0.000000,9.250000) -- (1.000000,9.250000);
\draw[very thick, black] (0.500000,10.000000) -- (0.500000,9.250000);
\draw[very thick, black] (0.250000,9.000000) -- (0.250000,9.250000);
\draw[very thick, black] (1.250000,9.000000) -- (1.250000,10.000000);
\draw[very thick, black] (1.000000,9.250000) -- (1.250000,9.250000);
\draw[very thick, black] (1.250000,9.500000) -- (2.000000,9.500000);
\draw[very thick, black] (2.000000,9.500000) -- (3.000000,9.500000);
\draw[very thick, black] (2.250000,10.000000) -- (2.250000,9.500000);
\draw[very thick, black] (2.750000,9.000000) -- (2.750000,9.500000);
\draw[very thick, black] (3.750000,9.000000) -- (3.750000,10.000000);
\draw[very thick, black] (3.000000,9.500000) -- (3.750000,9.500000);
\draw[very thick, black] (3.750000,9.750000) -- (4.000000,9.750000);
\draw[very thick, black] (4.000000,9.750000) -- (5.000000,9.750000);
\draw[very thick, black] (4.750000,10.000000) -- (4.750000,9.750000);
\draw[very thick, black] (4.500000,9.000000) -- (4.500000,9.750000);
\draw[very thick, black] (5.750000,9.000000) -- (5.750000,10.000000);
\draw[very thick, black] (5.000000,9.750000) -- (5.750000,9.750000);
\draw[very thick, black] (5.750000,9.500000) -- (6.000000,9.500000);
\draw[very thick, black] (6.000000,9.500000) -- (7.000000,9.500000);
\draw[very thick, black] (6.250000,10.000000) -- (6.250000,9.500000);
\draw[very thick, black] (6.750000,9.000000) -- (6.750000,9.500000);
\draw[very thick, black] (7.500000,9.000000) -- (7.500000,10.000000);
\draw[very thick, black] (7.000000,9.500000) -- (7.500000,9.500000);
\draw[very thick, black] (7.500000,9.250000) -- (8.000000,9.250000);
\draw[very thick, black] (8.000000,9.250000) -- (9.000000,9.250000);
\draw[very thick, black] (8.250000,10.000000) -- (8.250000,9.250000);
\draw[very thick, black] (8.750000,9.000000) -- (8.750000,9.250000);
\draw[very thick, black] (9.750000,9.000000) -- (9.750000,10.000000);
\draw[very thick, black] (9.000000,9.250000) -- (9.750000,9.250000);
\draw[very thick, black] (9.750000,9.750000) -- (10.000000,9.750000);
\draw[very thick, black] (10.000000,9.750000) -- (11.000000,9.750000);
\draw[very thick, black] (10.250000,10.000000) -- (10.250000,9.750000);
\draw[very thick, black] (10.500000,9.000000) -- (10.500000,9.750000);
\draw[very thick, black] (11.750000,9.000000) -- (11.750000,10.000000);
\draw[very thick, black] (11.000000,9.750000) -- (11.750000,9.750000);
\draw[very thick, black] (11.750000,9.500000) -- (12.000000,9.500000);
\draw[very thick, black] (0.500000,10.000000) -- (0.500000,11.000000);
\draw[very thick, black] (0.000000,10.250000) -- (0.500000,10.250000);
\draw[very thick, black] (0.500000,10.750000) -- (1.000000,10.750000);
\draw[very thick, black] (1.000000,10.750000) -- (2.000000,10.750000);
\draw[very thick, black] (1.500000,11.000000) -- (1.500000,10.750000);
\draw[very thick, black] (1.250000,10.000000) -- (1.250000,10.750000);
\draw[very thick, black] (2.250000,10.000000) -- (2.250000,11.000000);
\draw[very thick, black] (2.000000,10.750000) -- (2.250000,10.750000);
\draw[very thick, black] (2.250000,10.500000) -- (3.000000,10.500000);
\draw[very thick, black] (3.000000,10.500000) -- (4.000000,10.500000);
\draw[very thick, black] (3.500000,11.000000) -- (3.500000,10.500000);
\draw[very thick, black] (3.750000,10.000000) -- (3.750000,10.500000);
\draw[very thick, black] (4.750000,10.000000) -- (4.750000,11.000000);
\draw[very thick, black] (4.000000,10.500000) -- (4.750000,10.500000);
\draw[very thick, black] (4.750000,10.250000) -- (5.000000,10.250000);
\draw[very thick, black] (5.000000,10.250000) -- (6.000000,10.250000);
\draw[very thick, black] (5.250000,11.000000) -- (5.250000,10.250000);
\draw[very thick, black] (5.750000,10.000000) -- (5.750000,10.250000);
\draw[very thick, black] (6.250000,10.000000) -- (6.250000,11.000000);
\draw[very thick, black] (6.000000,10.250000) -- (6.250000,10.250000);
\draw[very thick, black] (6.250000,10.750000) -- (7.000000,10.750000);
\draw[very thick, black] (7.000000,10.750000) -- (8.000000,10.750000);
\draw[very thick, black] (7.750000,11.000000) -- (7.750000,10.750000);
\draw[very thick, black] (7.500000,10.000000) -- (7.500000,10.750000);
\draw[very thick, black] (8.250000,10.000000) -- (8.250000,11.000000);
\draw[very thick, black] (8.000000,10.750000) -- (8.250000,10.750000);
\draw[very thick, black] (8.250000,10.500000) -- (9.000000,10.500000);
\draw[very thick, black] (9.000000,10.500000) -- (10.000000,10.500000);
\draw[very thick, black] (9.500000,11.000000) -- (9.500000,10.500000);
\draw[very thick, black] (9.750000,10.000000) -- (9.750000,10.500000);
\draw[very thick, black] (10.250000,10.000000) -- (10.250000,11.000000);
\draw[very thick, black] (10.000000,10.500000) -- (10.250000,10.500000);
\draw[very thick, black] (10.250000,10.750000) -- (11.000000,10.750000);
\draw[very thick, black] (11.000000,10.750000) -- (12.000000,10.750000);
\draw[very thick, black] (11.500000,11.000000) -- (11.500000,10.750000);
\draw[very thick, black] (11.750000,10.000000) -- (11.750000,10.750000);
\draw[very thick, black] (0.000000,11.750000) -- (1.000000,11.750000);
\draw[very thick, black] (0.250000,12.000000) -- (0.250000,11.750000);
\draw[very thick, black] (0.500000,11.000000) -- (0.500000,11.750000);
\draw[very thick, black] (1.500000,11.000000) -- (1.500000,12.000000);
\draw[very thick, black] (1.000000,11.750000) -- (1.500000,11.750000);
\draw[very thick, black] (1.500000,11.250000) -- (2.000000,11.250000);
\draw[very thick, black] (2.000000,11.250000) -- (3.000000,11.250000);
\draw[very thick, black] (2.500000,12.000000) -- (2.500000,11.250000);
\draw[very thick, black] (2.250000,11.000000) -- (2.250000,11.250000);
\draw[very thick, black] (3.500000,11.000000) -- (3.500000,12.000000);
\draw[very thick, black] (3.000000,11.250000) -- (3.500000,11.250000);
\draw[very thick, black] (3.500000,11.500000) -- (4.000000,11.500000);
\draw[very thick, black] (4.000000,11.500000) -- (5.000000,11.500000);
\draw[very thick, black] (4.500000,12.000000) -- (4.500000,11.500000);
\draw[very thick, black] (4.750000,11.000000) -- (4.750000,11.500000);
\draw[very thick, black] (5.250000,11.000000) -- (5.250000,12.000000);
\draw[very thick, black] (5.000000,11.500000) -- (5.250000,11.500000);
\draw[very thick, black] (5.250000,11.250000) -- (6.000000,11.250000);
\draw[very thick, black] (6.000000,11.250000) -- (7.000000,11.250000);
\draw[very thick, black] (6.500000,12.000000) -- (6.500000,11.250000);
\draw[very thick, black] (6.250000,11.000000) -- (6.250000,11.250000);
\draw[very thick, black] (7.750000,11.000000) -- (7.750000,12.000000);
\draw[very thick, black] (7.000000,11.250000) -- (7.750000,11.250000);
\draw[very thick, black] (7.750000,11.750000) -- (8.000000,11.750000);
\draw[very thick, black] (8.000000,11.750000) -- (9.000000,11.750000);
\draw[very thick, black] (8.500000,12.000000) -- (8.500000,11.750000);
\draw[very thick, black] (8.250000,11.000000) -- (8.250000,11.750000);
\draw[very thick, black] (9.500000,11.000000) -- (9.500000,12.000000);
\draw[very thick, black] (9.000000,11.750000) -- (9.500000,11.750000);
\draw[very thick, black] (9.500000,11.500000) -- (10.000000,11.500000);
\draw[very thick, black] (10.000000,11.500000) -- (11.000000,11.500000);
\draw[very thick, black] (10.750000,12.000000) -- (10.750000,11.500000);
\draw[very thick, black] (10.250000,11.000000) -- (10.250000,11.500000);
\draw[very thick, black] (11.500000,11.000000) -- (11.500000,12.000000);
\draw[very thick, black] (11.000000,11.500000) -- (11.500000,11.500000);
\draw[very thick, black] (11.500000,11.250000) -- (12.000000,11.250000);
\draw[gray] (0,0) rectangle (12,12);
\end{tikzpicture}

%% file: result_n_3.tex
\begin{tikzpicture}[set style={{help lines}+=[dashed]}]
\draw[very thick, black] (0.250000,0.000000) -- (0.250000,1.000000);
\draw[very thick, black] (0.000000,0.500000) -- (0.250000,0.500000);
\draw[very thick, black] (0.250000,0.250000) -- (1.000000,0.250000);
\draw[very thick, black] (1.000000,0.250000) -- (2.000000,0.250000);
\draw[very thick, black] (1.750000,1.000000) -- (1.750000,0.250000);
\draw[very thick, black] (1.250000,0.000000) -- (1.250000,0.250000);
\draw[very thick, black] (2.750000,0.000000) -- (2.750000,1.000000);
\draw[very thick, black] (2.000000,0.250000) -- (2.750000,0.250000);
\draw[very thick, black] (2.750000,0.500000) -- (3.000000,0.500000);
\draw[very thick, black] (3.250000,0.000000) -- (3.250000,1.000000);
\draw[very thick, black] (3.000000,0.500000) -- (3.250000,0.500000);
\draw[very thick, black] (3.250000,0.250000) -- (4.000000,0.250000);
\draw[very thick, black] (4.250000,0.000000) -- (4.250000,1.000000);
\draw[very thick, black] (4.000000,0.250000) -- (4.250000,0.250000);
\draw[very thick, black] (4.250000,0.750000) -- (5.000000,0.750000);
\draw[very thick, black] (5.000000,0.750000) -- (6.000000,0.750000);
\draw[very thick, black] (5.750000,1.000000) -- (5.750000,0.750000);
\draw[very thick, black] (5.250000,0.000000) -- (5.250000,0.750000);
\draw[very thick, black] (6.000000,0.750000) -- (7.000000,0.750000);
\draw[very thick, black] (6.750000,1.000000) -- (6.750000,0.750000);
\draw[very thick, black] (6.250000,0.000000) -- (6.250000,0.750000);
\draw[very thick, black] (7.000000,0.750000) -- (8.000000,0.750000);
\draw[very thick, black] (7.500000,1.000000) -- (7.500000,0.750000);
\draw[very thick, black] (7.750000,0.000000) -- (7.750000,0.750000);
\draw[very thick, black] (8.750000,0.000000) -- (8.750000,1.000000);
\draw[very thick, black] (8.000000,0.750000) -- (8.750000,0.750000);
\draw[very thick, black] (8.750000,0.500000) -- (9.000000,0.500000);
\draw[very thick, black] (9.000000,0.500000) -- (10.000000,0.500000);
\draw[very thick, black] (9.750000,1.000000) -- (9.750000,0.500000);
\draw[very thick, black] (9.250000,0.000000) -- (9.250000,0.500000);
\draw[very thick, black] (10.750000,0.000000) -- (10.750000,1.000000);
\draw[very thick, black] (10.000000,0.500000) -- (10.750000,0.500000);
\draw[very thick, black] (10.750000,0.750000) -- (11.000000,0.750000);
\draw[very thick, black] (11.000000,0.750000) -- (12.000000,0.750000);
\draw[very thick, black] (11.500000,1.000000) -- (11.500000,0.750000);
\draw[very thick, black] (11.750000,0.000000) -- (11.750000,0.750000);
\draw[very thick, black] (0.000000,1.750000) -- (1.000000,1.750000);
\draw[very thick, black] (0.750000,2.000000) -- (0.750000,1.750000);
\draw[very thick, black] (0.250000,1.000000) -- (0.250000,1.750000);
\draw[very thick, black] (1.750000,1.000000) -- (1.750000,2.000000);
\draw[very thick, black] (1.000000,1.750000) -- (1.750000,1.750000);
\draw[very thick, black] (1.750000,1.500000) -- (2.000000,1.500000);
\draw[very thick, black] (2.000000,1.500000) -- (3.000000,1.500000);
\draw[very thick, black] (2.250000,2.000000) -- (2.250000,1.500000);
\draw[very thick, black] (2.750000,1.000000) -- (2.750000,1.500000);
\draw[very thick, black] (3.250000,1.000000) -- (3.250000,2.000000);
\draw[very thick, black] (3.000000,1.500000) -- (3.250000,1.500000);
\draw[very thick, black] (3.250000,1.750000) -- (4.000000,1.750000);
\draw[very thick, black] (4.000000,1.750000) -- (5.000000,1.750000);
\draw[very thick, black] (4.750000,2.000000) -- (4.750000,1.750000);
\draw[very thick, black] (4.250000,1.000000) -- (4.250000,1.750000);
\draw[very thick, black] (5.750000,1.000000) -- (5.750000,2.000000);
\draw[very thick, black] (5.000000,1.750000) -- (5.750000,1.750000);
\draw[very thick, black] (5.750000,1.500000) -- (6.000000,1.500000);
\draw[very thick, black] (6.750000,1.000000) -- (6.750000,2.000000);
\draw[very thick, black] (6.000000,1.500000) -- (6.750000,1.500000);
\draw[very thick, black] (6.750000,1.250000) -- (7.000000,1.250000);
\draw[very thick, black] (7.000000,1.250000) -- (8.000000,1.250000);
\draw[very thick, black] (7.750000,2.000000) -- (7.750000,1.250000);
\draw[very thick, black] (7.500000,1.000000) -- (7.500000,1.250000);
\draw[very thick, black] (8.000000,1.250000) -- (9.000000,1.250000);
\draw[very thick, black] (8.250000,2.000000) -- (8.250000,1.250000);
\draw[very thick, black] (8.750000,1.000000) -- (8.750000,1.250000);
\draw[very thick, black] (9.750000,1.000000) -- (9.750000,2.000000);
\draw[very thick, black] (9.000000,1.250000) -- (9.750000,1.250000);
\draw[very thick, black] (9.750000,1.750000) -- (10.000000,1.750000);
\draw[very thick, black] (10.000000,1.750000) -- (11.000000,1.750000);
\draw[very thick, black] (10.250000,2.000000) -- (10.250000,1.750000);
\draw[very thick, black] (10.750000,1.000000) -- (10.750000,1.750000);
\draw[very thick, black] (11.500000,1.000000) -- (11.500000,2.000000);
\draw[very thick, black] (11.000000,1.750000) -- (11.500000,1.750000);
\draw[very thick, black] (11.500000,1.500000) -- (12.000000,1.500000);
\draw[very thick, black] (0.000000,2.500000) -- (1.000000,2.500000);
\draw[very thick, black] (0.250000,3.000000) -- (0.250000,2.500000);
\draw[very thick, black] (0.750000,2.000000) -- (0.750000,2.500000);
\draw[very thick, black] (1.750000,2.000000) -- (1.750000,3.000000);
\draw[very thick, black] (1.000000,2.500000) -- (1.750000,2.500000);
\draw[very thick, black] (1.750000,2.750000) -- (2.000000,2.750000);
\draw[very thick, black] (2.250000,2.000000) -- (2.250000,3.000000);
\draw[very thick, black] (2.000000,2.750000) -- (2.250000,2.750000);
\draw[very thick, black] (2.250000,2.250000) -- (3.000000,2.250000);
\draw[very thick, black] (3.000000,2.250000) -- (4.000000,2.250000);
\draw[very thick, black] (3.750000,3.000000) -- (3.750000,2.250000);
\draw[very thick, black] (3.250000,2.000000) -- (3.250000,2.250000);
\draw[very thick, black] (4.000000,2.250000) -- (5.000000,2.250000);
\draw[very thick, black] (4.250000,3.000000) -- (4.250000,2.250000);
\draw[very thick, black] (4.750000,2.000000) -- (4.750000,2.250000);
\draw[very thick, black] (5.000000,2.250000) -- (6.000000,2.250000);
\draw[very thick, black] (5.250000,3.000000) -- (5.250000,2.250000);
\draw[very thick, black] (5.750000,2.000000) -- (5.750000,2.250000);
\draw[very thick, black] (6.750000,2.000000) -- (6.750000,3.000000);
\draw[very thick, black] (6.000000,2.250000) -- (6.750000,2.250000);
\draw[very thick, black] (6.750000,2.750000) -- (7.000000,2.750000);
\draw[very thick, black] (7.000000,2.750000) -- (8.000000,2.750000);
\draw[very thick, black] (7.500000,3.000000) -- (7.500000,2.750000);
\draw[very thick, black] (7.750000,2.000000) -- (7.750000,2.750000);
\draw[very thick, black] (8.250000,2.000000) -- (8.250000,3.000000);
\draw[very thick, black] (8.000000,2.750000) -- (8.250000,2.750000);
\draw[very thick, black] (8.250000,2.250000) -- (9.000000,2.250000);
\draw[very thick, black] (9.000000,2.250000) -- (10.000000,2.250000);
\draw[very thick, black] (9.500000,3.000000) -- (9.500000,2.250000);
\draw[very thick, black] (9.750000,2.000000) -- (9.750000,2.250000);
\draw[very thick, black] (10.250000,2.000000) -- (10.250000,3.000000);
\draw[very thick, black] (10.000000,2.250000) -- (10.250000,2.250000);
\draw[very thick, black] (10.250000,2.500000) -- (11.000000,2.500000);
\draw[very thick, black] (11.000000,2.500000) -- (12.000000,2.500000);
\draw[very thick, black] (11.750000,3.000000) -- (11.750000,2.500000);
\draw[very thick, black] (11.500000,2.000000) -- (11.500000,2.500000);
\draw[very thick, black] (0.000000,3.250000) -- (1.000000,3.250000);
\draw[very thick, black] (0.500000,4.000000) -- (0.500000,3.250000);
\draw[very thick, black] (0.250000,3.000000) -- (0.250000,3.250000);
\draw[very thick, black] (1.750000,3.000000) -- (1.750000,4.000000);
\draw[very thick, black] (1.000000,3.250000) -- (1.750000,3.250000);
\draw[very thick, black] (1.750000,3.500000) -- (2.000000,3.500000);
\draw[very thick, black] (2.000000,3.500000) -- (3.000000,3.500000);
\draw[very thick, black] (2.500000,4.000000) -- (2.500000,3.500000);
\draw[very thick, black] (2.250000,3.000000) -- (2.250000,3.500000);
\draw[very thick, black] (3.750000,3.000000) -- (3.750000,4.000000);
\draw[very thick, black] (3.000000,3.500000) -- (3.750000,3.500000);
\draw[very thick, black] (3.750000,3.250000) -- (4.000000,3.250000);
\draw[very thick, black] (4.250000,3.000000) -- (4.250000,4.000000);
\draw[very thick, black] (4.000000,3.250000) -- (4.250000,3.250000);
\draw[very thick, black] (4.250000,3.750000) -- (5.000000,3.750000);
\draw[very thick, black] (5.250000,3.000000) -- (5.250000,4.000000);
\draw[very thick, black] (5.000000,3.750000) -- (5.250000,3.750000);
\draw[very thick, black] (5.250000,3.500000) -- (6.000000,3.500000);
\draw[very thick, black] (6.000000,3.500000) -- (7.000000,3.500000);
\draw[very thick, black] (6.250000,4.000000) -- (6.250000,3.500000);
\draw[very thick, black] (6.750000,3.000000) -- (6.750000,3.500000);
\draw[very thick, black] (7.500000,3.000000) -- (7.500000,4.000000);
\draw[very thick, black] (7.000000,3.500000) -- (7.500000,3.500000);
\draw[very thick, black] (7.500000,3.250000) -- (8.000000,3.250000);
\draw[very thick, black] (8.250000,3.000000) -- (8.250000,4.000000);
\draw[very thick, black] (8.000000,3.250000) -- (8.250000,3.250000);
\draw[very thick, black] (8.250000,3.750000) -- (9.000000,3.750000);
\draw[very thick, black] (9.000000,3.750000) -- (10.000000,3.750000);
\draw[very thick, black] (9.750000,4.000000) -- (9.750000,3.750000);
\draw[very thick, black] (9.500000,3.000000) -- (9.500000,3.750000);
\draw[very thick, black] (10.000000,3.750000) -- (11.000000,3.750000);
\draw[very thick, black] (10.750000,4.000000) -- (10.750000,3.750000);
\draw[very thick, black] (10.250000,3.000000) -- (10.250000,3.750000);
\draw[very thick, black] (11.750000,3.000000) -- (11.750000,4.000000);
\draw[very thick, black] (11.000000,3.750000) -- (11.750000,3.750000);
\draw[very thick, black] (11.750000,3.500000) -- (12.000000,3.500000);
\draw[very thick, black] (0.500000,4.000000) -- (0.500000,5.000000);
\draw[very thick, black] (0.000000,4.250000) -- (0.500000,4.250000);
\draw[very thick, black] (0.500000,4.750000) -- (1.000000,4.750000);
\draw[very thick, black] (1.000000,4.750000) -- (2.000000,4.750000);
\draw[very thick, black] (1.250000,5.000000) -- (1.250000,4.750000);
\draw[very thick, black] (1.750000,4.000000) -- (1.750000,4.750000);
\draw[very thick, black] (2.500000,4.000000) -- (2.500000,5.000000);
\draw[very thick, black] (2.000000,4.750000) -- (2.500000,4.750000);
\draw[very thick, black] (2.500000,4.500000) -- (3.000000,4.500000);
\draw[very thick, black] (3.750000,4.000000) -- (3.750000,5.000000);
\draw[very thick, black] (3.000000,4.500000) -- (3.750000,4.500000);
\draw[very thick, black] (3.750000,4.750000) -- (4.000000,4.750000);
\draw[very thick, black] (4.250000,4.000000) -- (4.250000,5.000000);
\draw[very thick, black] (4.000000,4.750000) -- (4.250000,4.750000);
\draw[very thick, black] (4.250000,4.250000) -- (5.000000,4.250000);
\draw[very thick, black] (5.000000,4.250000) -- (6.000000,4.250000);
\draw[very thick, black] (5.500000,5.000000) -- (5.500000,4.250000);
\draw[very thick, black] (5.250000,4.000000) -- (5.250000,4.250000);
\draw[very thick, black] (6.250000,4.000000) -- (6.250000,5.000000);
\draw[very thick, black] (6.000000,4.250000) -- (6.250000,4.250000);
\draw[very thick, black] (6.250000,4.750000) -- (7.000000,4.750000);
\draw[very thick, black] (7.000000,4.750000) -- (8.000000,4.750000);
\draw[very thick, black] (7.750000,5.000000) -- (7.750000,4.750000);
\draw[very thick, black] (7.500000,4.000000) -- (7.500000,4.750000);
\draw[very thick, black] (8.250000,4.000000) -- (8.250000,5.000000);
\draw[very thick, black] (8.000000,4.750000) -- (8.250000,4.750000);
\draw[very thick, black] (8.250000,4.500000) -- (9.000000,4.500000);
\draw[very thick, black] (9.000000,4.500000) -- (10.000000,4.500000);
\draw[very thick, black] (9.250000,5.000000) -- (9.250000,4.500000);
\draw[very thick, black] (9.750000,4.000000) -- (9.750000,4.500000);
\draw[very thick, black] (10.750000,4.000000) -- (10.750000,5.000000);
\draw[very thick, black] (10.000000,4.500000) -- (10.750000,4.500000);
\draw[very thick, black] (10.750000,4.250000) -- (11.000000,4.250000);
\draw[very thick, black] (11.750000,4.000000) -- (11.750000,5.000000);
\draw[very thick, black] (11.000000,4.250000) -- (11.750000,4.250000);
\draw[very thick, black] (11.750000,4.750000) -- (12.000000,4.750000);
\draw[very thick, black] (0.000000,5.750000) -- (1.000000,5.750000);
\draw[very thick, black] (0.750000,6.000000) -- (0.750000,5.750000);
\draw[very thick, black] (0.500000,5.000000) -- (0.500000,5.750000);
\draw[very thick, black] (1.250000,5.000000) -- (1.250000,6.000000);
\draw[very thick, black] (1.000000,5.750000) -- (1.250000,5.750000);
\draw[very thick, black] (1.250000,5.250000) -- (2.000000,5.250000);
\draw[very thick, black] (2.000000,5.250000) -- (3.000000,5.250000);
\draw[very thick, black] (2.250000,6.000000) -- (2.250000,5.250000);
\draw[very thick, black] (2.500000,5.000000) -- (2.500000,5.250000);
\draw[very thick, black] (3.750000,5.000000) -- (3.750000,6.000000);
\draw[very thick, black] (3.000000,5.250000) -- (3.750000,5.250000);
\draw[very thick, black] (3.750000,5.750000) -- (4.000000,5.750000);
\draw[very thick, black] (4.000000,5.750000) -- (5.000000,5.750000);
\draw[very thick, black] (4.500000,6.000000) -- (4.500000,5.750000);
\draw[very thick, black] (4.250000,5.000000) -- (4.250000,5.750000);
\draw[very thick, black] (5.500000,5.000000) -- (5.500000,6.000000);
\draw[very thick, black] (5.000000,5.750000) -- (5.500000,5.750000);
\draw[very thick, black] (5.500000,5.250000) -- (6.000000,5.250000);
\draw[very thick, black] (6.250000,5.000000) -- (6.250000,6.000000);
\draw[very thick, black] (6.000000,5.250000) -- (6.250000,5.250000);
\draw[very thick, black] (6.250000,5.500000) -- (7.000000,5.500000);
\draw[very thick, black] (7.000000,5.500000) -- (8.000000,5.500000);
\draw[very thick, black] (7.500000,6.000000) -- (7.500000,5.500000);
\draw[very thick, black] (7.750000,5.000000) -- (7.750000,5.500000);
\draw[very thick, black] (8.000000,5.500000) -- (9.000000,5.500000);
\draw[very thick, black] (8.500000,6.000000) -- (8.500000,5.500000);
\draw[very thick, black] (8.250000,5.000000) -- (8.250000,5.500000);
\draw[very thick, black] (9.250000,5.000000) -- (9.250000,6.000000);
\draw[very thick, black] (9.000000,5.500000) -- (9.250000,5.500000);
\draw[very thick, black] (9.250000,5.750000) -- (10.000000,5.750000);
\draw[very thick, black] (10.000000,5.750000) -- (11.000000,5.750000);
\draw[very thick, black] (10.250000,6.000000) -- (10.250000,5.750000);
\draw[very thick, black] (10.750000,5.000000) -- (10.750000,5.750000);
\draw[very thick, black] (11.000000,5.750000) -- (12.000000,5.750000);
\draw[very thick, black] (11.500000,6.000000) -- (11.500000,5.750000);
\draw[very thick, black] (11.750000,5.000000) -- (11.750000,5.750000);
\draw[very thick, black] (0.000000,6.750000) -- (1.000000,6.750000);
\draw[very thick, black] (0.500000,7.000000) -- (0.500000,6.750000);
\draw[very thick, black] (0.750000,6.000000) -- (0.750000,6.750000);
\draw[very thick, black] (1.250000,6.000000) -- (1.250000,7.000000);
\draw[very thick, black] (1.000000,6.750000) -- (1.250000,6.750000);
\draw[very thick, black] (1.250000,6.500000) -- (2.000000,6.500000);
\draw[very thick, black] (2.250000,6.000000) -- (2.250000,7.000000);
\draw[very thick, black] (2.000000,6.500000) -- (2.250000,6.500000);
\draw[very thick, black] (2.250000,6.250000) -- (3.000000,6.250000);
\draw[very thick, black] (3.000000,6.250000) -- (4.000000,6.250000);
\draw[very thick, black] (3.250000,7.000000) -- (3.250000,6.250000);
\draw[very thick, black] (3.750000,6.000000) -- (3.750000,6.250000);
\draw[very thick, black] (4.500000,6.000000) -- (4.500000,7.000000);
\draw[very thick, black] (4.000000,6.250000) -- (4.500000,6.250000);
\draw[very thick, black] (4.500000,6.750000) -- (5.000000,6.750000);
\draw[very thick, black] (5.000000,6.750000) -- (6.000000,6.750000);
\draw[very thick, black] (5.250000,7.000000) -- (5.250000,6.750000);
\draw[very thick, black] (5.500000,6.000000) -- (5.500000,6.750000);
\draw[very thick, black] (6.250000,6.000000) -- (6.250000,7.000000);
\draw[very thick, black] (6.000000,6.750000) -- (6.250000,6.750000);
\draw[very thick, black] (6.250000,6.250000) -- (7.000000,6.250000);
\draw[very thick, black] (7.000000,6.250000) -- (8.000000,6.250000);
\draw[very thick, black] (7.750000,7.000000) -- (7.750000,6.250000);
\draw[very thick, black] (7.500000,6.000000) -- (7.500000,6.250000);
\draw[very thick, black] (8.000000,6.250000) -- (9.000000,6.250000);
\draw[very thick, black] (8.250000,7.000000) -- (8.250000,6.250000);
\draw[very thick, black] (8.500000,6.000000) -- (8.500000,6.250000);
\draw[very thick, black] (9.250000,6.000000) -- (9.250000,7.000000);
\draw[very thick, black] (9.000000,6.250000) -- (9.250000,6.250000);
\draw[very thick, black] (9.250000,6.500000) -- (10.000000,6.500000);
\draw[very thick, black] (10.250000,6.000000) -- (10.250000,7.000000);
\draw[very thick, black] (10.000000,6.500000) -- (10.250000,6.500000);
\draw[very thick, black] (10.250000,6.250000) -- (11.000000,6.250000);
\draw[very thick, black] (11.000000,6.250000) -- (12.000000,6.250000);
\draw[very thick, black] (11.750000,7.000000) -- (11.750000,6.250000);
\draw[very thick, black] (11.500000,6.000000) -- (11.500000,6.250000);
\draw[very thick, black] (0.000000,7.250000) -- (1.000000,7.250000);
\draw[very thick, black] (0.250000,8.000000) -- (0.250000,7.250000);
\draw[very thick, black] (0.500000,7.000000) -- (0.500000,7.250000);
\draw[very thick, black] (1.000000,7.250000) -- (2.000000,7.250000);
\draw[very thick, black] (1.500000,8.000000) -- (1.500000,7.250000);
\draw[very thick, black] (1.250000,7.000000) -- (1.250000,7.250000);
\draw[very thick, black] (2.000000,7.250000) -- (3.000000,7.250000);
\draw[very thick, black] (2.750000,8.000000) -- (2.750000,7.250000);
\draw[very thick, black] (2.250000,7.000000) -- (2.250000,7.250000);
\draw[very thick, black] (3.250000,7.000000) -- (3.250000,8.000000);
\draw[very thick, black] (3.000000,7.250000) -- (3.250000,7.250000);
\draw[very thick, black] (3.250000,7.750000) -- (4.000000,7.750000);
\draw[very thick, black] (4.500000,7.000000) -- (4.500000,8.000000);
\draw[very thick, black] (4.000000,7.750000) -- (4.500000,7.750000);
\draw[very thick, black] (4.500000,7.500000) -- (5.000000,7.500000);
\draw[very thick, black] (5.250000,7.000000) -- (5.250000,8.000000);
\draw[very thick, black] (5.000000,7.500000) -- (5.250000,7.500000);
\draw[very thick, black] (5.250000,7.750000) -- (6.000000,7.750000);
\draw[very thick, black] (6.000000,7.750000) -- (7.000000,7.750000);
\draw[very thick, black] (6.500000,8.000000) -- (6.500000,7.750000);
\draw[very thick, black] (6.250000,7.000000) -- (6.250000,7.750000);
\draw[very thick, black] (7.750000,7.000000) -- (7.750000,8.000000);
\draw[very thick, black] (7.000000,7.750000) -- (7.750000,7.750000);
\draw[very thick, black] (7.750000,7.500000) -- (8.000000,7.500000);
\draw[very thick, black] (8.000000,7.500000) -- (9.000000,7.500000);
\draw[very thick, black] (8.750000,8.000000) -- (8.750000,7.500000);
\draw[very thick, black] (8.250000,7.000000) -- (8.250000,7.500000);
\draw[very thick, black] (9.250000,7.000000) -- (9.250000,8.000000);
\draw[very thick, black] (9.000000,7.500000) -- (9.250000,7.500000);
\draw[very thick, black] (9.250000,7.250000) -- (10.000000,7.250000);
\draw[very thick, black] (10.000000,7.250000) -- (11.000000,7.250000);
\draw[very thick, black] (10.500000,8.000000) -- (10.500000,7.250000);
\draw[very thick, black] (10.250000,7.000000) -- (10.250000,7.250000);
\draw[very thick, black] (11.750000,7.000000) -- (11.750000,8.000000);
\draw[very thick, black] (11.000000,7.250000) -- (11.750000,7.250000);
\draw[very thick, black] (11.750000,7.500000) -- (12.000000,7.500000);
\draw[very thick, black] (0.250000,8.000000) -- (0.250000,9.000000);
\draw[very thick, black] (0.000000,8.750000) -- (0.250000,8.750000);
\draw[very thick, black] (0.250000,8.250000) -- (1.000000,8.250000);
\draw[very thick, black] (1.000000,8.250000) -- (2.000000,8.250000);
\draw[very thick, black] (1.250000,9.000000) -- (1.250000,8.250000);
\draw[very thick, black] (1.500000,8.000000) -- (1.500000,8.250000);
\draw[very thick, black] (2.750000,8.000000) -- (2.750000,9.000000);
\draw[very thick, black] (2.000000,8.250000) -- (2.750000,8.250000);
\draw[very thick, black] (2.750000,8.750000) -- (3.000000,8.750000);
\draw[very thick, black] (3.000000,8.750000) -- (4.000000,8.750000);
\draw[very thick, black] (3.500000,9.000000) -- (3.500000,8.750000);
\draw[very thick, black] (3.250000,8.000000) -- (3.250000,8.750000);
\draw[very thick, black] (4.500000,8.000000) -- (4.500000,9.000000);
\draw[very thick, black] (4.000000,8.750000) -- (4.500000,8.750000);
\draw[very thick, black] (4.500000,8.250000) -- (5.000000,8.250000);
\draw[very thick, black] (5.000000,8.250000) -- (6.000000,8.250000);
\draw[very thick, black] (5.500000,9.000000) -- (5.500000,8.250000);
\draw[very thick, black] (5.250000,8.000000) -- (5.250000,8.250000);
\draw[very thick, black] (6.000000,8.250000) -- (7.000000,8.250000);
\draw[very thick, black] (6.750000,9.000000) -- (6.750000,8.250000);
\draw[very thick, black] (6.500000,8.000000) -- (6.500000,8.250000);
\draw[very thick, black] (7.000000,8.250000) -- (8.000000,8.250000);
\draw[very thick, black] (7.250000,9.000000) -- (7.250000,8.250000);
\draw[very thick, black] (7.750000,8.000000) -- (7.750000,8.250000);
\draw[very thick, black] (8.750000,8.000000) -- (8.750000,9.000000);
\draw[very thick, black] (8.000000,8.250000) -- (8.750000,8.250000);
\draw[very thick, black] (8.750000,8.750000) -- (9.000000,8.750000);
\draw[very thick, black] (9.000000,8.750000) -- (10.000000,8.750000);
\draw[very thick, black] (9.500000,9.000000) -- (9.500000,8.750000);
\draw[very thick, black] (9.250000,8.000000) -- (9.250000,8.750000);
\draw[very thick, black] (10.000000,8.750000) -- (11.000000,8.750000);
\draw[very thick, black] (10.250000,9.000000) -- (10.250000,8.750000);
\draw[very thick, black] (10.500000,8.000000) -- (10.500000,8.750000);
\draw[very thick, black] (11.750000,8.000000) -- (11.750000,9.000000);
\draw[very thick, black] (11.000000,8.750000) -- (11.750000,8.750000);
\draw[very thick, black] (11.750000,8.500000) -- (12.000000,8.500000);
\draw[very thick, black] (0.000000,9.250000) -- (1.000000,9.250000);
\draw[very thick, black] (0.500000,10.000000) -- (0.500000,9.250000);
\draw[very thick, black] (0.250000,9.000000) -- (0.250000,9.250000);
\draw[very thick, black] (1.250000,9.000000) -- (1.250000,10.000000);
\draw[very thick, black] (1.000000,9.250000) -- (1.250000,9.250000);
\draw[very thick, black] (1.250000,9.500000) -- (2.000000,9.500000);
\draw[very thick, black] (2.000000,9.500000) -- (3.000000,9.500000);
\draw[very thick, black] (2.500000,10.000000) -- (2.500000,9.500000);
\draw[very thick, black] (2.750000,9.000000) -- (2.750000,9.500000);
\draw[very thick, black] (3.000000,9.500000) -- (4.000000,9.500000);
\draw[very thick, black] (3.750000,10.000000) -- (3.750000,9.500000);
\draw[very thick, black] (3.500000,9.000000) -- (3.500000,9.500000);
\draw[very thick, black] (4.000000,9.500000) -- (5.000000,9.500000);
\draw[very thick, black] (4.750000,10.000000) -- (4.750000,9.500000);
\draw[very thick, black] (4.500000,9.000000) -- (4.500000,9.500000);
\draw[very thick, black] (5.500000,9.000000) -- (5.500000,10.000000);
\draw[very thick, black] (5.000000,9.500000) -- (5.500000,9.500000);
\draw[very thick, black] (5.500000,9.750000) -- (6.000000,9.750000);
\draw[very thick, black] (6.000000,9.750000) -- (7.000000,9.750000);
\draw[very thick, black] (6.250000,10.000000) -- (6.250000,9.750000);
\draw[very thick, black] (6.750000,9.000000) -- (6.750000,9.750000);
\draw[very thick, black] (7.250000,9.000000) -- (7.250000,10.000000);
\draw[very thick, black] (7.000000,9.750000) -- (7.250000,9.750000);
\draw[very thick, black] (7.250000,9.250000) -- (8.000000,9.250000);
\draw[very thick, black] (8.000000,9.250000) -- (9.000000,9.250000);
\draw[very thick, black] (8.250000,10.000000) -- (8.250000,9.250000);
\draw[very thick, black] (8.750000,9.000000) -- (8.750000,9.250000);
\draw[very thick, black] (9.500000,9.000000) -- (9.500000,10.000000);
\draw[very thick, black] (9.000000,9.250000) -- (9.500000,9.250000);
\draw[very thick, black] (9.500000,9.750000) -- (10.000000,9.750000);
\draw[very thick, black] (10.000000,9.750000) -- (11.000000,9.750000);
\draw[very thick, black] (10.750000,10.000000) -- (10.750000,9.750000);
\draw[very thick, black] (10.250000,9.000000) -- (10.250000,9.750000);
\draw[very thick, black] (11.750000,9.000000) -- (11.750000,10.000000);
\draw[very thick, black] (11.000000,9.750000) -- (11.750000,9.750000);
\draw[very thick, black] (11.750000,9.500000) -- (12.000000,9.500000);
\draw[very thick, black] (0.000000,10.750000) -- (1.000000,10.750000);
\draw[very thick, black] (0.750000,11.000000) -- (0.750000,10.750000);
\draw[very thick, black] (0.500000,10.000000) -- (0.500000,10.750000);
\draw[very thick, black] (1.250000,10.000000) -- (1.250000,11.000000);
\draw[very thick, black] (1.000000,10.750000) -- (1.250000,10.750000);
\draw[very thick, black] (1.250000,10.250000) -- (2.000000,10.250000);
\draw[very thick, black] (2.500000,10.000000) -- (2.500000,11.000000);
\draw[very thick, black] (2.000000,10.250000) -- (2.500000,10.250000);
\draw[very thick, black] (2.500000,10.500000) -- (3.000000,10.500000);
\draw[very thick, black] (3.000000,10.500000) -- (4.000000,10.500000);
\draw[very thick, black] (3.500000,11.000000) -- (3.500000,10.500000);
\draw[very thick, black] (3.750000,10.000000) -- (3.750000,10.500000);
\draw[very thick, black] (4.750000,10.000000) -- (4.750000,11.000000);
\draw[very thick, black] (4.000000,10.500000) -- (4.750000,10.500000);
\draw[very thick, black] (4.750000,10.750000) -- (5.000000,10.750000);
\draw[very thick, black] (5.500000,10.000000) -- (5.500000,11.000000);
\draw[very thick, black] (5.000000,10.750000) -- (5.500000,10.750000);
\draw[very thick, black] (5.500000,10.250000) -- (6.000000,10.250000);
\draw[very thick, black] (6.250000,10.000000) -- (6.250000,11.000000);
\draw[very thick, black] (6.000000,10.250000) -- (6.250000,10.250000);
\draw[very thick, black] (6.250000,10.500000) -- (7.000000,10.500000);
\draw[very thick, black] (7.000000,10.500000) -- (8.000000,10.500000);
\draw[very thick, black] (7.750000,11.000000) -- (7.750000,10.500000);
\draw[very thick, black] (7.250000,10.000000) -- (7.250000,10.500000);
\draw[very thick, black] (8.250000,10.000000) -- (8.250000,11.000000);
\draw[very thick, black] (8.000000,10.500000) -- (8.250000,10.500000);
\draw[very thick, black] (8.250000,10.250000) -- (9.000000,10.250000);
\draw[very thick, black] (9.500000,10.000000) -- (9.500000,11.000000);
\draw[very thick, black] (9.000000,10.250000) -- (9.500000,10.250000);
\draw[very thick, black] (9.500000,10.750000) -- (10.000000,10.750000);
\draw[very thick, black] (10.750000,10.000000) -- (10.750000,11.000000);
\draw[very thick, black] (10.000000,10.750000) -- (10.750000,10.750000);
\draw[very thick, black] (10.750000,10.250000) -- (11.000000,10.250000);
\draw[very thick, black] (11.000000,10.250000) -- (12.000000,10.250000);
\draw[very thick, black] (11.250000,11.000000) -- (11.250000,10.250000);
\draw[very thick, black] (11.750000,10.000000) -- (11.750000,10.250000);
\draw[very thick, black] (0.000000,11.500000) -- (1.000000,11.500000);
\draw[very thick, black] (0.250000,12.000000) -- (0.250000,11.500000);
\draw[very thick, black] (0.750000,11.000000) -- (0.750000,11.500000);
\draw[very thick, black] (1.000000,11.500000) -- (2.000000,11.500000);
\draw[very thick, black] (1.500000,12.000000) -- (1.500000,11.500000);
\draw[very thick, black] (1.250000,11.000000) -- (1.250000,11.500000);
\draw[very thick, black] (2.000000,11.500000) -- (3.000000,11.500000);
\draw[very thick, black] (2.250000,12.000000) -- (2.250000,11.500000);
\draw[very thick, black] (2.500000,11.000000) -- (2.500000,11.500000);
\draw[very thick, black] (3.500000,11.000000) -- (3.500000,12.000000);
\draw[very thick, black] (3.000000,11.500000) -- (3.500000,11.500000);
\draw[very thick, black] (3.500000,11.750000) -- (4.000000,11.750000);
\draw[very thick, black] (4.750000,11.000000) -- (4.750000,12.000000);
\draw[very thick, black] (4.000000,11.750000) -- (4.750000,11.750000);
\draw[very thick, black] (4.750000,11.500000) -- (5.000000,11.500000);
\draw[very thick, black] (5.500000,11.000000) -- (5.500000,12.000000);
\draw[very thick, black] (5.000000,11.500000) -- (5.500000,11.500000);
\draw[very thick, black] (5.500000,11.750000) -- (6.000000,11.750000);
\draw[very thick, black] (6.000000,11.750000) -- (7.000000,11.750000);
\draw[very thick, black] (6.500000,12.000000) -- (6.500000,11.750000);
\draw[very thick, black] (6.250000,11.000000) -- (6.250000,11.750000);
\draw[very thick, black] (7.750000,11.000000) -- (7.750000,12.000000);
\draw[very thick, black] (7.000000,11.750000) -- (7.750000,11.750000);
\draw[very thick, black] (7.750000,11.250000) -- (8.000000,11.250000);
\draw[very thick, black] (8.000000,11.250000) -- (9.000000,11.250000);
\draw[very thick, black] (8.500000,12.000000) -- (8.500000,11.250000);
\draw[very thick, black] (8.250000,11.000000) -- (8.250000,11.250000);
\draw[very thick, black] (9.500000,11.000000) -- (9.500000,12.000000);
\draw[very thick, black] (9.000000,11.250000) -- (9.500000,11.250000);
\draw[very thick, black] (9.500000,11.500000) -- (10.000000,11.500000);
\draw[very thick, black] (10.000000,11.500000) -- (11.000000,11.500000);
\draw[very thick, black] (10.250000,12.000000) -- (10.250000,11.500000);
\draw[very thick, black] (10.750000,11.000000) -- (10.750000,11.500000);
\draw[very thick, black] (11.250000,11.000000) -- (11.250000,12.000000);
\draw[very thick, black] (11.000000,11.500000) -- (11.250000,11.500000);
\draw[very thick, black] (11.250000,11.250000) -- (12.000000,11.250000);
\draw[gray] (0,0) rectangle (12,12);
\end{tikzpicture}

%% file: result_n_5.tex
\begin{tikzpicture}[set style={{help lines}+=[dashed]}]
\draw[very thick, black] (0.250000,0.000000) -- (0.250000,1.000000);
\draw[very thick, black] (0.000000,0.500000) -- (0.250000,0.500000);
\draw[very thick, black] (0.250000,0.250000) -- (1.000000,0.250000);
\draw[very thick, black] (1.000000,0.250000) -- (2.000000,0.250000);
\draw[very thick, black] (1.750000,1.000000) -- (1.750000,0.250000);
\draw[very thick, black] (1.250000,0.000000) -- (1.250000,0.250000);
\draw[very thick, black] (2.750000,0.000000) -- (2.750000,1.000000);
\draw[very thick, black] (2.000000,0.250000) -- (2.750000,0.250000);
\draw[very thick, black] (2.750000,0.500000) -- (3.000000,0.500000);
\draw[very thick, black] (3.250000,0.000000) -- (3.250000,1.000000);
\draw[very thick, black] (3.000000,0.500000) -- (3.250000,0.500000);
\draw[very thick, black] (3.250000,0.250000) -- (4.000000,0.250000);
\draw[very thick, black] (4.500000,0.000000) -- (4.500000,1.000000);
\draw[very thick, black] (4.000000,0.250000) -- (4.500000,0.250000);
\draw[very thick, black] (4.500000,0.750000) -- (5.000000,0.750000);
\draw[very thick, black] (5.000000,0.750000) -- (6.000000,0.750000);
\draw[very thick, black] (5.500000,1.000000) -- (5.500000,0.750000);
\draw[very thick, black] (5.750000,0.000000) -- (5.750000,0.750000);
\draw[very thick, black] (6.500000,0.000000) -- (6.500000,1.000000);
\draw[very thick, black] (6.000000,0.750000) -- (6.500000,0.750000);
\draw[very thick, black] (6.500000,0.250000) -- (7.000000,0.250000);
\draw[very thick, black] (7.000000,0.250000) -- (8.000000,0.250000);
\draw[very thick, black] (7.750000,1.000000) -- (7.750000,0.250000);
\draw[very thick, black] (7.250000,0.000000) -- (7.250000,0.250000);
\draw[very thick, black] (8.250000,0.000000) -- (8.250000,1.000000);
\draw[very thick, black] (8.000000,0.250000) -- (8.250000,0.250000);
\draw[very thick, black] (8.250000,0.750000) -- (9.000000,0.750000);
\draw[very thick, black] (9.250000,0.000000) -- (9.250000,1.000000);
\draw[very thick, black] (9.000000,0.750000) -- (9.250000,0.750000);
\draw[very thick, black] (9.250000,0.500000) -- (10.000000,0.500000);
\draw[very thick, black] (10.750000,0.000000) -- (10.750000,1.000000);
\draw[very thick, black] (10.000000,0.500000) -- (10.750000,0.500000);
\draw[very thick, black] (10.750000,0.250000) -- (11.000000,0.250000);
\draw[very thick, black] (11.000000,0.250000) -- (12.000000,0.250000);
\draw[very thick, black] (11.500000,1.000000) -- (11.500000,0.250000);
\draw[very thick, black] (11.250000,0.000000) -- (11.250000,0.250000);
\draw[very thick, black] (0.000000,1.750000) -- (1.000000,1.750000);
\draw[very thick, black] (0.750000,2.000000) -- (0.750000,1.750000);
\draw[very thick, black] (0.250000,1.000000) -- (0.250000,1.750000);
\draw[very thick, black] (1.750000,1.000000) -- (1.750000,2.000000);
\draw[very thick, black] (1.000000,1.750000) -- (1.750000,1.750000);
\draw[very thick, black] (1.750000,1.500000) -- (2.000000,1.500000);
\draw[very thick, black] (2.000000,1.500000) -- (3.000000,1.500000);
\draw[very thick, black] (2.250000,2.000000) -- (2.250000,1.500000);
\draw[very thick, black] (2.750000,1.000000) -- (2.750000,1.500000);
\draw[very thick, black] (3.250000,1.000000) -- (3.250000,2.000000);
\draw[very thick, black] (3.000000,1.500000) -- (3.250000,1.500000);
\draw[very thick, black] (3.250000,1.750000) -- (4.000000,1.750000);
\draw[very thick, black] (4.000000,1.750000) -- (5.000000,1.750000);
\draw[very thick, black] (4.250000,2.000000) -- (4.250000,1.750000);
\draw[very thick, black] (4.500000,1.000000) -- (4.500000,1.750000);
\draw[very thick, black] (5.500000,1.000000) -- (5.500000,2.000000);
\draw[very thick, black] (5.000000,1.750000) -- (5.500000,1.750000);
\draw[very thick, black] (5.500000,1.500000) -- (6.000000,1.500000);
\draw[very thick, black] (6.000000,1.500000) -- (7.000000,1.500000);
\draw[very thick, black] (6.250000,2.000000) -- (6.250000,1.500000);
\draw[very thick, black] (6.500000,1.000000) -- (6.500000,1.500000);
\draw[very thick, black] (7.750000,1.000000) -- (7.750000,2.000000);
\draw[very thick, black] (7.000000,1.500000) -- (7.750000,1.500000);
\draw[very thick, black] (7.750000,1.750000) -- (8.000000,1.750000);
\draw[very thick, black] (8.250000,1.000000) -- (8.250000,2.000000);
\draw[very thick, black] (8.000000,1.750000) -- (8.250000,1.750000);
\draw[very thick, black] (8.250000,1.250000) -- (9.000000,1.250000);
\draw[very thick, black] (9.250000,1.000000) -- (9.250000,2.000000);
\draw[very thick, black] (9.000000,1.250000) -- (9.250000,1.250000);
\draw[very thick, black] (9.250000,1.750000) -- (10.000000,1.750000);
\draw[very thick, black] (10.000000,1.750000) -- (11.000000,1.750000);
\draw[very thick, black] (10.250000,2.000000) -- (10.250000,1.750000);
\draw[very thick, black] (10.750000,1.000000) -- (10.750000,1.750000);
\draw[very thick, black] (11.500000,1.000000) -- (11.500000,2.000000);
\draw[very thick, black] (11.000000,1.750000) -- (11.500000,1.750000);
\draw[very thick, black] (11.500000,1.500000) -- (12.000000,1.500000);
\draw[very thick, black] (0.000000,2.500000) -- (1.000000,2.500000);
\draw[very thick, black] (0.500000,3.000000) -- (0.500000,2.500000);
\draw[very thick, black] (0.750000,2.000000) -- (0.750000,2.500000);
\draw[very thick, black] (1.750000,2.000000) -- (1.750000,3.000000);
\draw[very thick, black] (1.000000,2.500000) -- (1.750000,2.500000);
\draw[very thick, black] (1.750000,2.750000) -- (2.000000,2.750000);
\draw[very thick, black] (2.250000,2.000000) -- (2.250000,3.000000);
\draw[very thick, black] (2.000000,2.750000) -- (2.250000,2.750000);
\draw[very thick, black] (2.250000,2.500000) -- (3.000000,2.500000);
\draw[very thick, black] (3.000000,2.500000) -- (4.000000,2.500000);
\draw[very thick, black] (3.750000,3.000000) -- (3.750000,2.500000);
\draw[very thick, black] (3.250000,2.000000) -- (3.250000,2.500000);
\draw[very thick, black] (4.000000,2.500000) -- (5.000000,2.500000);
\draw[very thick, black] (4.750000,3.000000) -- (4.750000,2.500000);
\draw[very thick, black] (4.250000,2.000000) -- (4.250000,2.500000);
\draw[very thick, black] (5.000000,2.500000) -- (6.000000,2.500000);
\draw[very thick, black] (5.250000,3.000000) -- (5.250000,2.500000);
\draw[very thick, black] (5.500000,2.000000) -- (5.500000,2.500000);
\draw[very thick, black] (6.000000,2.500000) -- (7.000000,2.500000);
\draw[very thick, black] (6.750000,3.000000) -- (6.750000,2.500000);
\draw[very thick, black] (6.250000,2.000000) -- (6.250000,2.500000);
\draw[very thick, black] (7.750000,2.000000) -- (7.750000,3.000000);
\draw[very thick, black] (7.000000,2.500000) -- (7.750000,2.500000);
\draw[very thick, black] (7.750000,2.250000) -- (8.000000,2.250000);
\draw[very thick, black] (8.250000,2.000000) -- (8.250000,3.000000);
\draw[very thick, black] (8.000000,2.250000) -- (8.250000,2.250000);
\draw[very thick, black] (8.250000,2.500000) -- (9.000000,2.500000);
\draw[very thick, black] (9.000000,2.500000) -- (10.000000,2.500000);
\draw[very thick, black] (9.750000,3.000000) -- (9.750000,2.500000);
\draw[very thick, black] (9.250000,2.000000) -- (9.250000,2.500000);
\draw[very thick, black] (10.000000,2.500000) -- (11.000000,2.500000);
\draw[very thick, black] (10.750000,3.000000) -- (10.750000,2.500000);
\draw[very thick, black] (10.250000,2.000000) -- (10.250000,2.500000);
\draw[very thick, black] (11.000000,2.500000) -- (12.000000,2.500000);
\draw[very thick, black] (11.750000,3.000000) -- (11.750000,2.500000);
\draw[very thick, black] (11.500000,2.000000) -- (11.500000,2.500000);
\draw[very thick, black] (0.000000,3.750000) -- (1.000000,3.750000);
\draw[very thick, black] (0.750000,4.000000) -- (0.750000,3.750000);
\draw[very thick, black] (0.500000,3.000000) -- (0.500000,3.750000);
\draw[very thick, black] (1.750000,3.000000) -- (1.750000,4.000000);
\draw[very thick, black] (1.000000,3.750000) -- (1.750000,3.750000);
\draw[very thick, black] (1.750000,3.500000) -- (2.000000,3.500000);
\draw[very thick, black] (2.000000,3.500000) -- (3.000000,3.500000);
\draw[very thick, black] (2.500000,4.000000) -- (2.500000,3.500000);
\draw[very thick, black] (2.250000,3.000000) -- (2.250000,3.500000);
\draw[very thick, black] (3.750000,3.000000) -- (3.750000,4.000000);
\draw[very thick, black] (3.000000,3.500000) -- (3.750000,3.500000);
\draw[very thick, black] (3.750000,3.250000) -- (4.000000,3.250000);
\draw[very thick, black] (4.750000,3.000000) -- (4.750000,4.000000);
\draw[very thick, black] (4.000000,3.250000) -- (4.750000,3.250000);
\draw[very thick, black] (4.750000,3.500000) -- (5.000000,3.500000);
\draw[very thick, black] (5.250000,3.000000) -- (5.250000,4.000000);
\draw[very thick, black] (5.000000,3.500000) -- (5.250000,3.500000);
\draw[very thick, black] (5.250000,3.250000) -- (6.000000,3.250000);
\draw[very thick, black] (6.750000,3.000000) -- (6.750000,4.000000);
\draw[very thick, black] (6.000000,3.250000) -- (6.750000,3.250000);
\draw[very thick, black] (6.750000,3.500000) -- (7.000000,3.500000);
\draw[very thick, black] (7.750000,3.000000) -- (7.750000,4.000000);
\draw[very thick, black] (7.000000,3.500000) -- (7.750000,3.500000);
\draw[very thick, black] (7.750000,3.250000) -- (8.000000,3.250000);
\draw[very thick, black] (8.000000,3.250000) -- (9.000000,3.250000);
\draw[very thick, black] (8.500000,4.000000) -- (8.500000,3.250000);
\draw[very thick, black] (8.250000,3.000000) -- (8.250000,3.250000);
\draw[very thick, black] (9.750000,3.000000) -- (9.750000,4.000000);
\draw[very thick, black] (9.000000,3.250000) -- (9.750000,3.250000);
\draw[very thick, black] (9.750000,3.500000) -- (10.000000,3.500000);
\draw[very thick, black] (10.000000,3.500000) -- (11.000000,3.500000);
\draw[very thick, black] (10.500000,4.000000) -- (10.500000,3.500000);
\draw[very thick, black] (10.750000,3.000000) -- (10.750000,3.500000);
\draw[very thick, black] (11.750000,3.000000) -- (11.750000,4.000000);
\draw[very thick, black] (11.000000,3.500000) -- (11.750000,3.500000);
\draw[very thick, black] (11.750000,3.250000) -- (12.000000,3.250000);
\draw[very thick, black] (0.750000,4.000000) -- (0.750000,5.000000);
\draw[very thick, black] (0.000000,4.250000) -- (0.750000,4.250000);
\draw[very thick, black] (0.750000,4.500000) -- (1.000000,4.500000);
\draw[very thick, black] (1.750000,4.000000) -- (1.750000,5.000000);
\draw[very thick, black] (1.000000,4.500000) -- (1.750000,4.500000);
\draw[very thick, black] (1.750000,4.250000) -- (2.000000,4.250000);
\draw[very thick, black] (2.500000,4.000000) -- (2.500000,5.000000);
\draw[very thick, black] (2.000000,4.250000) -- (2.500000,4.250000);
\draw[very thick, black] (2.500000,4.500000) -- (3.000000,4.500000);
\draw[very thick, black] (3.750000,4.000000) -- (3.750000,5.000000);
\draw[very thick, black] (3.000000,4.500000) -- (3.750000,4.500000);
\draw[very thick, black] (3.750000,4.750000) -- (4.000000,4.750000);
\draw[very thick, black] (4.750000,4.000000) -- (4.750000,5.000000);
\draw[very thick, black] (4.000000,4.750000) -- (4.750000,4.750000);
\draw[very thick, black] (4.750000,4.250000) -- (5.000000,4.250000);
\draw[very thick, black] (5.000000,4.250000) -- (6.000000,4.250000);
\draw[very thick, black] (5.500000,5.000000) -- (5.500000,4.250000);
\draw[very thick, black] (5.250000,4.000000) -- (5.250000,4.250000);
\draw[very thick, black] (6.750000,4.000000) -- (6.750000,5.000000);
\draw[very thick, black] (6.000000,4.250000) -- (6.750000,4.250000);
\draw[very thick, black] (6.750000,4.500000) -- (7.000000,4.500000);
\draw[very thick, black] (7.000000,4.500000) -- (8.000000,4.500000);
\draw[very thick, black] (7.250000,5.000000) -- (7.250000,4.500000);
\draw[very thick, black] (7.750000,4.000000) -- (7.750000,4.500000);
\draw[very thick, black] (8.500000,4.000000) -- (8.500000,5.000000);
\draw[very thick, black] (8.000000,4.500000) -- (8.500000,4.500000);
\draw[very thick, black] (8.500000,4.250000) -- (9.000000,4.250000);
\draw[very thick, black] (9.000000,4.250000) -- (10.000000,4.250000);
\draw[very thick, black] (9.500000,5.000000) -- (9.500000,4.250000);
\draw[very thick, black] (9.750000,4.000000) -- (9.750000,4.250000);
\draw[very thick, black] (10.500000,4.000000) -- (10.500000,5.000000);
\draw[very thick, black] (10.000000,4.250000) -- (10.500000,4.250000);
\draw[very thick, black] (10.500000,4.500000) -- (11.000000,4.500000);
\draw[very thick, black] (11.000000,4.500000) -- (12.000000,4.500000);
\draw[very thick, black] (11.250000,5.000000) -- (11.250000,4.500000);
\draw[very thick, black] (11.750000,4.000000) -- (11.750000,4.500000);
\draw[very thick, black] (0.000000,5.500000) -- (1.000000,5.500000);
\draw[very thick, black] (0.250000,6.000000) -- (0.250000,5.500000);
\draw[very thick, black] (0.750000,5.000000) -- (0.750000,5.500000);
\draw[very thick, black] (1.750000,5.000000) -- (1.750000,6.000000);
\draw[very thick, black] (1.000000,5.500000) -- (1.750000,5.500000);
\draw[very thick, black] (1.750000,5.250000) -- (2.000000,5.250000);
\draw[very thick, black] (2.000000,5.250000) -- (3.000000,5.250000);
\draw[very thick, black] (2.250000,6.000000) -- (2.250000,5.250000);
\draw[very thick, black] (2.500000,5.000000) -- (2.500000,5.250000);
\draw[very thick, black] (3.750000,5.000000) -- (3.750000,6.000000);
\draw[very thick, black] (3.000000,5.250000) -- (3.750000,5.250000);
\draw[very thick, black] (3.750000,5.750000) -- (4.000000,5.750000);
\draw[very thick, black] (4.000000,5.750000) -- (5.000000,5.750000);
\draw[very thick, black] (4.250000,6.000000) -- (4.250000,5.750000);
\draw[very thick, black] (4.750000,5.000000) -- (4.750000,5.750000);
\draw[very thick, black] (5.500000,5.000000) -- (5.500000,6.000000);
\draw[very thick, black] (5.000000,5.750000) -- (5.500000,5.750000);
\draw[very thick, black] (5.500000,5.250000) -- (6.000000,5.250000);
\draw[very thick, black] (6.750000,5.000000) -- (6.750000,6.000000);
\draw[very thick, black] (6.000000,5.250000) -- (6.750000,5.250000);
\draw[very thick, black] (6.750000,5.750000) -- (7.000000,5.750000);
\draw[very thick, black] (7.000000,5.750000) -- (8.000000,5.750000);
\draw[very thick, black] (7.500000,6.000000) -- (7.500000,5.750000);
\draw[very thick, black] (7.250000,5.000000) -- (7.250000,5.750000);
\draw[very thick, black] (8.500000,5.000000) -- (8.500000,6.000000);
\draw[very thick, black] (8.000000,5.750000) -- (8.500000,5.750000);
\draw[very thick, black] (8.500000,5.500000) -- (9.000000,5.500000);
\draw[very thick, black] (9.000000,5.500000) -- (10.000000,5.500000);
\draw[very thick, black] (9.250000,6.000000) -- (9.250000,5.500000);
\draw[very thick, black] (9.500000,5.000000) -- (9.500000,5.500000);
\draw[very thick, black] (10.000000,5.500000) -- (11.000000,5.500000);
\draw[very thick, black] (10.250000,6.000000) -- (10.250000,5.500000);
\draw[very thick, black] (10.500000,5.000000) -- (10.500000,5.500000);
\draw[very thick, black] (11.250000,5.000000) -- (11.250000,6.000000);
\draw[very thick, black] (11.000000,5.500000) -- (11.250000,5.500000);
\draw[very thick, black] (11.250000,5.250000) -- (12.000000,5.250000);
\draw[very thick, black] (0.250000,6.000000) -- (0.250000,7.000000);
\draw[very thick, black] (0.000000,6.500000) -- (0.250000,6.500000);
\draw[very thick, black] (0.250000,6.250000) -- (1.000000,6.250000);
\draw[very thick, black] (1.000000,6.250000) -- (2.000000,6.250000);
\draw[very thick, black] (1.250000,7.000000) -- (1.250000,6.250000);
\draw[very thick, black] (1.750000,6.000000) -- (1.750000,6.250000);
\draw[very thick, black] (2.000000,6.250000) -- (3.000000,6.250000);
\draw[very thick, black] (2.750000,7.000000) -- (2.750000,6.250000);
\draw[very thick, black] (2.250000,6.000000) -- (2.250000,6.250000);
\draw[very thick, black] (3.000000,6.250000) -- (4.000000,6.250000);
\draw[very thick, black] (3.500000,7.000000) -- (3.500000,6.250000);
\draw[very thick, black] (3.750000,6.000000) -- (3.750000,6.250000);
\draw[very thick, black] (4.250000,6.000000) -- (4.250000,7.000000);
\draw[very thick, black] (4.000000,6.250000) -- (4.250000,6.250000);
\draw[very thick, black] (4.250000,6.750000) -- (5.000000,6.750000);
\draw[very thick, black] (5.000000,6.750000) -- (6.000000,6.750000);
\draw[very thick, black] (5.750000,7.000000) -- (5.750000,6.750000);
\draw[very thick, black] (5.500000,6.000000) -- (5.500000,6.750000);
\draw[very thick, black] (6.750000,6.000000) -- (6.750000,7.000000);
\draw[very thick, black] (6.000000,6.750000) -- (6.750000,6.750000);
\draw[very thick, black] (6.750000,6.250000) -- (7.000000,6.250000);
\draw[very thick, black] (7.000000,6.250000) -- (8.000000,6.250000);
\draw[very thick, black] (7.750000,7.000000) -- (7.750000,6.250000);
\draw[very thick, black] (7.500000,6.000000) -- (7.500000,6.250000);
\draw[very thick, black] (8.000000,6.250000) -- (9.000000,6.250000);
\draw[very thick, black] (8.250000,7.000000) -- (8.250000,6.250000);
\draw[very thick, black] (8.500000,6.000000) -- (8.500000,6.250000);
\draw[very thick, black] (9.250000,6.000000) -- (9.250000,7.000000);
\draw[very thick, black] (9.000000,6.250000) -- (9.250000,6.250000);
\draw[very thick, black] (9.250000,6.750000) -- (10.000000,6.750000);
\draw[very thick, black] (10.000000,6.750000) -- (11.000000,6.750000);
\draw[very thick, black] (10.500000,7.000000) -- (10.500000,6.750000);
\draw[very thick, black] (10.250000,6.000000) -- (10.250000,6.750000);
\draw[very thick, black] (11.250000,6.000000) -- (11.250000,7.000000);
\draw[very thick, black] (11.000000,6.750000) -- (11.250000,6.750000);
\draw[very thick, black] (11.250000,6.500000) -- (12.000000,6.500000);
\draw[very thick, black] (0.000000,7.250000) -- (1.000000,7.250000);
\draw[very thick, black] (0.750000,8.000000) -- (0.750000,7.250000);
\draw[very thick, black] (0.250000,7.000000) -- (0.250000,7.250000);
\draw[very thick, black] (1.250000,7.000000) -- (1.250000,8.000000);
\draw[very thick, black] (1.000000,7.250000) -- (1.250000,7.250000);
\draw[very thick, black] (1.250000,7.500000) -- (2.000000,7.500000);
\draw[very thick, black] (2.750000,7.000000) -- (2.750000,8.000000);
\draw[very thick, black] (2.000000,7.500000) -- (2.750000,7.500000);
\draw[very thick, black] (2.750000,7.250000) -- (3.000000,7.250000);
\draw[very thick, black] (3.000000,7.250000) -- (4.000000,7.250000);
\draw[very thick, black] (3.750000,8.000000) -- (3.750000,7.250000);
\draw[very thick, black] (3.500000,7.000000) -- (3.500000,7.250000);
\draw[very thick, black] (4.250000,7.000000) -- (4.250000,8.000000);
\draw[very thick, black] (4.000000,7.250000) -- (4.250000,7.250000);
\draw[very thick, black] (4.250000,7.500000) -- (5.000000,7.500000);
\draw[very thick, black] (5.750000,7.000000) -- (5.750000,8.000000);
\draw[very thick, black] (5.000000,7.500000) -- (5.750000,7.500000);
\draw[very thick, black] (5.750000,7.750000) -- (6.000000,7.750000);
\draw[very thick, black] (6.000000,7.750000) -- (7.000000,7.750000);
\draw[very thick, black] (6.250000,8.000000) -- (6.250000,7.750000);
\draw[very thick, black] (6.750000,7.000000) -- (6.750000,7.750000);
\draw[very thick, black] (7.750000,7.000000) -- (7.750000,8.000000);
\draw[very thick, black] (7.000000,7.750000) -- (7.750000,7.750000);
\draw[very thick, black] (7.750000,7.250000) -- (8.000000,7.250000);
\draw[very thick, black] (8.000000,7.250000) -- (9.000000,7.250000);
\draw[very thick, black] (8.750000,8.000000) -- (8.750000,7.250000);
\draw[very thick, black] (8.250000,7.000000) -- (8.250000,7.250000);
\draw[very thick, black] (9.250000,7.000000) -- (9.250000,8.000000);
\draw[very thick, black] (9.000000,7.250000) -- (9.250000,7.250000);
\draw[very thick, black] (9.250000,7.750000) -- (10.000000,7.750000);
\draw[very thick, black] (10.000000,7.750000) -- (11.000000,7.750000);
\draw[very thick, black] (10.250000,8.000000) -- (10.250000,7.750000);
\draw[very thick, black] (10.500000,7.000000) -- (10.500000,7.750000);
\draw[very thick, black] (11.250000,7.000000) -- (11.250000,8.000000);
\draw[very thick, black] (11.000000,7.750000) -- (11.250000,7.750000);
\draw[very thick, black] (11.250000,7.250000) -- (12.000000,7.250000);
\draw[very thick, black] (0.750000,8.000000) -- (0.750000,9.000000);
\draw[very thick, black] (0.000000,8.250000) -- (0.750000,8.250000);
\draw[very thick, black] (0.750000,8.750000) -- (1.000000,8.750000);
\draw[very thick, black] (1.000000,8.750000) -- (2.000000,8.750000);
\draw[very thick, black] (1.750000,9.000000) -- (1.750000,8.750000);
\draw[very thick, black] (1.250000,8.000000) -- (1.250000,8.750000);
\draw[very thick, black] (2.750000,8.000000) -- (2.750000,9.000000);
\draw[very thick, black] (2.000000,8.750000) -- (2.750000,8.750000);
\draw[very thick, black] (2.750000,8.500000) -- (3.000000,8.500000);
\draw[very thick, black] (3.000000,8.500000) -- (4.000000,8.500000);
\draw[very thick, black] (3.500000,9.000000) -- (3.500000,8.500000);
\draw[very thick, black] (3.750000,8.000000) -- (3.750000,8.500000);
\draw[very thick, black] (4.250000,8.000000) -- (4.250000,9.000000);
\draw[very thick, black] (4.000000,8.500000) -- (4.250000,8.500000);
\draw[very thick, black] (4.250000,8.750000) -- (5.000000,8.750000);
\draw[very thick, black] (5.750000,8.000000) -- (5.750000,9.000000);
\draw[very thick, black] (5.000000,8.750000) -- (5.750000,8.750000);
\draw[very thick, black] (5.750000,8.250000) -- (6.000000,8.250000);
\draw[very thick, black] (6.000000,8.250000) -- (7.000000,8.250000);
\draw[very thick, black] (6.500000,9.000000) -- (6.500000,8.250000);
\draw[very thick, black] (6.250000,8.000000) -- (6.250000,8.250000);
\draw[very thick, black] (7.000000,8.250000) -- (8.000000,8.250000);
\draw[very thick, black] (7.500000,9.000000) -- (7.500000,8.250000);
\draw[very thick, black] (7.750000,8.000000) -- (7.750000,8.250000);
\draw[very thick, black] (8.750000,8.000000) -- (8.750000,9.000000);
\draw[very thick, black] (8.000000,8.250000) -- (8.750000,8.250000);
\draw[very thick, black] (8.750000,8.750000) -- (9.000000,8.750000);
\draw[very thick, black] (9.000000,8.750000) -- (10.000000,8.750000);
\draw[very thick, black] (9.750000,9.000000) -- (9.750000,8.750000);
\draw[very thick, black] (9.250000,8.000000) -- (9.250000,8.750000);
\draw[very thick, black] (10.000000,8.750000) -- (11.000000,8.750000);
\draw[very thick, black] (10.500000,9.000000) -- (10.500000,8.750000);
\draw[very thick, black] (10.250000,8.000000) -- (10.250000,8.750000);
\draw[very thick, black] (11.250000,8.000000) -- (11.250000,9.000000);
\draw[very thick, black] (11.000000,8.750000) -- (11.250000,8.750000);
\draw[very thick, black] (11.250000,8.250000) -- (12.000000,8.250000);
\draw[very thick, black] (0.000000,9.250000) -- (1.000000,9.250000);
\draw[very thick, black] (0.250000,10.000000) -- (0.250000,9.250000);
\draw[very thick, black] (0.750000,9.000000) -- (0.750000,9.250000);
\draw[very thick, black] (1.750000,9.000000) -- (1.750000,10.000000);
\draw[very thick, black] (1.000000,9.250000) -- (1.750000,9.250000);
\draw[very thick, black] (1.750000,9.750000) -- (2.000000,9.750000);
\draw[very thick, black] (2.000000,9.750000) -- (3.000000,9.750000);
\draw[very thick, black] (2.250000,10.000000) -- (2.250000,9.750000);
\draw[very thick, black] (2.750000,9.000000) -- (2.750000,9.750000);
\draw[very thick, black] (3.500000,9.000000) -- (3.500000,10.000000);
\draw[very thick, black] (3.000000,9.750000) -- (3.500000,9.750000);
\draw[very thick, black] (3.500000,9.500000) -- (4.000000,9.500000);
\draw[very thick, black] (4.000000,9.500000) -- (5.000000,9.500000);
\draw[very thick, black] (4.750000,10.000000) -- (4.750000,9.500000);
\draw[very thick, black] (4.250000,9.000000) -- (4.250000,9.500000);
\draw[very thick, black] (5.000000,9.500000) -- (6.000000,9.500000);
\draw[very thick, black] (5.250000,10.000000) -- (5.250000,9.500000);
\draw[very thick, black] (5.750000,9.000000) -- (5.750000,9.500000);
\draw[very thick, black] (6.000000,9.500000) -- (7.000000,9.500000);
\draw[very thick, black] (6.250000,10.000000) -- (6.250000,9.500000);
\draw[very thick, black] (6.500000,9.000000) -- (6.500000,9.500000);
\draw[very thick, black] (7.500000,9.000000) -- (7.500000,10.000000);
\draw[very thick, black] (7.000000,9.500000) -- (7.500000,9.500000);
\draw[very thick, black] (7.500000,9.250000) -- (8.000000,9.250000);
\draw[very thick, black] (8.000000,9.250000) -- (9.000000,9.250000);
\draw[very thick, black] (8.250000,10.000000) -- (8.250000,9.250000);
\draw[very thick, black] (8.750000,9.000000) -- (8.750000,9.250000);
\draw[very thick, black] (9.750000,9.000000) -- (9.750000,10.000000);
\draw[very thick, black] (9.000000,9.250000) -- (9.750000,9.250000);
\draw[very thick, black] (9.750000,9.500000) -- (10.000000,9.500000);
\draw[very thick, black] (10.000000,9.500000) -- (11.000000,9.500000);
\draw[very thick, black] (10.750000,10.000000) -- (10.750000,9.500000);
\draw[very thick, black] (10.500000,9.000000) -- (10.500000,9.500000);
\draw[very thick, black] (11.250000,9.000000) -- (11.250000,10.000000);
\draw[very thick, black] (11.000000,9.500000) -- (11.250000,9.500000);
\draw[very thick, black] (11.250000,9.750000) -- (12.000000,9.750000);
\draw[very thick, black] (0.000000,10.250000) -- (1.000000,10.250000);
\draw[very thick, black] (0.500000,11.000000) -- (0.500000,10.250000);
\draw[very thick, black] (0.250000,10.000000) -- (0.250000,10.250000);
\draw[very thick, black] (1.750000,10.000000) -- (1.750000,11.000000);
\draw[very thick, black] (1.000000,10.250000) -- (1.750000,10.250000);
\draw[very thick, black] (1.750000,10.500000) -- (2.000000,10.500000);
\draw[very thick, black] (2.250000,10.000000) -- (2.250000,11.000000);
\draw[very thick, black] (2.000000,10.500000) -- (2.250000,10.500000);
\draw[very thick, black] (2.250000,10.750000) -- (3.000000,10.750000);
\draw[very thick, black] (3.500000,10.000000) -- (3.500000,11.000000);
\draw[very thick, black] (3.000000,10.750000) -- (3.500000,10.750000);
\draw[very thick, black] (3.500000,10.500000) -- (4.000000,10.500000);
\draw[very thick, black] (4.750000,10.000000) -- (4.750000,11.000000);
\draw[very thick, black] (4.000000,10.500000) -- (4.750000,10.500000);
\draw[very thick, black] (4.750000,10.750000) -- (5.000000,10.750000);
\draw[very thick, black] (5.000000,10.750000) -- (6.000000,10.750000);
\draw[very thick, black] (5.500000,11.000000) -- (5.500000,10.750000);
\draw[very thick, black] (5.250000,10.000000) -- (5.250000,10.750000);
\draw[very thick, black] (6.000000,10.750000) -- (7.000000,10.750000);
\draw[very thick, black] (6.750000,11.000000) -- (6.750000,10.750000);
\draw[very thick, black] (6.250000,10.000000) -- (6.250000,10.750000);
\draw[very thick, black] (7.500000,10.000000) -- (7.500000,11.000000);
\draw[very thick, black] (7.000000,10.750000) -- (7.500000,10.750000);
\draw[very thick, black] (7.500000,10.250000) -- (8.000000,10.250000);
\draw[very thick, black] (8.250000,10.000000) -- (8.250000,11.000000);
\draw[very thick, black] (8.000000,10.250000) -- (8.250000,10.250000);
\draw[very thick, black] (8.250000,10.500000) -- (9.000000,10.500000);
\draw[very thick, black] (9.750000,10.000000) -- (9.750000,11.000000);
\draw[very thick, black] (9.000000,10.500000) -- (9.750000,10.500000);
\draw[very thick, black] (9.750000,10.750000) -- (10.000000,10.750000);
\draw[very thick, black] (10.750000,10.000000) -- (10.750000,11.000000);
\draw[very thick, black] (10.000000,10.750000) -- (10.750000,10.750000);
\draw[very thick, black] (10.750000,10.500000) -- (11.000000,10.500000);
\draw[very thick, black] (11.000000,10.500000) -- (12.000000,10.500000);
\draw[very thick, black] (11.500000,11.000000) -- (11.500000,10.500000);
\draw[very thick, black] (11.250000,10.000000) -- (11.250000,10.500000);
\draw[very thick, black] (0.000000,11.250000) -- (1.000000,11.250000);
\draw[very thick, black] (0.250000,12.000000) -- (0.250000,11.250000);
\draw[very thick, black] (0.500000,11.000000) -- (0.500000,11.250000);
\draw[very thick, black] (1.000000,11.250000) -- (2.000000,11.250000);
\draw[very thick, black] (1.500000,12.000000) -- (1.500000,11.250000);
\draw[very thick, black] (1.750000,11.000000) -- (1.750000,11.250000);
\draw[very thick, black] (2.000000,11.250000) -- (3.000000,11.250000);
\draw[very thick, black] (2.500000,12.000000) -- (2.500000,11.250000);
\draw[very thick, black] (2.250000,11.000000) -- (2.250000,11.250000);
\draw[very thick, black] (3.500000,11.000000) -- (3.500000,12.000000);
\draw[very thick, black] (3.000000,11.250000) -- (3.500000,11.250000);
\draw[very thick, black] (3.500000,11.500000) -- (4.000000,11.500000);
\draw[very thick, black] (4.000000,11.500000) -- (5.000000,11.500000);
\draw[very thick, black] (4.500000,12.000000) -- (4.500000,11.500000);
\draw[very thick, black] (4.750000,11.000000) -- (4.750000,11.500000);
\draw[very thick, black] (5.500000,11.000000) -- (5.500000,12.000000);
\draw[very thick, black] (5.000000,11.500000) -- (5.500000,11.500000);
\draw[very thick, black] (5.500000,11.750000) -- (6.000000,11.750000);
\draw[very thick, black] (6.000000,11.750000) -- (7.000000,11.750000);
\draw[very thick, black] (6.500000,12.000000) -- (6.500000,11.750000);
\draw[very thick, black] (6.750000,11.000000) -- (6.750000,11.750000);
\draw[very thick, black] (7.000000,11.750000) -- (8.000000,11.750000);
\draw[very thick, black] (7.750000,12.000000) -- (7.750000,11.750000);
\draw[very thick, black] (7.500000,11.000000) -- (7.500000,11.750000);
\draw[very thick, black] (8.000000,11.750000) -- (9.000000,11.750000);
\draw[very thick, black] (8.750000,12.000000) -- (8.750000,11.750000);
\draw[very thick, black] (8.250000,11.000000) -- (8.250000,11.750000);
\draw[very thick, black] (9.750000,11.000000) -- (9.750000,12.000000);
\draw[very thick, black] (9.000000,11.750000) -- (9.750000,11.750000);
\draw[very thick, black] (9.750000,11.500000) -- (10.000000,11.500000);
\draw[very thick, black] (10.000000,11.500000) -- (11.000000,11.500000);
\draw[very thick, black] (10.250000,12.000000) -- (10.250000,11.500000);
\draw[very thick, black] (10.750000,11.000000) -- (10.750000,11.500000);
\draw[very thick, black] (11.500000,11.000000) -- (11.500000,12.000000);
\draw[very thick, black] (11.000000,11.500000) -- (11.500000,11.500000);
\draw[very thick, black] (11.500000,11.250000) -- (12.000000,11.250000);
\draw[gray] (0,0) rectangle (12,12);
\end{tikzpicture}

%% file: border_constraint.tex
\begin{tikzpicture}[set style={{help lines}+=[dashed]}]
\draw (0,0) grid (10,5);

\draw[fill=red]   (0,0) -- (0.5,0.5) -- (1,0);
\draw[fill=blue]  (1,0) -- (1.5,0.5) -- (2,0);
\draw[fill=green] (2,0) -- (2.5,0.5) -- (3,0);
\draw[fill=green] (3,0) -- (3.5,0.5) -- (4,0);
\draw[fill=red]   (4,0) -- (4.5,0.5) -- (5,0);
\draw[fill=green] (5,0) -- (5.5,0.5) -- (6,0);
\draw[fill=blue]  (6,0) -- (6.5,0.5) -- (7,0);
\draw[fill=blue]  (7,0) -- (7.5,0.5) -- (8,0);
\draw[fill=red]   (8,0) -- (8.5,0.5) -- (9,0);
\draw[fill=blue]  (9,0) -- (9.5,0.5) -- (10,0);

\draw[fill=red]   (0,5) -- (0.5,4.5) -- (1,5);
\draw[fill=blue]  (1,5) -- (1.5,4.5) -- (2,5);
\draw[fill=green] (2,5) -- (2.5,4.5) -- (3,5);
\draw[fill=green] (3,5) -- (3.5,4.5) -- (4,5);
\draw[fill=red]   (4,5) -- (4.5,4.5) -- (5,5);
\draw[fill=green] (5,5) -- (5.5,4.5) -- (6,5);
\draw[fill=blue]  (6,5) -- (6.5,4.5) -- (7,5);
\draw[fill=blue]  (7,5) -- (7.5,4.5) -- (8,5);
\draw[fill=red]   (8,5) -- (8.5,4.5) -- (9,5);
\draw[fill=blue]  (9,5) -- (9.5,4.5) -- (10,5);

\draw[fill=blue]  (0,0) -- (0.5,0.5) -- (0,1);
\draw[fill=red]   (0,1) -- (0.5,1.5) -- (0,2);
\draw[fill=green] (0,2) -- (0.5,2.5) -- (0,3);
\draw[fill=red]   (0,3) -- (0.5,3.5) -- (0,4);
\draw[fill=green] (0,4) -- (0.5,4.5) -- (0,5);

\draw[fill=blue]  (10,0) -- (9.5,0.5) -- (10,1);
\draw[fill=red]   (10,1) -- (9.5,1.5) -- (10,2);
\draw[fill=green] (10,2) -- (9.5,2.5) -- (10,3);
\draw[fill=red]   (10,3) -- (9.5,3.5) -- (10,4);
\draw[fill=green] (10,4) -- (9.5,4.5) -- (10,5);

\foreach \i in {0,...,9}
{
	\foreach \j in {0,...,4}
	{
		\draw (\i,\j) -- (\i+1,\j+1);
		\draw (\i,\j+1) -- (\i+1,\j);		
	}
}
\end{tikzpicture}

%% file: row_last_2_vert.tex
\begin{tikzpicture}[set style={{help lines}+=[dashed]}, scale=1.5]
\draw[ultra thick] (0,0) -- (5,0) -- (5,1) -- (0,1);

\draw[ultra thick,fill=red!50] (1,0) rectangle (2,1);
\draw[ultra thick] (1.5,0) -- (1.5,1);
\draw[ultra thick,fill=red!50] (3,0) rectangle (4,1);
\draw[ultra thick] (3.5,0) -- (3.5,1);

\draw[dashed, ultra thick] (0.2, 0.5) -- (0.8, 0.5);
\draw[dashed, ultra thick] (2.2, 0.5) -- (2.8, 0.5);
\draw[dashed, ultra thick] (4.2, 0.5) -- (4.8, 0.5);

\draw [decorate,decoration={brace,amplitude=3pt,mirror},yshift=-2pt,xshift=0pt, ultra thick]
(2,0) -- (3,0);

\draw [decorate,decoration={brace,amplitude=3pt,mirror},yshift=-2pt,xshift=0pt, ultra thick]
(4,0) -- (5,0);

\draw[ultra thick,arrows={latex-latex}] (2.5, -0.2) -- (2.5, -0.5) -- (4.5, -0.5) node [midway, below] {Horizontal tiles} -- (4.5, -0.2);

\draw[ultra thick,arrows={latex-latex}] (1.5, 1.1) -- (1.5, 1.4) -- (3.5, 1.4) node [midway, above] {Last vertical tiles} -- (3.5, 1.1);

\node[xshift=11pt] at (1, 0.75) {$V_{sl,j}$};
\node[xshift=16pt] at (2, 0.75) {$V_{sl+1,j}$};

\node[xshift=10pt] at (3, 0.75) {$V_{l,j}$};
\node[xshift=15pt] at (4, 0.75) {$V_{l+1,j}$};

\node[xshift=5pt] at (5, 0.75) {$c$};

\node[below] at (1.5, 0) {$(sl,j)$};
\node[below] at (3.5, 0) {$(l,j)$};

\end{tikzpicture}